\documentclass[12pt]{article}
\usepackage{amssymb,amsmath}
\usepackage{epsfig}
\usepackage{epstopdf}
\usepackage{latexsym}
\usepackage{graphicx}
\usepackage{booktabs}

\usepackage[margin=20pt,small]{caption}
\usepackage{subcaption}

\usepackage[toc]{appendix}

\usepackage{color}
\usepackage{datetime}
\usepackage[
      colorlinks=false,
      linkcolor=darkblue,  
      urlcolor=blue,    
      filecolor=blue,     
      citecolor=red,
linktocpage=true,
      pdfstartview=FitV,
      bookmarksopen=true,
	  hidelinks
      ]{hyperref}

\ifpdf
\fi

\newcommand*{\boxedcolor}{red}
\makeatletter
\renewcommand{\boxed}[1]{\textcolor{\boxedcolor}{%
  \fbox{\normalcolor\m@th$\displaystyle#1$}}}
\makeatother

\definecolor{cardinal}{rgb}{0.6,0,0}
\definecolor{darkgreen}{rgb}{0,0.5,0}
\definecolor{golden}{rgb}{0.92, 0.7, 0}
\definecolor{midnight}{rgb}{0, 0, 0.5}
\definecolor{darkblue}{rgb}{0.2, 0, 0.8}

\def\be{\begin{equation}}
\def\ee{\end{equation}}

\begin{document}  

\begin{titlepage}

\begin{center}

{\Large \bf  The NUTs and Bolts of Squashed Holography}

\bigskip
\bigskip
\bigskip
\bigskip
\bigskip

{\bf Nikolay Bobev, Thomas Hertog, and Yannick Vreys \\ }
\bigskip
\bigskip
Institute for Theoretical Physics, KU Leuven \\
Celestijnenlaan 200D, B-3001 Leuven, Belgium

\bigskip
\bigskip

\texttt{nikolay.bobev,~thomas.hertog,~yannick.vreys~@kuleuven.be } \\
\end{center}

\bigskip
\bigskip
\bigskip
\bigskip

\begin{abstract}

\noindent  
\end{abstract}

\noindent We evaluate the partition function of the free $O(N)$ model on a two-parameter family of squashed three spheres. We also find new solutions of general relativity with negative cosmological constant and the same double squashed boundary geometry and analyse their thermodynamic properties. Remarkably, both systems exhibit a qualitatively similar behaviour over the entire configuration space of boundary geometries. Recent formulations of dS/CFT enable one to interpret the field theory partition function as a function of the two squashing parameters as the Hartle-Hawking wave function in a minisuperspace model of anisotropic deformations of de Sitter space. The resulting probability distribution is normalisable and globally peaked at the round three sphere, with a low amplitude of boundary geometries with negative scalar curvature.

\end{titlepage}


\setcounter{tocdepth}{2}
\tableofcontents

\section{Introduction}

Studying conformal field theories (CFTs) on curved Euclidean manifolds is a fruitful enterprise that has led to many interesting insights into the structure of quantum field theory. The metric of the manifold on which the CFT is defined can be viewed as a background field which naturally couples to the energy momentum tensor operator of the CFT. From this point of view deformations away from the flat metric have the potential to reveal universal properties in CFTs.

A different vantage point to CFTs on curved manifolds is offered by AdS/CFT. The boundary of AdS can be flat Euclidean space or the round sphere since the two are related by a conformal transformation. Any deformation of the boundary metric which takes it away from this conformally flat class gives rise to a new bulk geometry which differs from AdS. Gauge/gravity duality thus allows one to use classical general relativity in asymptotically locally AdS spaces to study CFTs on a range of curved backgrounds or, alternatively, to study aspects of quantum gravity by using dual CFTs defined on curved spaces.

A third application of field theories on curved manifolds arises from dS/CFT. The dS/CFT correspondence \cite{Balasubramanian2001,Strominger2001,Maldacena2002} relates the Hartle-Hawking wave function of the universe \cite{Hartle:1983ai} in asymptotically de Sitter (dS) space to the partition function of deformations of Euclidean CFTs defined on the future boundary. In this context, the values of the sources in the deformation of the CFT correspond to the argument of the asymptotic wave function. Partition functions of CFTs defined on deformed, i.e. non-conformally flat, boundary geometries can thus be used to evaluate the wave function on histories that differ from dS, including histories that are initially singular.

In this paper we consider CFTs and their holographic duals on a two-parameter family of squashed three spheres whose metric can be written as,
\begin{align}
ds^2= \frac{r_0^2}{4} \left((\sigma_1)^2 +\frac{1}{1+\beta} (\sigma_2)^2 + \frac{1}{1+\alpha}(\sigma_3)^2 \right) 
\label{eqn:metric}\;,
\end{align}
where $r_0$ is an overall radius for which we choose the normalisation $r_0=1$, and $\sigma_i$, with $i=1,2,3$, are the left-invariant one-forms of $SU(2)$ given by
\begin{equation}\label{eqn:left1forms}
 \sigma_1=-\sin\psi d\theta+\cos\psi \sin \theta d\phi \ , \qquad
 \sigma_2=\cos\psi d\theta+\sin\psi \sin \theta d\phi \ ,\qquad
 \sigma_3=d\psi+\cos \theta d\phi \ ,
\end{equation} 
with $0\leq \theta\leq \pi$, $0\leq \phi \leq 2\pi$ and $0\leq \psi \leq 4\pi$. In particular we evaluate the partition function of the free $O(N)$ vector model as a function of the two squashing parameters $\alpha$ and $\beta$ in \eqref{eqn:metric}. In the limit when one of the squashing parameters vanishes we recover the results of \cite{Anninos:2012ft}.

An interesting CFT to study is the three-dimensional $O(N)$ vector model, for which it is well-known that it is dual to Vasiliev higher-spin gravity in $AdS_4$ \cite{Sezgin:2002rt,Klebanov:2002ja,Giombi:2009wh}. It was further conjectured in \cite{Anninos2011} that the dual of higher-spin gravity in $dS_4$ is given by the non-unitarity $Sp(N)$ vector model which can be thought of as being obtained from the $O(N)$ model via an analytic continuation in $N$. Here we will not consider higher-spin gravitational theories directly. Instead we will aim for a qualitative comparison between the physics of the $O(N)$ model on the squashed sphere in \eqref{eqn:metric} and Einstein gravity with either AdS or dS boundary conditions. To do so we first numerically construct new solutions of general relativity with a negative cosmological constant that are everywhere regular and have a double squashed sphere of the form \eqref{eqn:metric} as their boundary. Our solutions are generalisations of the well known AdS Taub-NUT and Taub-Bolt solutions \cite{Taub1951, Newman1963}. Comparing the thermodynamic properties of these new solutions with the partition function of the free $O(N)$ model as a function of the two squashing parameters $\alpha$ and $\beta$ we find that both systems exhibit a qualitatively similar behaviour over the entire configuration space of boundary geometries. On the other hand they differ in specific features such as the NUT to Bolt transition at large positive values of the squashing parameters, which is evidently absent in the free dual theory.   
 
In the context of dS/CFT the squashed spheres \eqref{eqn:metric} enter as the future boundary of homogeneous but anisotropic deformations of de Sitter space. Through dS/CFT the partition function of the free $O(N)$ model as a function of $\alpha$ and $\beta$ provides a toy model calculation of the Hartle-Hawking wave function in a minisuperspace model consisting of this set of anisotropic cosmologies. We compute the probability distribution over histories in this model, which turns out to be normalisable and globally peaked at the round three sphere. 

\begin{figure}[ht!]
\centering
\includegraphics[width=0.45\textwidth]{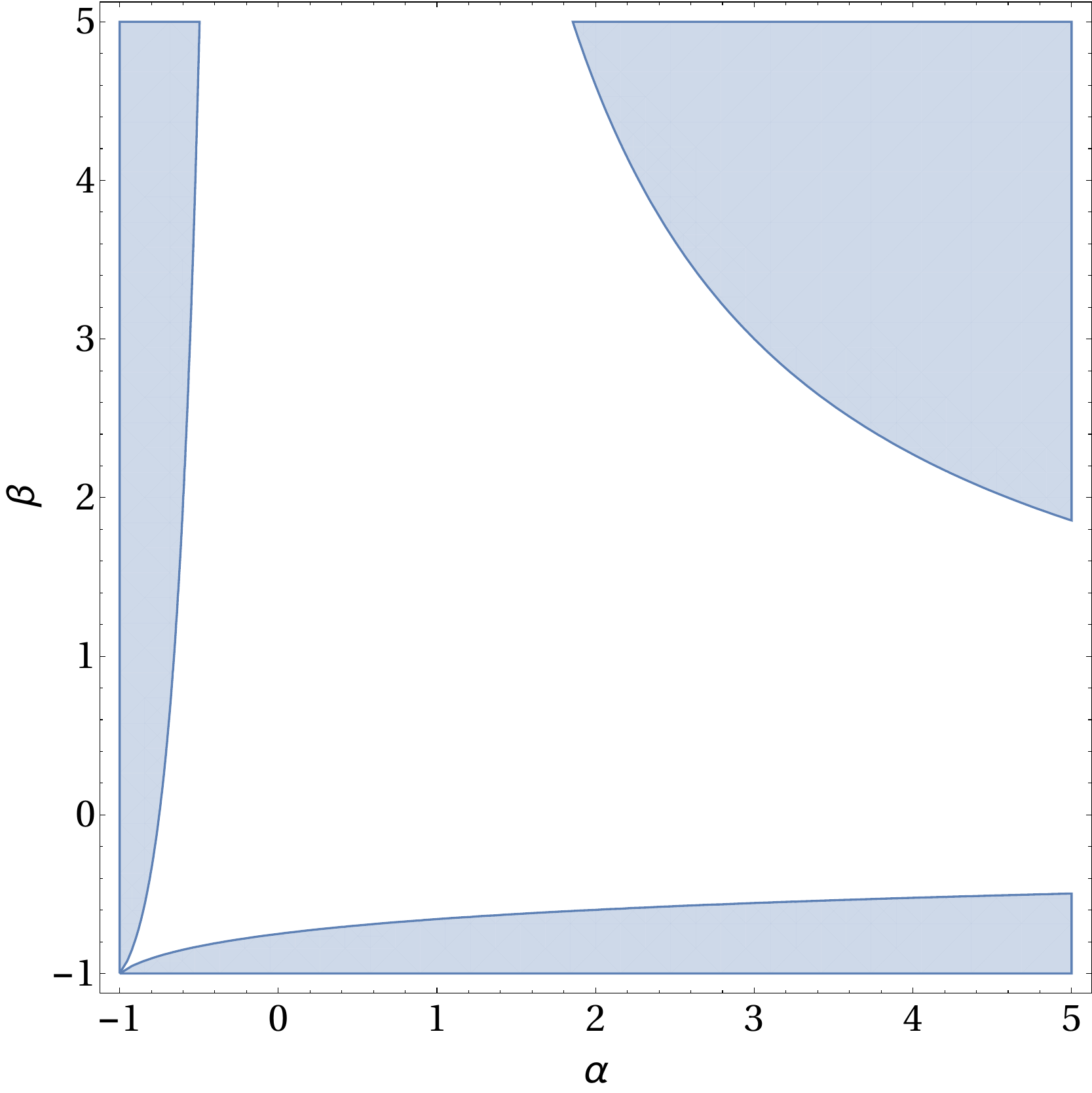}
\caption{The shaded blue region is where the Ricci scalar, $R$, becomes negative. For one squashing there is only one place where $R$ changes sign, but away from these points there are for given $\alpha$ or $\beta$ two regions where $R$ is negative.}\label{fig:ricci}
\end{figure}

The region of superspace where the Ricci scalar on the boundary is negative is particularly intriguing. In general the Ricci scalar of a double squashed three sphere of the form \eqref{eqn:metric} is given by
\begin{align}
 R=\frac{6+8\alpha+8\beta+2\alpha \beta (6-\alpha \beta)}{(1+\alpha)(1+\beta)},
 \label{Ricci}
\end{align}
which is symmetric in $\alpha$ and $\beta$. For $\beta=0$ there is a single region $\alpha < -3/4$ where $R$ is negative. Adding a second squashing, however, leads to an additional $R<0$ region associated with large positive values of both $\alpha$ and $\beta$. This is illustrated in Figure \ref{fig:ricci}. In the minisuperspace model we consider in this paper the overall amplitude of boundary geometries with a negative scalar curvature is exponentially small. However this dramatically changes when other directions in superspace are included, raising new questions about the normalisability of the wave function. We return to this point below in Section \ref{sec:Cosmology}.

Before we proceed with the details of our analysis, it is worth mentioning that supersymmetric CFTs on the squashed $S^3$ have been studied recently in the context of supersymmetric localisation \cite{Hama:2011ea,Imamura:2011uw,Imamura:2011wg} (see also \cite{Closset:2012ru,Nishioka:2013gza} for some applications of these results as well as \cite{Martelli:2011fu,Martelli:2011fw,Martelli:2012sz} for holographic studies in this context). We emphasize that in this approach, in order to preserve supersymmetry, one has to turn on background gauge fields for the R-symmetry in addition to the curved metric. This differs from our construction here since we are not concerned with supersymmetric theories and thus in general we do not have an R-symmetry at our disposal and the curved metric in \eqref{eqn:metric} is the only background field. In particular our gravitational solutions are different from the supergravity backgrounds constructed in \cite{Martelli:2013aqa}, since those solutions are supersymmetric and have non-trivial electromagnetic fields.

Our paper is organised as follows. After a brief review of the class of known solutions of general relativity with a single squashed $S^3$ boundary we find new solutions with a double squashed $S^3$ boundary in Section \ref{sec:AdS}. In Section \ref{sec:TBTD} we study the thermodynamic properties of these new solutions. In Section \ref{sec:CFT} we switch gears and analyse the free $O(N)$ model on the double squashed $S^3$ and compare its behaviour with the thermodynamics of our new gravity solutions. By a simple analytic continuation in $N$ our results also apply to the non-unitary $Sp(N)$ model and are therefore applicable in the context of dS/CFT. Using this we compute the Hartle-Hawking wave function in anisotropic minisuperspace in Section \ref{sec:Cosmology}. We conclude with a discussion and some avenues for future work in Section \ref{sec:discussion}. The two appendices are devoted to a summary of the technical aspects of the calculation leading to the spectrum of the Laplacian on the squashed $S^3$ and to some results on the linearised expansion of the equations of motion of general relativity.

\section{Double squashed AdS Taub-NUT/Bolt}
\label{sec:AdS}

In this section we discuss four-dimensional solutions of general relativity with a negative cosmological constant with the action
\begin{equation}\label{eqn:GRaction}
S=-\frac{1}{16\pi G} \int_{\mathcal{M}}d^4x \sqrt{g} (R-2\Lambda)\;,
\end{equation}
 that have a squashed sphere as their asymptotic boundary. We work exclusively with Euclidean signature. In the case of one squashing, i.e. $\beta = 0$ in \eqref{eqn:metric}, these backgrounds are well-known and can be thought of as extensions of the usual asymptotically flat Taub-NUT and Taub-Bolt solutions \cite{Taub1951, Newman1963}.  We review these backgrounds in Section \ref{subsec:AdSTNB} and present their generalisation in Section \ref{subsec:AdS2squash}.

\subsection{Single squashed AdS Taub-NUT/Bolt}
\label{subsec:AdSTNB}

The AdS Taub-NUT/Bolt solutions are a family of solutions that are asymptotically AdS for which the metric is given by \cite{Emparan:1999pm} (see also \cite{Stephani:2003tm})
\begin{align}\label{eqn:metricTaubNUTBolt}
ds^2 = 4n^2V(\rho) (\sigma_3)^2 + \frac{d\rho^2}{V(\rho)} +(\rho^2 -n^2)(\sigma_1^2+\sigma_2^2) \ , 
\end{align}
where the one-forms $\sigma_i$ are defined in \eqref{eqn:left1forms} and $V$ is given by
\begin{align}
V\equiv \frac{ (\rho^2+n^2)- 2m \rho + l^{-2} (\rho^4-6n^2\rho^2-3n^4)}{\rho^2-n^2} \ ,
\end{align}
with $n$ denoting the NUT charge, $m$ the generalised mass and $l$ the AdS length scale ($l^2=-\Lambda/3$). The asymptotic behaviour for $\rho \to \infty$ of the metric in \eqref{eqn:metricTaubNUTBolt} is locally the same as the one for $AdS_4$. The only difference being that the boundary is a squashed $S^3$ with a single squashing parameter. Comparing the boundary metric with the metric on the squashed sphere in \eqref{eqn:metric} one finds that 
\begin{align}
\frac{1}{4(1+\alpha)}= \frac{ n^2}{l^2}\;, \qquad\qquad \beta = 0\, .
\end{align}

There are now two subclasses of topologically distinct solutions. The first set consists of the NUT solutions, which are defined by requiring that there is a zero dimensional fixed point set. Furthermore the Dirac-Misner string should be unobservable and there should be no conical defect around $\rho=n$. These requirements restrict the mass parameter $m$ to be \cite{Misner1963,Emparan:1999pm}
\begin{equation}\label{eqn:mNUT}
m_n=n-\frac{4 n^3}{l^2} \ .
\end{equation}
This mass parameter makes the space around $\rho=n$ look like the origin of a smooth $\mathbb{R}^4$. Notice that there is a special case when we put $n=l/2$, the squashing at the boundary disappears and we recover the usual $AdS_4$ space. Appropriately in this case we find that the mass $m_n$ in \eqref{eqn:mNUT} vanishes.

The second set of solutions is called the Bolt solutions, these are characterised by a two dimensional fixed point set. This is achieved by requiring that there is a Bolt, i.e. a topological $S^2$, at $\rho=\rho_b >n$, and no conical singularities. These two conditions lead to the following identities
\begin{align}
V(\rho_b)=0, \qquad V'(\rho_b)= \frac{1}{2n} \ .
\label{eqn:Boltcond}
\end{align}
From the first condition in \eqref{eqn:Boltcond} one finds that the mass of the Bolt should satisfy
\begin{align}
m_b= \frac{\rho_b^2+n^2}{2\rho_b} + \frac{1}{2l^2} \left( \rho_b^3-6n^2 \rho_b- 3\frac{n^4}{\rho_b} \right) \ . \label{eqn:massBolt}
\end{align}
The second condition in \eqref{eqn:Boltcond} yields a relation between $\rho_b$ and $n$ and $l$:
\begin{align}
\rho_{b\pm} =\frac{l^2}{12n} \left(1\pm \sqrt{1-48 \frac{n^2}{l^2}+144\frac{n^4}{l^4}}\right)\ . \label{eqn:radiusBolt}
\end{align}
Therefore there are two branches of real solutions, if the discriminant is positive. This positiviy condition combined with the requirement that $\rho_b>n$ restricts the existence of the Bolt solutions to the region of parameter space where \footnote{Expressed in terms of the parameter $\alpha$ this range is $\alpha > \alpha_{\text{crit}} = 5+3\sqrt{3}$.}
\begin{equation}
0< \frac{n^2}{l^2} \leq \left(\frac{n^2}{l^2}\right)_{\text{crit}}  = \frac{2-\sqrt{3}}{12}\approx 0.089\;.
\end{equation}
In particular this implies that the AdS Taub-Bolt solutions do not exist in the same region of parameter space with Euclidean $AdS_4$ which is obtained by setting $n=l/2$.

There is a Hawking-Page-type topology changing phase transition which occurs at\footnote{In terms of the squashing parameter $\alpha$ this value is at $\alpha_{\text{HP}} = 6+2\sqrt{10}$.}
\begin{equation}
\left(\frac{n^2}{l^2}\right)_{\text{HP}} =\frac{1}{28+8\sqrt{10}}\approx 0.0188\;.
\end{equation}
For values of $n^2/l^2$ lower than this critical value the Taub-Bolt solution is the dominant one, whereas for larger values of  $n^2/l^2$ the Taub-NUT solution is thermodynamically preferred. The thermodynamic properties of these solutions are reviewed in more detail in Section \ref{subsec:AdSTNB-therom} below.

\subsection{Double squashings}
\label{subsec:AdS2squash}

Having reviewed the well-known AdS-Taub-NUT and Taub-Bolt solutions we are ready to tackle the more serious problem of finding asymptotically locally $AdS_4$ backgrounds which have the same NUT/Bolt topology but asymptote to the squashed sphere in \eqref{eqn:metric} with two non-vanishing squashing parameters.

We start by imposing a metric Ansatz compatible with the isometries of the sphere at the asymptotic boundary\footnote{We assume that the full four-dimensional solutions have the same isometries as the asymptotic boundary.}  
\begin{equation}\label{eqn:Ansatz2sq}
ds^2= l_0(r)dr^2 +l_1(r) \sigma_1^2+ l_2(r) \sigma_2^2+l_3(r) \sigma_3^2   \ .
\end{equation}
We then plug this Ansatz into the equations of motion derived from the action in \eqref{eqn:GRaction} and derive a system of non-linear second order differential equations for the metric functions $l_a(r)$. Since we were not able to solve these equations analytically we resorted to a perturbative analysis near the asymptotic boundary and near the NUT/Bolt locus. In addition to that we exhibit numerical solutions which interpolate between these two asymptotic regions. Some technical details pertaining to this analysis are presented in Appendix \ref{App:AdSexpansions}.

To avoid confusion we emphasize that in order to construct numerical solutions to the equations of motion we have found it easier to  choose a different gauge, $l_0(r)=1$, for the radial coordinate as compared to the analytic solution in \eqref{eqn:metricTaubNUTBolt}. We adopt this gauge for most of the following discussion.

\subsubsection{UV expansion}

We start by considering an expansion at large values of $r$ which, employing holographic terminology, we call UV expansion. The UV expansion will be the same for both the NUT and Bolt solutions since in both cases the non-trivial information is encoded in the interior of the solutions, i.e. the IR. The UV expansion is of the standard Fefferman-Graham type and thus we are dealing with asymptotically locally $AdS_4$ solutions.  The leading order terms in the metric for large $r$ are given by 
\begin{equation}
ds^2= dr^2 + e^{2r} \left(A_0 \sigma_1^2+ B_0 \sigma_2^2+ C_0 \sigma_3^2 \right) \label{eqn:UVmetric} \ .
\end{equation}
Notice that we have implemented the gauge $l_0(r)=1$ and from now on we fix the cosmological constant to be $\Lambda=-3$ or alternatively the $AdS_4$ length scale $l=1$. Taking this as the starting point the UV expansion at large $r$ of the metric functions takes the form
\begin{equation}\label{eqn:genUV}
\begin{split} 
l_1(r)=A_0 e^{2r} + A_k e^{(2-k) r} \;, \qquad 
l_2(r)=B_0 e^{2r} + B_k e^{(2-k) r} \;, \qquad
l_3(r)=C_0 e^{2r} + C_k e^{(2-k) r} \;, 
\end{split}
\end{equation}
where the sum over $k$ goes over all positive integers. 

With this Ansatz at hand one can plug the series expansion \eqref{eqn:genUV} into the Einstein equations  and solve them order by order in powers of $e^r$. The results of this procedure are summarised in Appendix \ref{App:AdSexpansions}, see Equation \eqref{eqn:UVexpansion}. The important upshot of this analysis is that the UV expansion is controlled by five independent parameters $\{A_0,B_0,C_0,A_3,B_3\}$. It turns out that the Einstein equations are invariant under constant shifts of $r$ and we can use this freedom to eliminate one of the five parameters. We make the choice 
\begin{equation}
A_0=\frac{1}{4} \ .
\end{equation}
Comparing the asymptotic form of the metric with the metric \eqref{eqn:metric} on the double squashed sphere one can find the following relation between the squashing parameters $\alpha$ and $\beta$ and the leading order coefficients $B_0$ and $C_0$
\begin{equation}\label{eqn:abBC}
\alpha = \frac{1}{4C_0} - 1\;, \qquad\qquad \beta = \frac{1}{4B_0} - 1\;.
\end{equation}
The remaining independent subleading coefficients, $A_3$ and $B_3$, remain undetermined. As we discuss in Appendix \ref{App:AdSexpansions} their values are ultimately fixed by imposing regularity conditions (either a NUT or a Bolt) in the bulk of the full solution of the nonlinear equations of motion. To understand how to do this we now move on to the analysis of the two possible regular IR expansions.

\subsubsection{NUTs}
\label{subsubsec:NUTs}

From the analytic AdS-Taub-NUT solution presented in Section \ref{subsec:AdSTNB} we know that close to the NUT locus the space should look like the origin of $\mathbb{R}^4$ without any conical singularity.  We will impose the same condition when looking for new solutions. However in our IR expansion we will not impose any other restrictions on the metric functions $l_i(r)$, i.e.  we will look for solutions where all $l_i$ are distinct. If we take $r^*$ to indicate the location of the NUT, the metric around this point, to leading order in $(r-r^*)$, should be
\begin{align}
 ds^2= dr^2+\frac{(r-r^*)^2}{4}(\sigma_1^2+\sigma^2_2+\sigma_3^2)\ .
\end{align}
This defines the first order terms of the IR expansion.  The gauge choice, $l_0(r)=1$, that we used in the UV expansion is globally well-defined and is already implemented in the above metric.  The Ansatz for the IR expansion then becomes
\begin{equation}\label{eqn:genIRNUT}
\begin{split}
 l_1(r)&=\frac{1}{4}( r-r^*)^2  + \beta_{k+2} ( r-r^*)^{k+2}\;, \\
 l_2(r)&=\frac{1}{4}( r-r^*)^2  + \gamma_{k+2} ( r-r^*)^{k+2} \;, \\
 l_3(r)&=\frac{1}{4}( r-r^*)^2  + \delta_{k+2} ( r-r^*)^{k+2}\;, 
\end{split}
\end{equation}
where $k$ runs from $1$ to $\infty$. Plugging this into the Einstein equations and solving them order by order in $(r-r^*)$ leads to a consistent series solution which is controlled by two real constants $\gamma_4$ and $\beta_4$. All other constants $\beta_k$, $\gamma_k$ and $\delta_k$, can be expressed in terms of these two parameters. In particular we find that all coefficients of odd powers of  $(r-r^*)$ vanish. The parameter $r^*$ is spurious since it can be shifted to any value by a constant shift of the radial variable $r$.

The IR expansion in \eqref{eqn:genIRNUT} can be used as an initial condition to integrate the equations of motion numerically to the UV. We will thus get a two-parameter family of solutions that are controlled by $\gamma_4$ and $\beta_4$. There are two distinct classes of solutions. The first class consists of regular solutions for which the metric functions $l_{1,2,3}(r)$ grow exponentially and the boundary metric is a sphere with two non-trivial squashing parameters as in \eqref{eqn:metric}. We also find a class of singular solutions for which one or more of the metric functions $l_{1,2,3}(r)$ vanish at some finite value of $r$ which leads to a curvature singularity. We will ignore the second class of solutions since they do not seem to be of physical relevance.

\begin{figure}[ht!]
\centering
\includegraphics[width=0.45\textwidth]{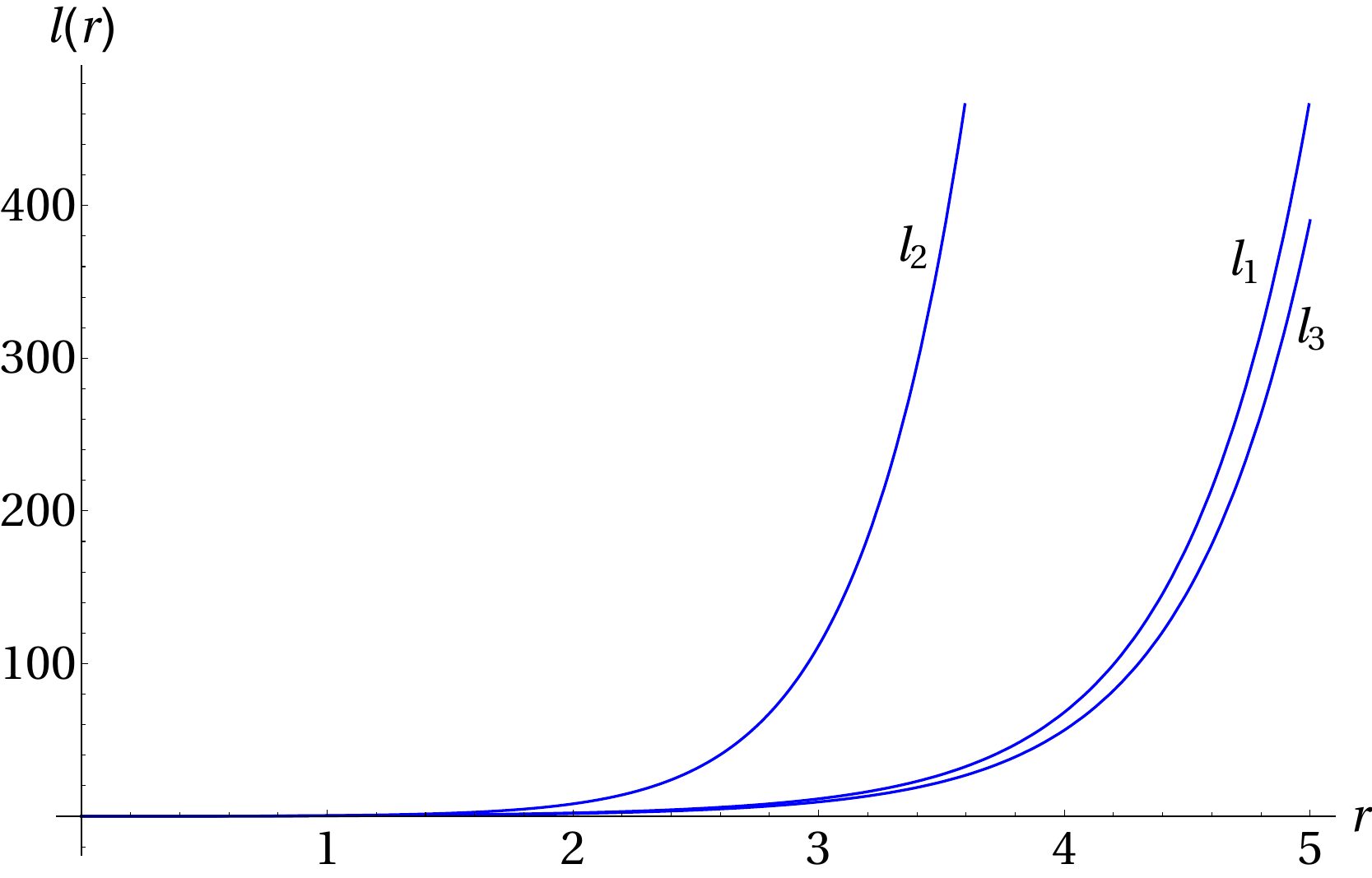}
\includegraphics[width=0.45\textwidth]{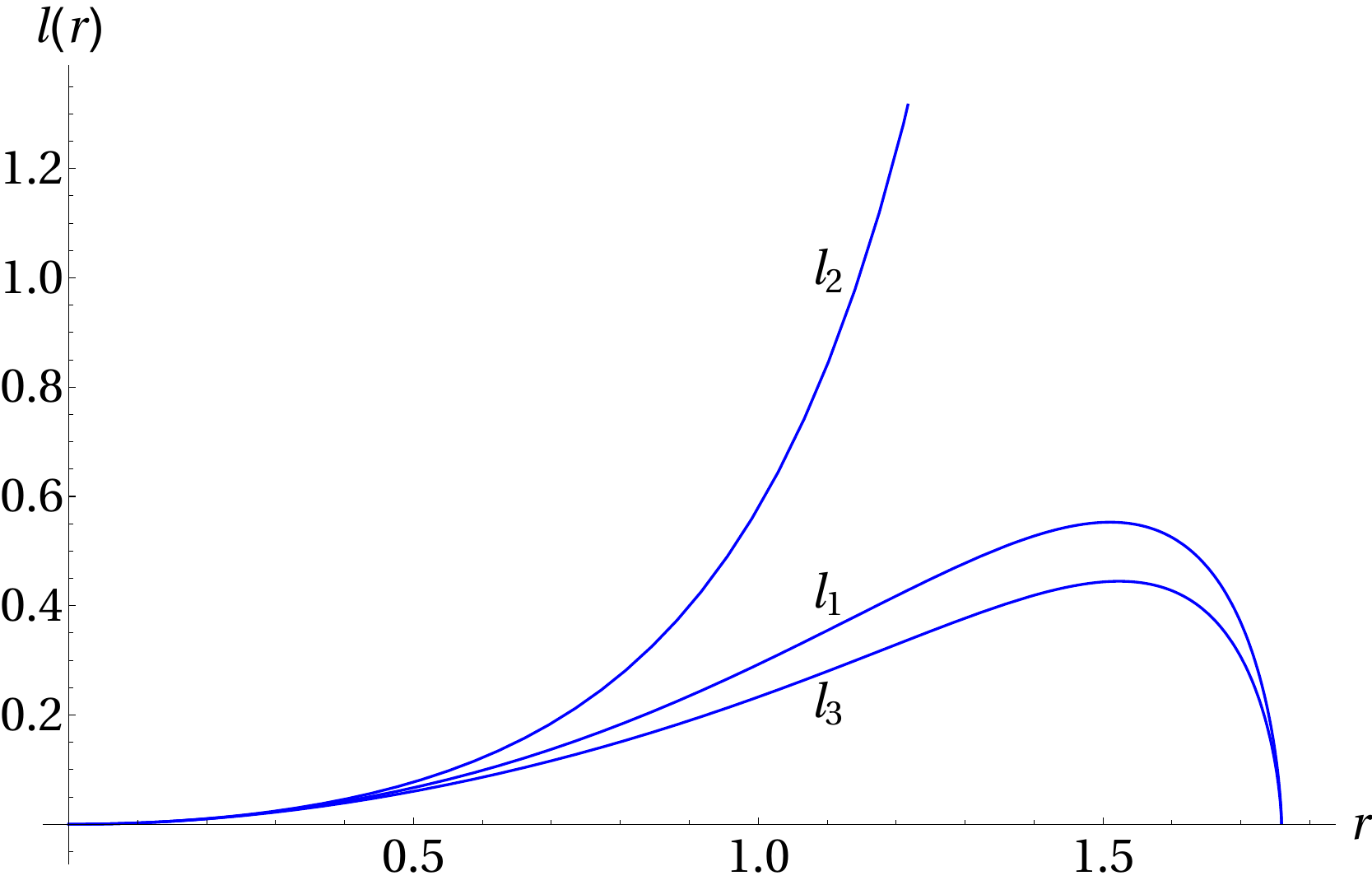}
\caption{Two typical solutions with a NUT in the IR. Left: a solution with $\beta_4=1/12$ and $\gamma_4=1/6$ for which all the $l_i$ keep on growing exponentially. This is an AdS-Taub-NUT solution with a double squashed sphere boundary geometry. Right: a solution with $\beta_4=1/12$ and $\gamma_4 = 3/14$, for which there is a singularity where $l_1(r)=l_3(r)=0$ at a finite value of $r$. Here we have chosen $r^*=0$.}\label{fig:genericSolutions}
\end{figure}

Two representative examples of these classes of solutions are shown in Figure \ref{fig:genericSolutions}. In the right panel of the figure we present an example of a singular solution. In the left panel of the figure is an example of a regular AdS-Taub-NUT solution with two different squashing parameters of the boundary $S^3$. We have constructed numerous such numerical solutions and we will analyse their properties in Section \ref{sec:TBTD} and Section \ref{sec:CFT}. For more details on the construction of these solutions we refer to Appendix \ref{App:AdSexpansions}. 

 \subsubsection{Bolts}
 
To find generalisations of the AdS-Taub-Bolt solutions from Section \ref{subsec:AdSTNB} we have to impose that there is a two dimensional fixed point set of the Killing vector $\partial_{\psi}$. We take this locus to be at $r=r^*$. The geometry in the neighbourhood of $r^*$ is determined by a metric on $\mathbb{R}^2\times S^2$. Therefore the metric around $r^*$ should take the form
\begin{align}
  ds^2= dr^2+\frac{(r-r^*)^2}{4}\sigma_3^2 +\beta_0\sigma_1^2+\gamma_0 \sigma^2_2\ .
\end{align}
We have again implemented the gauge choice $l_0(r)=1$. Our Ansatz for the IR expansion of the Bolt solutions thus becomes
\begin{equation}\label{eqn:genIRBolt}
\begin{split}
 l_1(r)&=\beta_0 + \beta_k ( r-r^*)^k  \;, \\
 l_2(r)&=\gamma_0 + \gamma_k ( r-r^*)^k\;, \\
 l_3(r)&=\frac{1}{4}(r-r^*)^2  + \delta_{k+2} ( r-r^*)^{k+2} \;,
\end{split}
\end{equation}
where the integer $k$ goes from $1$ to $\infty$. With this at hand we proceed as before. We substitute the expansion \eqref{eqn:genIRBolt} in the Einstein equations and solve them order by order in $(r-r^*)$. The details of this procedure are summarised in Appendix \ref{App:AdSexpansions}, see in particular equation \eqref{eqn:IRBoltexpansion}. The leading order analysis leads to $\beta_0=\gamma_0$ and one finds non-vanishing coefficients only for the even powers of $(r-r^*)$. The rest of the expansion coefficients in \eqref{eqn:genIRBolt} are determined in terms of two parameters which can be chosen to be $\gamma_0$ and $\gamma_4$.

\begin{figure}[ht!]
\centering
\includegraphics[width=0.45\textwidth]{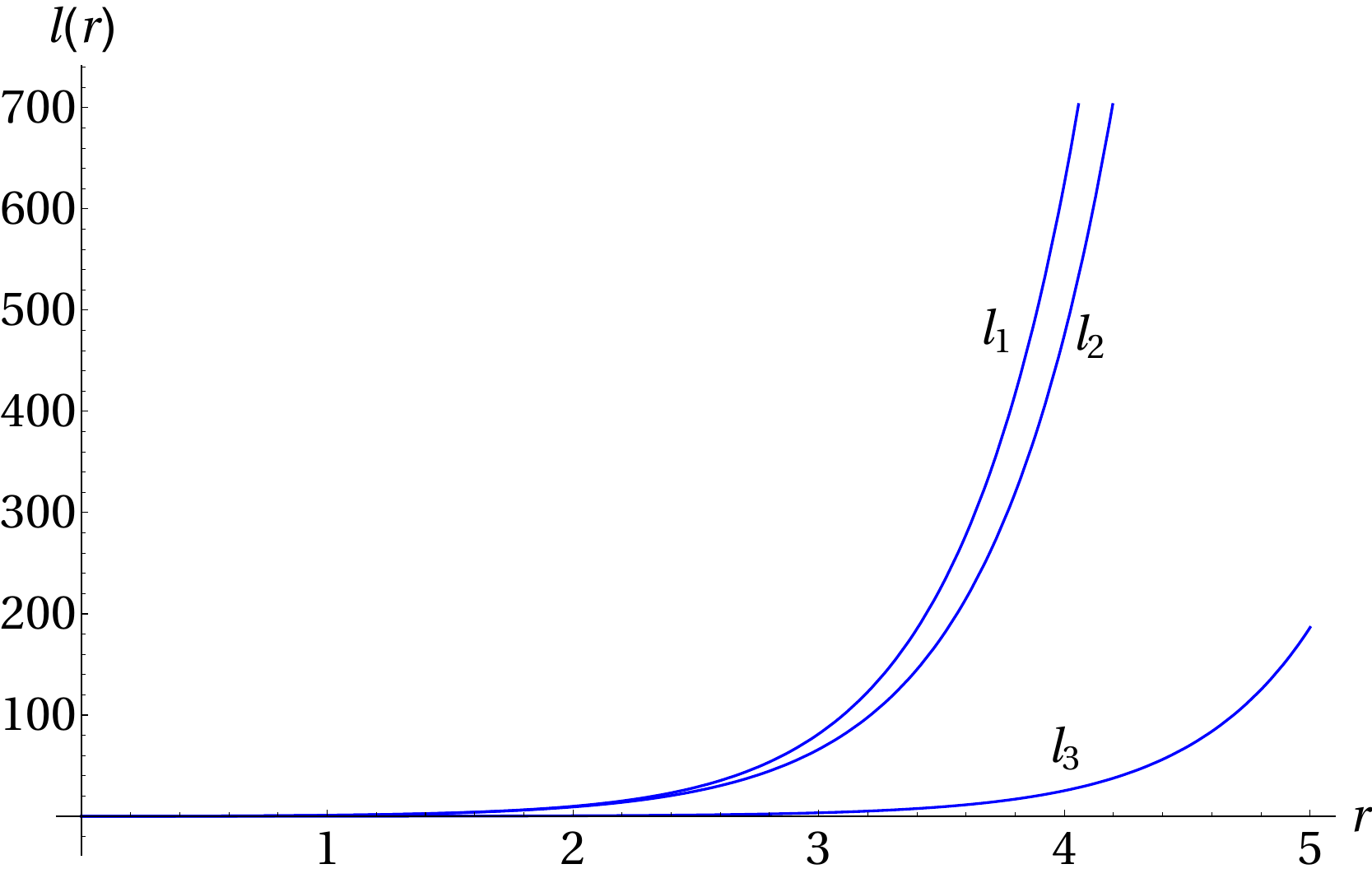}
\includegraphics[width=0.45\textwidth]{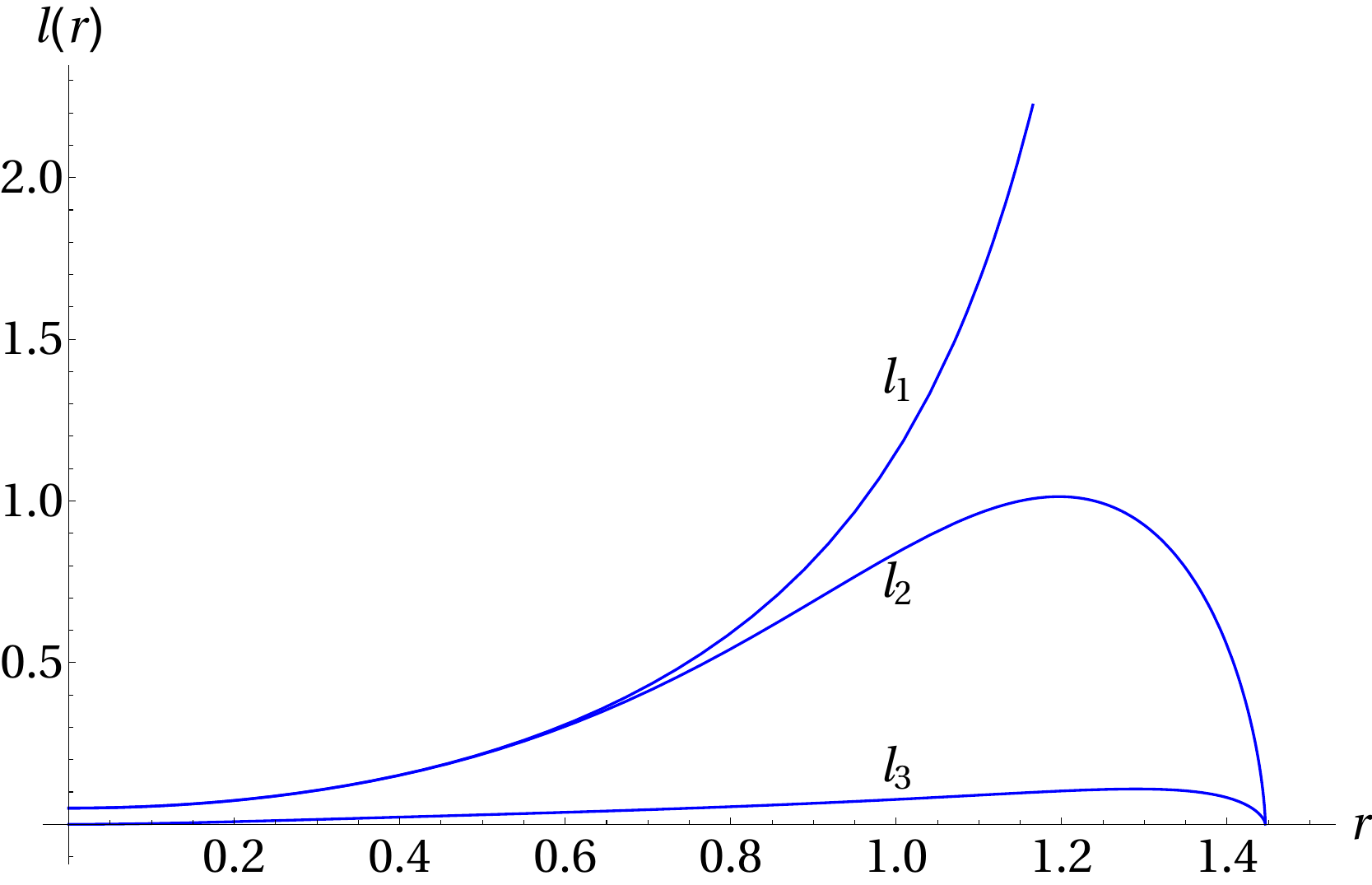}
\caption{Two representative solutions with a Bolt in the IR. Left: a solution with $\gamma_0=1/20$ and $\gamma_4=0.382281$ for which all the $l_i$ keep on growing exponentially. This is an AdS-Taub-Bolt solution. Right: a solution with $\gamma_0= 1/20$ and $\gamma_4 = 0.380208$, for which there is a singularity where $l_2(r)=l_3(r)=0$ at a finite value of $r$. We have chosen $r^*=0$.}\label{fig:genericSolutionsBolt}
\end{figure}

Constructing numerical AdS-Taub-Bolt solutions of the nonlinear equations of motion is very similar to the NUT solutions discussed in Section \ref{subsubsec:NUTs}. One has to use the series expansion in \eqref{eqn:genIRBolt} with $\gamma_0$ and $\gamma_4$ as independent integration parameters. One again finds two classes of solutions. The regular ones are the generalisations of the AdS-Taub-Bolt solutions we are after and one can read off the squashing parameters $\alpha$ and $\beta$ from the UV expansion of the numerical solutions after using the relations \eqref{eqn:abBC}. A typical regular AdS-Taub-Bolt solution with two non-vanishing squashing parameters is presented in the left panel of Figure \ref{fig:genericSolutionsBolt}. There are also singular solutions which we ignore from now on. An example of the latter is shown in the right panel of Figure \ref{fig:genericSolutionsBolt}.

\begin{figure}[ht!]
\centering
\includegraphics[width=0.49\textwidth]{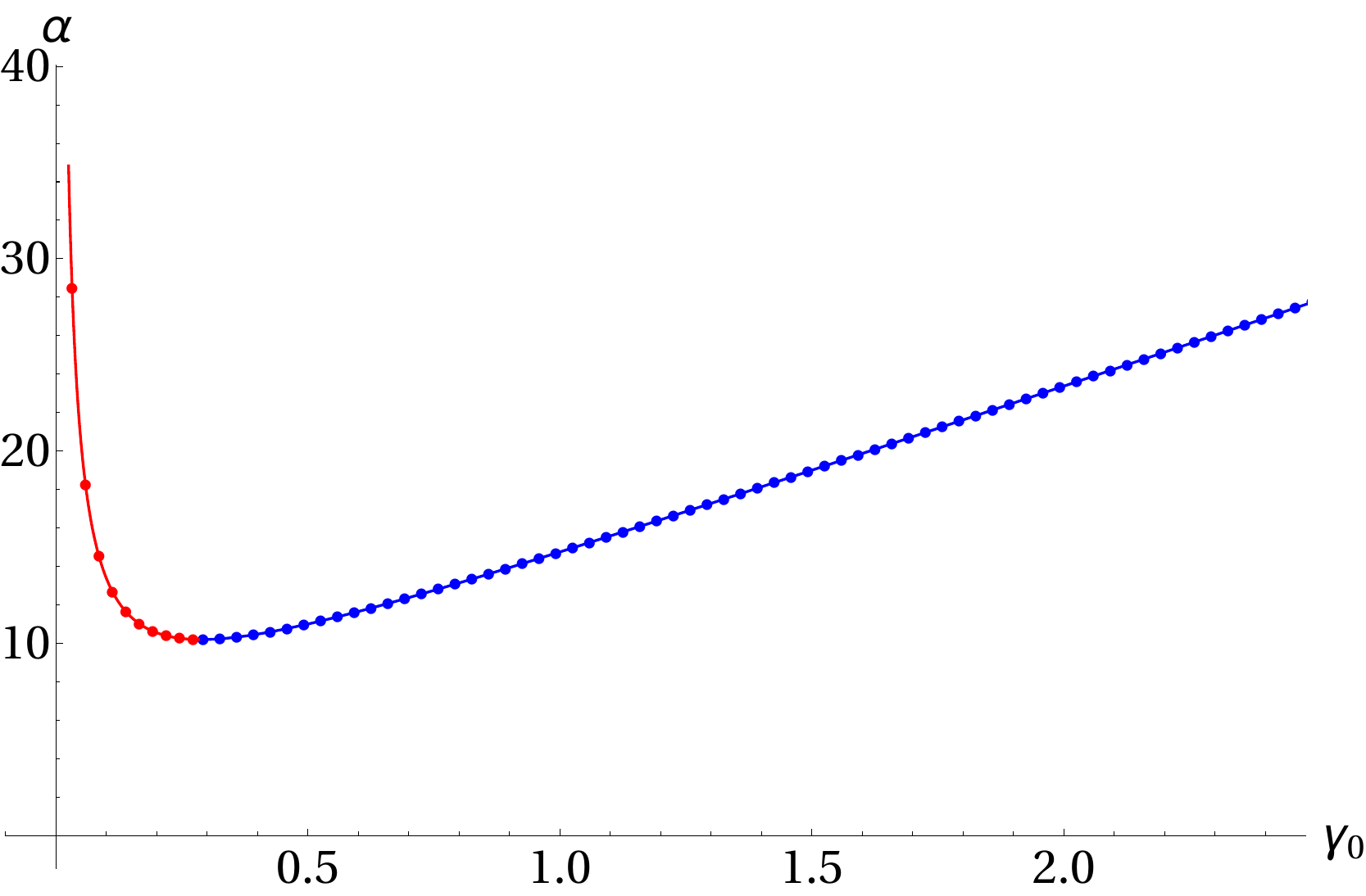}
\includegraphics[width=0.49\textwidth]{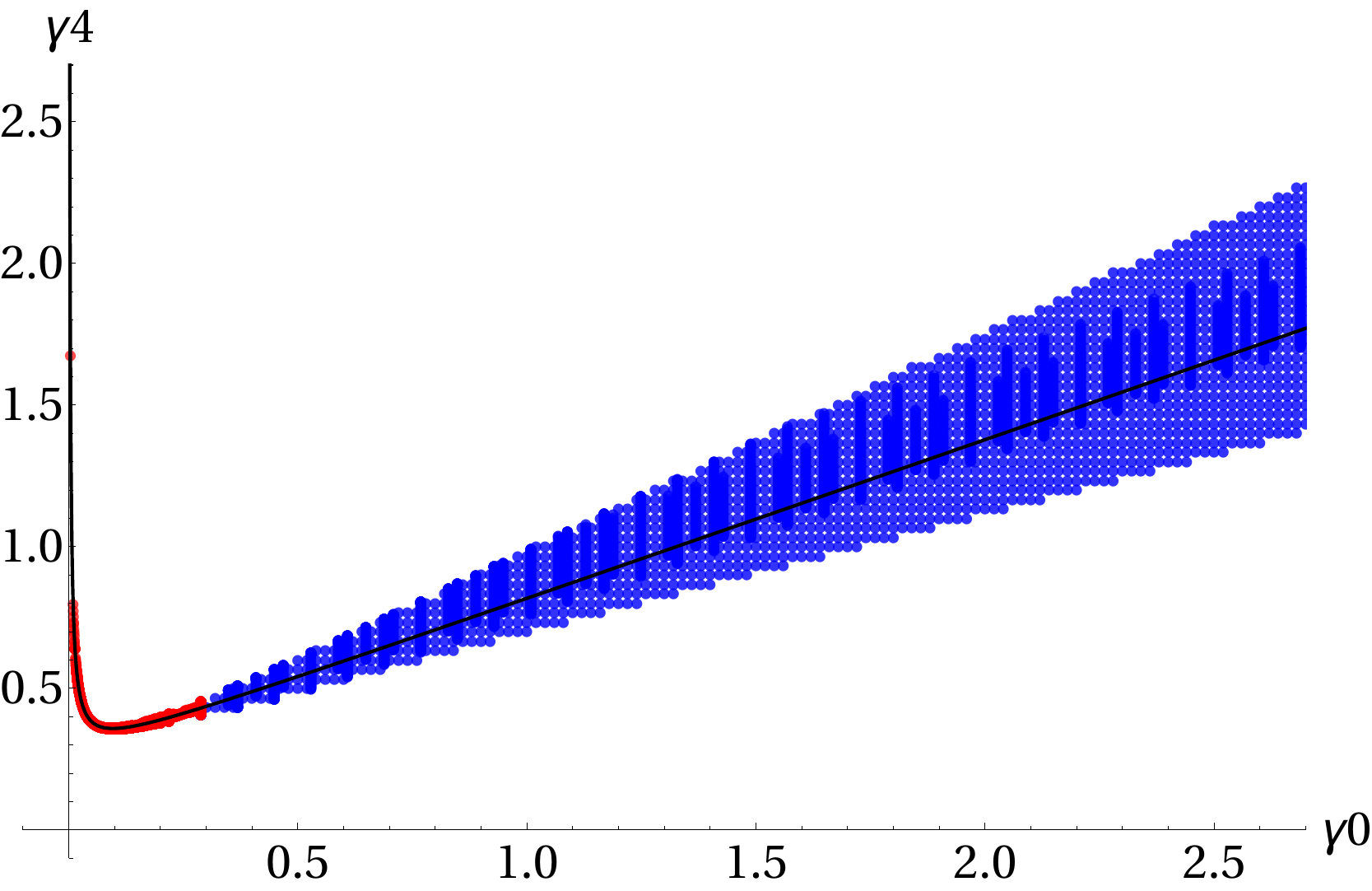}
\caption{Left: The squashing parameter $\alpha$ as a function of $\gamma_0$ for the single squashed AdS-Taub-Bolt solutions; the solid curve is the analytic result in \eqref{eqn:agam0}. Right: the range of values of $(\gamma_0,\gamma_4)$ for which there is a regular AdS-Taub-Bolt solution with two nontrivial squashing parameters; the solid curve is the relationship in \eqref{eqn:gam0gam4}.}\label{fig:oneSquashingBolt}
\end{figure}

Despite their similarity with the NUT solutions, the regular AdS-Taub-Bolt backgrounds are more subtle because there are two branches of such solutions. To understand these branches better it is worthwhile to focus briefly on the solutions with a single squashing parameter by setting $\beta=0$. This condition imposes the following relation between the two independent IR parameters
\begin{equation}\label{eqn:gam0gam4}
 \gamma_4=\frac{1+48\gamma_0 +108 \gamma_0^2}{192 \gamma_0} \ .
\end{equation}
Since we have an analytic solution for $\beta=0$, namely the AdS-Taub-Bolt background of Section \ref{subsec:AdSTNB}, we can also express the squashing parameter $\alpha$ as a function of $\gamma_0$.
\begin{equation}\label{eqn:agam0}
\alpha = 9\gamma_0 + \frac{3}{4\gamma_0} +5\;.
\end{equation}
It is now clear that for every positive value of $\alpha$ there are two different values of $\gamma_0$. This is illustrated in the left panel of Figure \ref{fig:oneSquashingBolt}. At the critical value $\gamma_0=\frac{1}{2\sqrt{3}}\approx0.288676$ one finds a minimum of the function in \eqref{eqn:agam0} and the value $\alpha=5+ 3\sqrt{3}\approx 10.1962 $ at this point is precisely the minimum value of $\alpha$ below which there are no AdS-Taub-Bolt with a single squashing parameter. The branch of solutions with a value of $\gamma_0$ greater/smaller than the critical value will be dubbed ``positive"/``negative" respectively. To construct numerical solutions with non-vanishing $\beta$ we proceed as follows. We choose values of $\gamma_0$ and $\gamma_4$ that lie on the analytic curve \eqref{eqn:gam0gam4} characterising solutions with $\beta=0$ and we change $\gamma_4$ to explore the full parameter space in the $(\gamma_0,\gamma_4)$ plane. In this manner we can always keep track of which branch of solutions the resulting numerical solution belongs to. In the right panel of Figure \ref{fig:oneSquashingBolt} we show part of the region in the $(\gamma_0,\gamma_4)$ plane where there are regular AdS-Taub-Bolt solutions. The ``negative" branch solutions are depicted in red while the ``positive" branch is in blue.

\section{From NUTs to Bolts: Thermodynamics and phase transitions }
\label{sec:TBTD}

In Section \ref{sec:AdS} we found a new class of solutions of the Einstein equations which extends the known AdS-Taub-NUT/Bolt solutions. In this section we study their thermodynamic behaviour by evaluating their regularised Euclidean on-shell action. This also sets the stage for a holographic comparison with the field theory results for the free energy in Section \ref{sec:CFT}.

 \subsection{Renormalising the action}
 \label{subsec:Renormaction}
 
The Euclidean gravitational action in \eqref{eqn:GRaction} has to be supplemented with the Gibbons-Hawking boundary term in order to have a well-defined variational principle \cite{Gibbons:1976ue}. The resulting action is
\begin{equation}
 S=-\frac{1}{16\pi G} \int_{\mathcal{M}}d^4x \sqrt{g} (R-2\Lambda)- \frac{1}{8\pi G} \int_{\partial \mathcal{M}} d^3 \sqrt{h} K \  ,\label{eqn:EHactionGH}
\end{equation}
where $h$ is the determinant of the induced metric on the boundary, $h_{ij}$, and $K$ is the trace of the extrinsic curvature.

As usual for asymptotically locally AdS space, the value of the on-shell action diverges, and one needs to implement a regularisation procedure. We apply the usual tools of holographic renormalisation which were used for the NUT/Bolt solutions of Section \ref{subsec:AdSTNB} in \cite{Emparan:1999pm} (see also \cite{Mann:1999pc} and \cite{Skenderis:2002wp} for a review). This procedure amounts to adding infinite counterterms to the action in \eqref{eqn:EHactionGH}  that make it finite on-shell. These counterterms are universal for a given gravitational theory and thus we can simply apply the results of \cite{Emparan:1999pm} to our setup. The counterterms are given by 
\begin{align}
 S_{ct}=\frac{1}{8\pi G} \int_{\partial \mathcal{M}} d^3x \sqrt{h}\left(2+\frac{\mathcal{R}}{2} \right) \   , \label{eqn:actionct}
\end{align}
where $\mathcal{R}$ is the scalar curvature of the boundary metric $h_{ij}$. 
Evaluating this counterterm action with the Ansatz in \eqref{eqn:Ansatz2sq} (with $l_0=1$) yields
\begin{align}
	S_{ct}=\pi\frac{2(l_1l_2+l_2l_3+l_1l_3)+8l_1l_2l_3-l_1^2-l_2^2-l_3^2}{2G \sqrt{l_1 l_2 l_3}}\;.
\end{align}
Substituting our asymptotic expansions of the functions $l_i(r)$, eq. \eqref{eqn:genUV}, gives
\begin{align}\label{eqn:Scter}
S_{ct}= \frac{\pi}{G}\left(4\sqrt{A_0 B_0 C_0} e^{3 r} -(3+2\Lambda)\frac{A_0^2+B_0^2+C_0^2-2B_0 C_0 -2A_0 B_0 -2A_0 C_0}{4\Lambda \sqrt{A_0B_0C_0}}e^{r}+\mathcal{O}(e^{-r})\right)\;.
\end{align}
The asymptotic form of the original on-shell gravitational action in \eqref{eqn:EHactionGH} reads
\begin{align}\label{eqn:Ser}
S= -\frac{\pi}{G}\left(4\sqrt{A_0 B_0 C_0} e^{3r}-(3+2\Lambda)\frac{A_0^2+B_0^2+C_0^2-2B_0 C_0 -2A_0 B_0 -2A_0 C_0}{4\Lambda\sqrt{A_0B_0C_0}}e^{r}+\mathcal{O}(1)\right)\;.
\end{align}
As expected the sum 
\begin{equation}
S_{ren} = S+S_{ct}\;,
\end{equation}
remains finite in the $r\to \infty $ limit and thus this sum can serve as a good regularised on-shell action.  

Since our gravitational solutions are constructed numerically, evaluating the regularised on-shell action $S_{ren}$ is tricky. The difficulty comes from the fact that one has to add a large positive and a large negative number and this could lead to numerical instabilities. To remedy this we found it useful to employ the following strategy. From \eqref{eqn:Ser} we know how the on-shell action diverges at large values of $r$. We can thus evaluate numerically this on-shell action at large but finite values of $r$ and fit the resulting values to the function
\begin{align}\label{eqn:fitfun}
f= A e^{3r_c} + B e^{2r_c}+ C e^{r_c} + D + E e^{-r_c} + F e^{-2r_c} \ .
\end{align}
We can then read of the coefficients $A$, $B$, and $C$ and use the first three terms in \eqref{eqn:fitfun} as our numerical counterterm action that should be added to $S$ to produce a finite result. 

As a consistency check of our numerical results we should find that the coefficient $B$ in \eqref{eqn:fitfun} is approximately $0$. In addition the coefficients $A$ and $C$ should agree with the coefficients of the first two terms in \eqref{eqn:Ser}. In Table \ref{tbl:action} we provide some representative values emerging from our numerical analysis which convincingly show that the numerical regularisation procedure described above is very accurate.

\begin{table}
  \centering
  \begin{tabular}{c|c||c|c|c|c|c} 
   $\beta_4$  &  $\gamma_4$ & $A$  & $B$ & $C$  &  {$A_{\textrm{analytic}}$} &  {$C_{\textrm{analytic}}$}
   \\
    \hline \hline 
{$-0.011$} & $0.131$ & $-0.192952$&  {$3.10 \times 10^{-11} $}&  {$-0.535429$} &  {$-0.192952$} & $-0.5354289 $\\ \hline
$0.182 $& $0.212$ &{$-0.127241 $}&  {$-2.09 \times 10^{-6}$}&  {$151.219$} &  {$ -0.1272407$} & $151.1883$  \\ \hline
$ 0.76$ & $0.7600015$ &  {$-0.149134$}&  {$-3.14\times 10^{-10}$}&  {$-0.27225$} &  {$-0.149134$} & $-0.272254$\\ \hline
$1.18$& $1.18000523$&  {$-0.0823926$}&  {$4.3 \times 10^{-4}$}&  {$1682.7 $}&  {$-0.0823926$} & $1687.5$\\ \hline
$-0.109$ & $0.186$& {$-0.185742 $}&  {$-5.74 \times 10^{-10} $}&  {$-0.298644$} &  {$ -0.185742$} & $-0.298349$ \end{tabular}
  \caption{For different values of the initial conditions we present the first three coefficients of the numerical fit of the action in \eqref{eqn:fitfun} and compare them with the ``analytically" obtained equivalent values from \eqref{eqn:Ser}. As discussed in the text the ``analytic" value of the coefficient $B$ is 0. We fixed $G=1$ and $\Lambda=-3$.}
\label{tbl:action}   
\end{table}   

\subsection{Single squashed NUTs and Bolts}
\label{subsec:AdSTNB-therom}

To test our numerical methods further, we compare them with the known, analytic results for the regularised on-shell action of the AdS-Taub-NUT and Bolt solutions with a single squashed $S^3$ on the boundary \cite{Emparan:1999pm}. 

The regularised on-shell action for the Taub-NUT/Bolt solutions of Section \ref{subsec:AdSTNB} can be found by plugging the metric \eqref{eqn:metricTaubNUTBolt} into the action \eqref{eqn:EHactionGH} and adding the counterterms given in \eqref{eqn:actionct}. This yields the following result \cite{Emparan:1999pm}:
\begin{align}
	S_{\text{on-shell}}= \frac{4\pi n }{Gl^2}(l^2 m + 3n^2 \rho_+ -\rho^3_+) \ ,
\end{align}
In this formula one has to plug in the value of $m$ corresponding to the NUT \eqref{eqn:mNUT} or Bolt \eqref{eqn:massBolt} solution. The value for $\rho_+$ corresponds to the minimum possible value of $\rho$, i.e. the location of the fixed point of the Killing vector $\partial_{\psi}$.

\begin{figure}[ht!]
\centering
\includegraphics[width=0.45\textwidth]{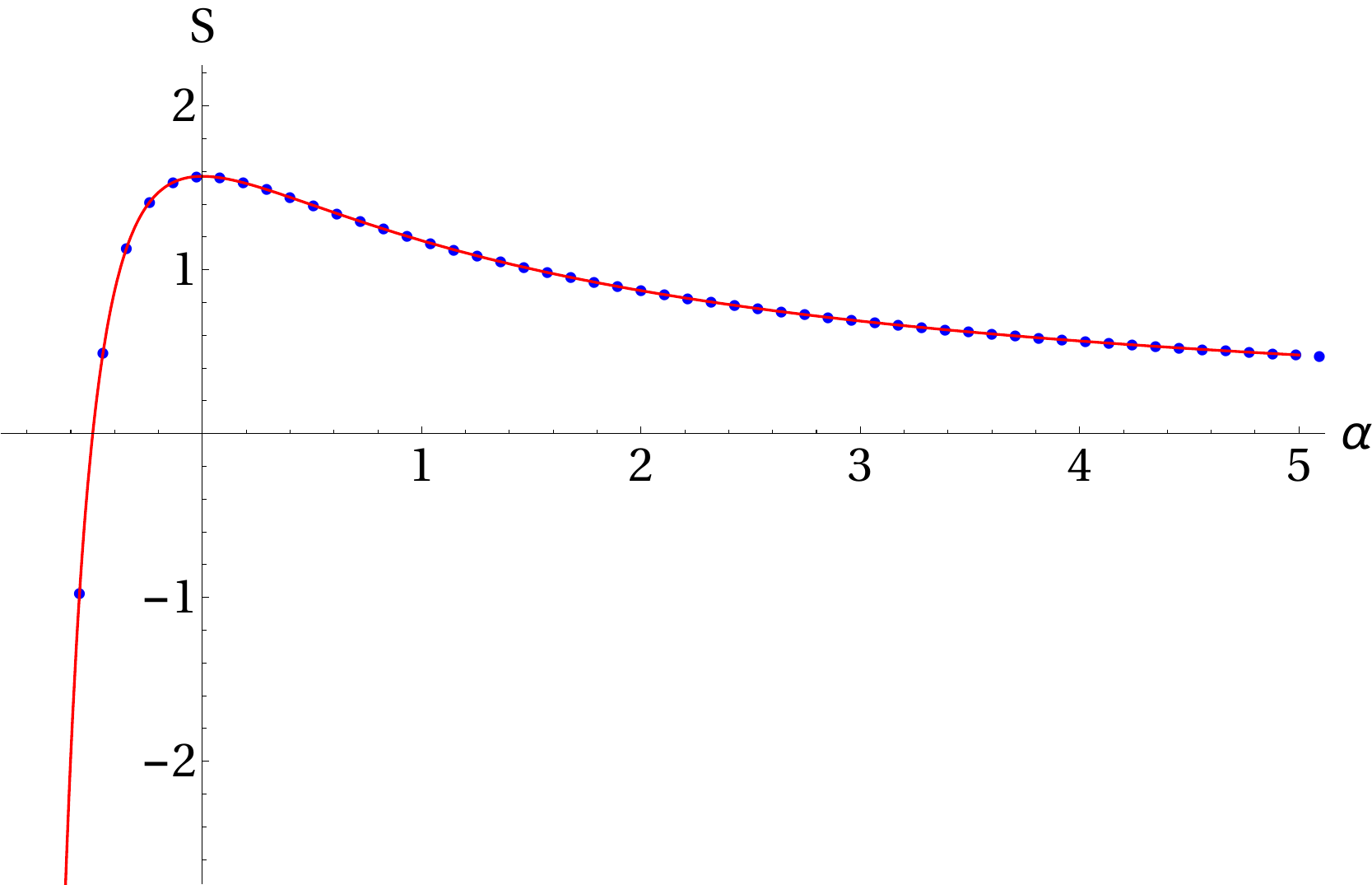}
\includegraphics[width=0.45\textwidth]{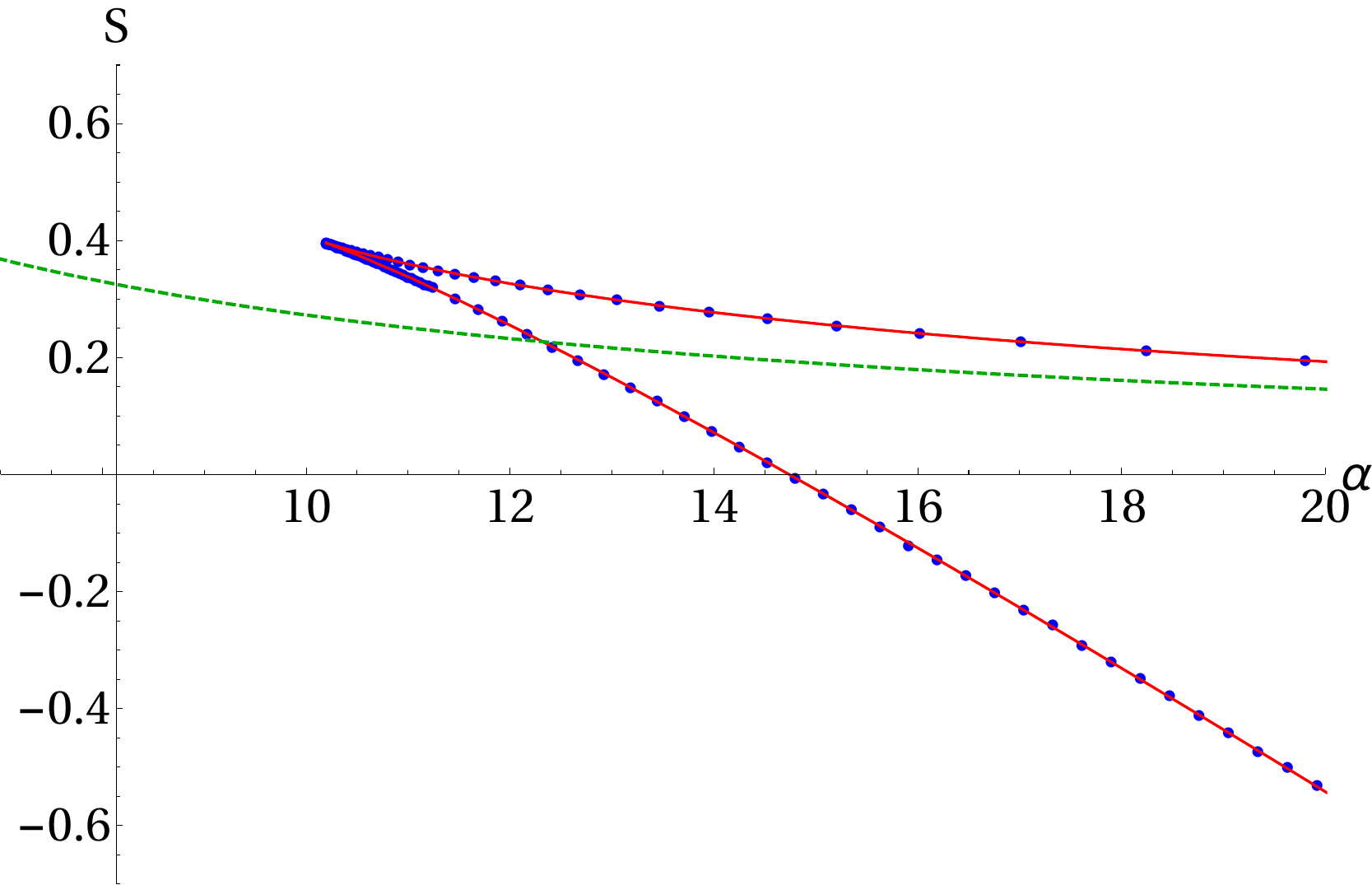}
\caption{Left: The analytic Taub-NUT action for a single squashing at the boundary in red compared with the results from our numerical techniques in blue. Rght: Idem for the Taub-Bolt action.}\label{fig:NUTAction}
\end{figure}

For the NUT solutions we have to substitute $\rho_+=n$ and $m=m_n$ \eqref{eqn:mNUT}. This gives a general expression for the NUT actions which we, from now on, write as a function of $\alpha$
\begin{align}
	S_{\text{NUT}}=\frac{\pi(1+2 \alpha)}{2(1+\alpha)^2} \ .
\end{align}
We have set $G=1$ and $l=1$ in the formula above. In the left plot of Figure \ref{fig:NUTAction} we compare the analytic single squashed, Taub-NUT action (in red) as a function of $\alpha$ with the action, in blue dots, we obtain from our methods described in Section \ref{subsec:Renormaction} above. It is clear from this figure that the numerical procedure reproduces the analytical results with a very good accuracy.

We can repeat this comparison for the Bolt solutions. To find the analytic results we just have to plug in the Bolt mass parameter $m_b$ from \eqref{eqn:massBolt} and the positive or negative Bolt radius $\rho_{b\pm}$ from \eqref{eqn:radiusBolt} into $m$ and $\rho_+$ respectively. This gives two different values for the Bolt action 
\begin{equation}
 S_{\pm}=\frac{\pi  \left(17-\left(1\pm\sqrt{\frac{(\alpha -10) \alpha -2}{(\alpha +1)^2}}\right) \alpha ^2+2 \left(8\pm5 \sqrt{\frac{(\alpha -10) \alpha -2}{(\alpha +1)^2}}\right) \alpha \pm2 \sqrt{\frac{(\alpha -10) \alpha
   -2}{(\alpha +1)^2}}\right)}{54 (\alpha +1)} \ ,
\end{equation}
where $S_+$ corresponds to the positive branch and $S_-$ to the negative branch of solutions. For both of these branches, the analytic (full red line) and the numerical results (blue dots), are shown in the right plot of Figure \ref{fig:NUTAction}. In the same figure we also plot the on-shell action of the NUT solution (dashed green curve) for comparison. Again there is excellent agreement between the numerical and analytical calculations, giving us confidence in our numerical techniques.

It is clear from Figure \ref{fig:NUTAction} that there is a phase transition from the NUT to the Bolt solutions as one increases the value of $\alpha$. This phase transition is similar to the Hawking-Page one and to find the precise value of the squashing parameter at which it occurs one has to compare the regularised on-shell action for the two types of solutions, i.e. the solution with the lower on-shell action is the thermodynamically preferred one. For $\alpha < \alpha_{\text{crit}} = 5+3\sqrt{3}\approx 10.2$ there are only NUT solutions. The Bolt solutions with higher on-shell action are the ones from the ``negative" branch with action $S_-$. They are never thermodynamically preferred. The ``positive" branch Bolt solutions with action $S_+$ become thermodynamically preferred for $\alpha > \alpha_{\text{HP}}\equiv 6+2\sqrt{10}\approx 12.3$. The precise value $\alpha_{\text{HP}}$ is obtained by solving the algebraic equation $S_+=S_{\text{NUT}}$. 

\subsection{Double squashed NUTs and Bolts}
\label{subsec:DS}
\begin{figure}[ht!]
\centering
    \includegraphics[width=0.49\textwidth]{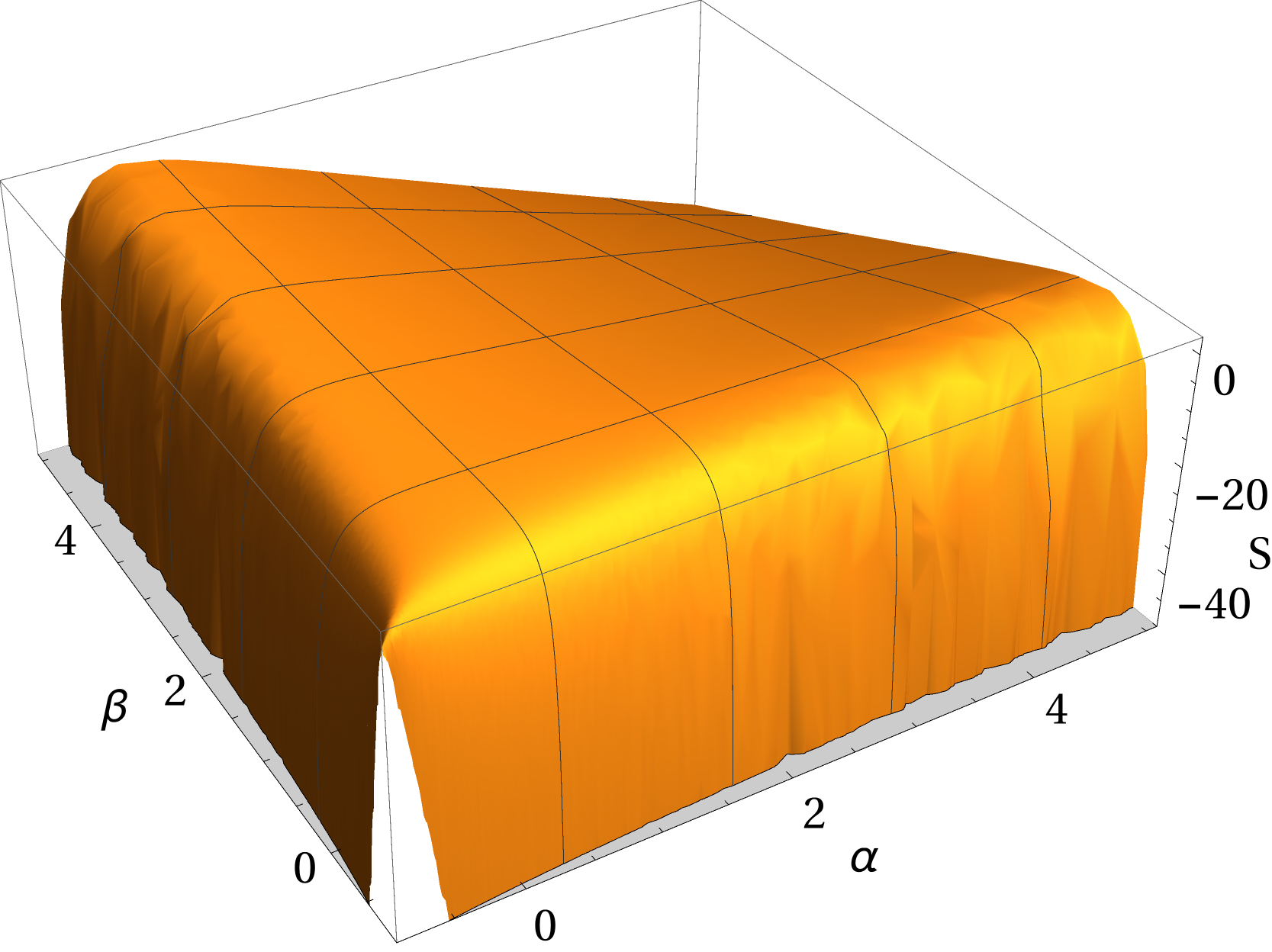}
\caption{The regularised on-shell action of the AdS-Taub-NUT solutions as a function of the two squashing parameters.}\label{fig:actionsAdStwosquashings}
\end{figure}

\begin{figure}[ht!]
\centering
  \begin{subfigure}[t]{0.32\textwidth}
    \includegraphics[width=\textwidth]{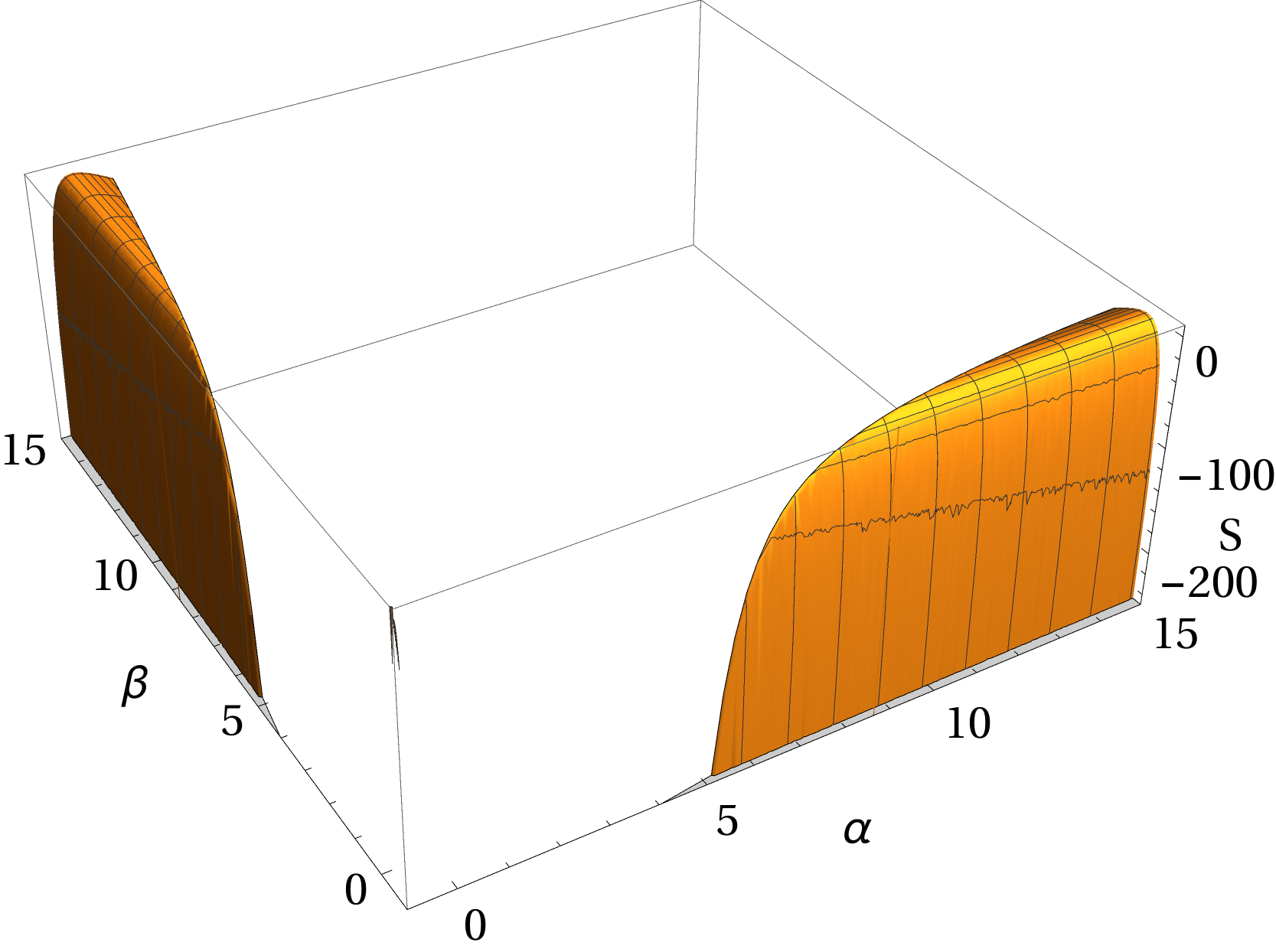}
    \caption{The positive branch Bolt solutions}
    \label{fig:actionboltptwo}
  \end{subfigure}
    \begin{subfigure}[t]{0.32\textwidth}
    \includegraphics[width=\textwidth]{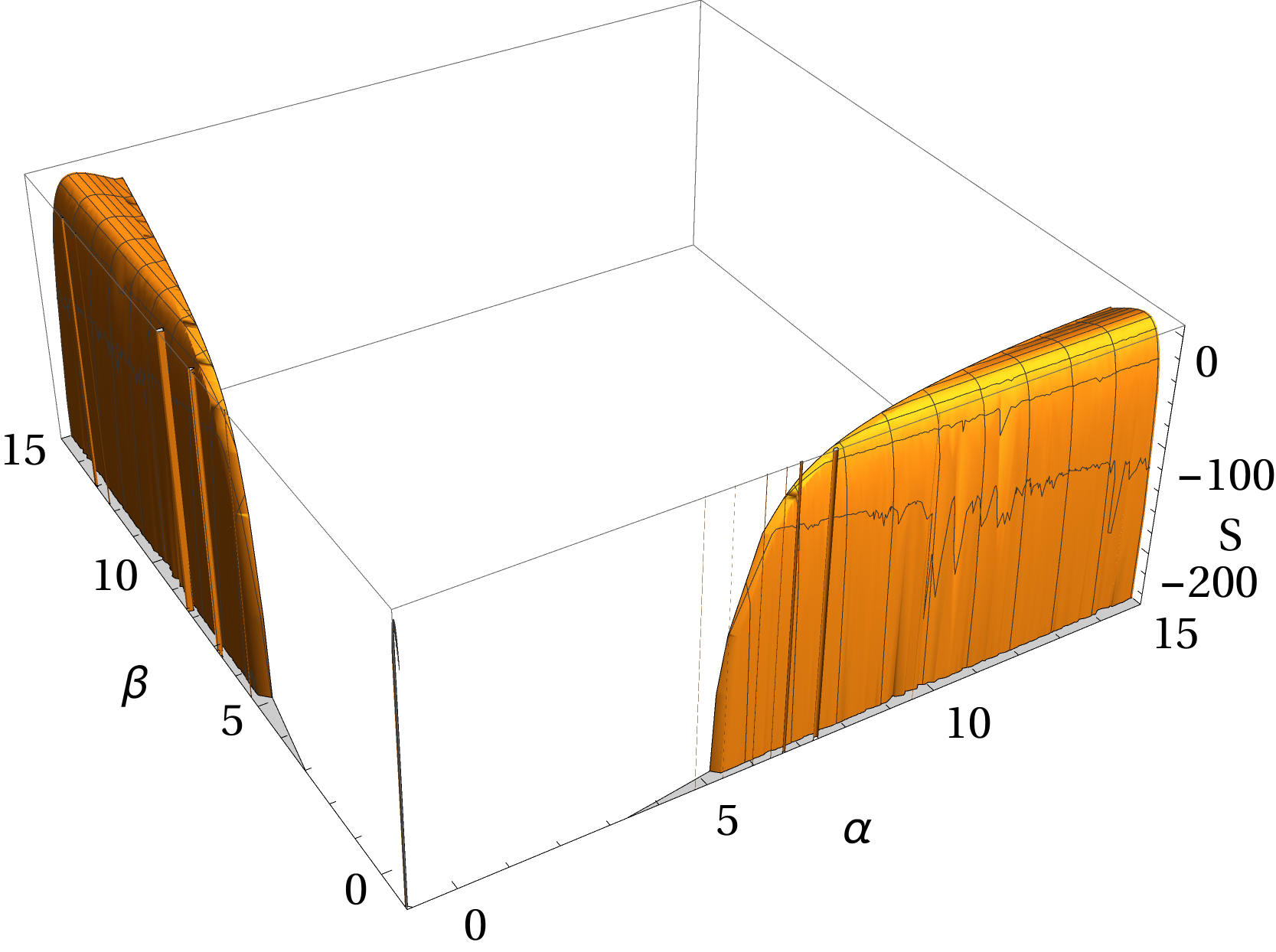}
    \caption{The negative branch Bolt solutions}
    \label{fig:actionboltntwo}
  \end{subfigure}
	    \begin{subfigure}[t]{0.32\textwidth}
    \includegraphics[width=\textwidth]{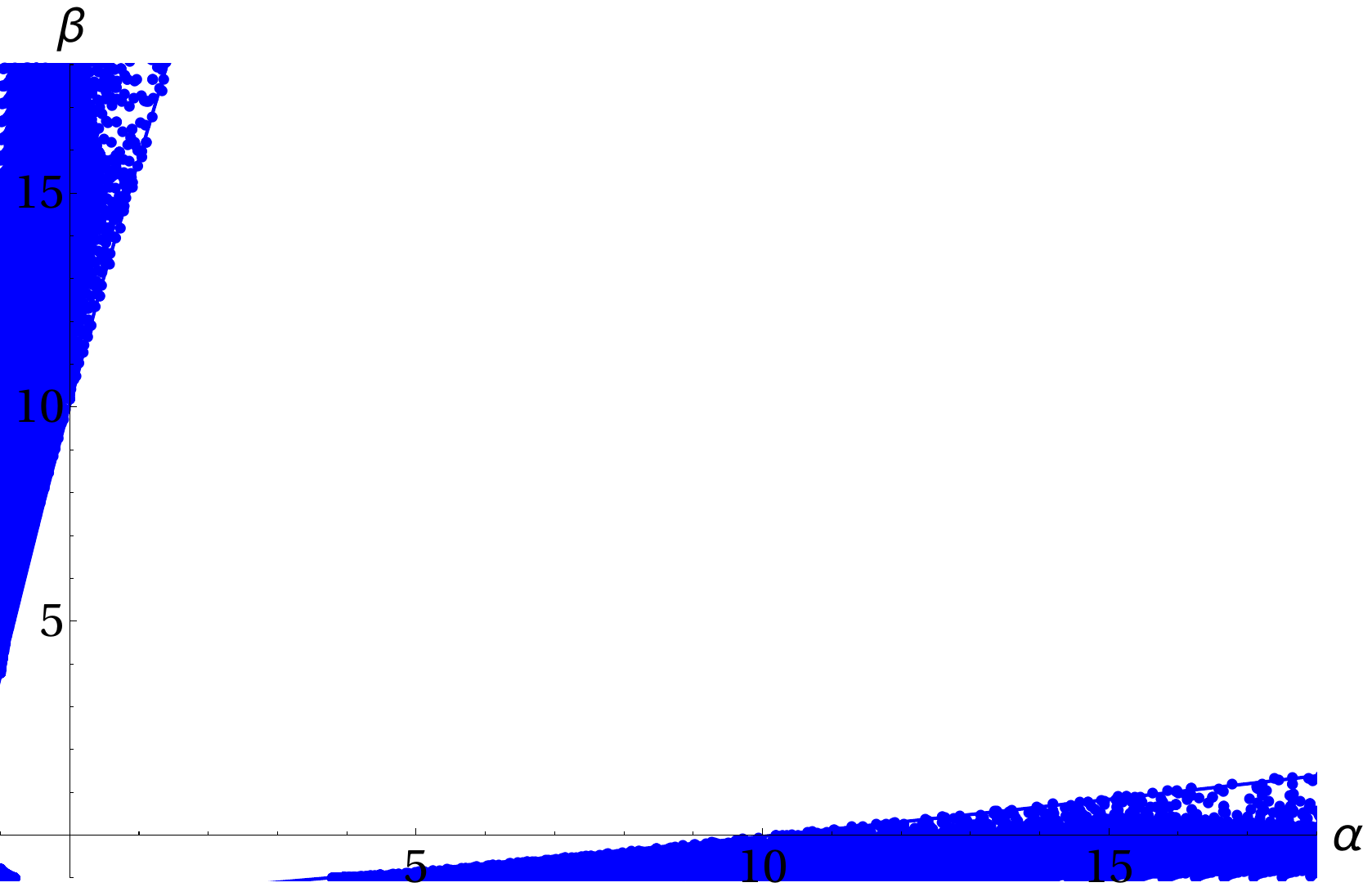}
    \caption{The region in the $(\alpha,\beta)$ plane where the Bolt solutions exist.}
    \label{fig:regionBolt}
  \end{subfigure}
\caption{ The regularised on-shell action of the two branches of the AdS-Taub-Bolt solutions as a function of the two squashing parameters and the region in the $(\alpha,\beta)$ plane for which the solutions exist.}\label{fig:actionsAdStwosquashingsB}
\end{figure}

%
Having some faith in our numerical techniques, it is time to apply them to the new asymptotically $AdS_4$ solutions with two squashing parameters that we constructed in Section \ref{subsec:AdS2squash}. Since we do not have analytic solutions we evaluate the regularised on-shell action numerically as described in Section \ref{subsec:Renormaction}. The resulting on-shell action for the AdS-Taub-NUT solutions is plotted in Figure \ref{fig:actionsAdStwosquashings}. It is clear from the plot  that the on-shell action exhibits a global maximum at $\alpha=\beta=0$. If one considers slices of constant $\beta$ there is a maximum around $\alpha=0$ for positive $\beta$ and at $\alpha=\beta$ for negative $\beta$ and vice versa for slices of constant $\alpha$. A similar analysis can be done for the AdS-Taub-Bolt solutions with two squashings. As discussed in Section \ref{subsec:AdS2squash} in this case we have two branches of solutions both of which exist only in a limited region in the $(\alpha,\beta)$ plane. The regularised on-shell action for the two branches of solutions is plotted in Figure \ref{fig:actionsAdStwosquashingsB}.
%
%

%
\begin{figure}[ht!]
\centering
  \begin{subfigure}[t]{0.32\textwidth}
    \includegraphics[width=\textwidth]{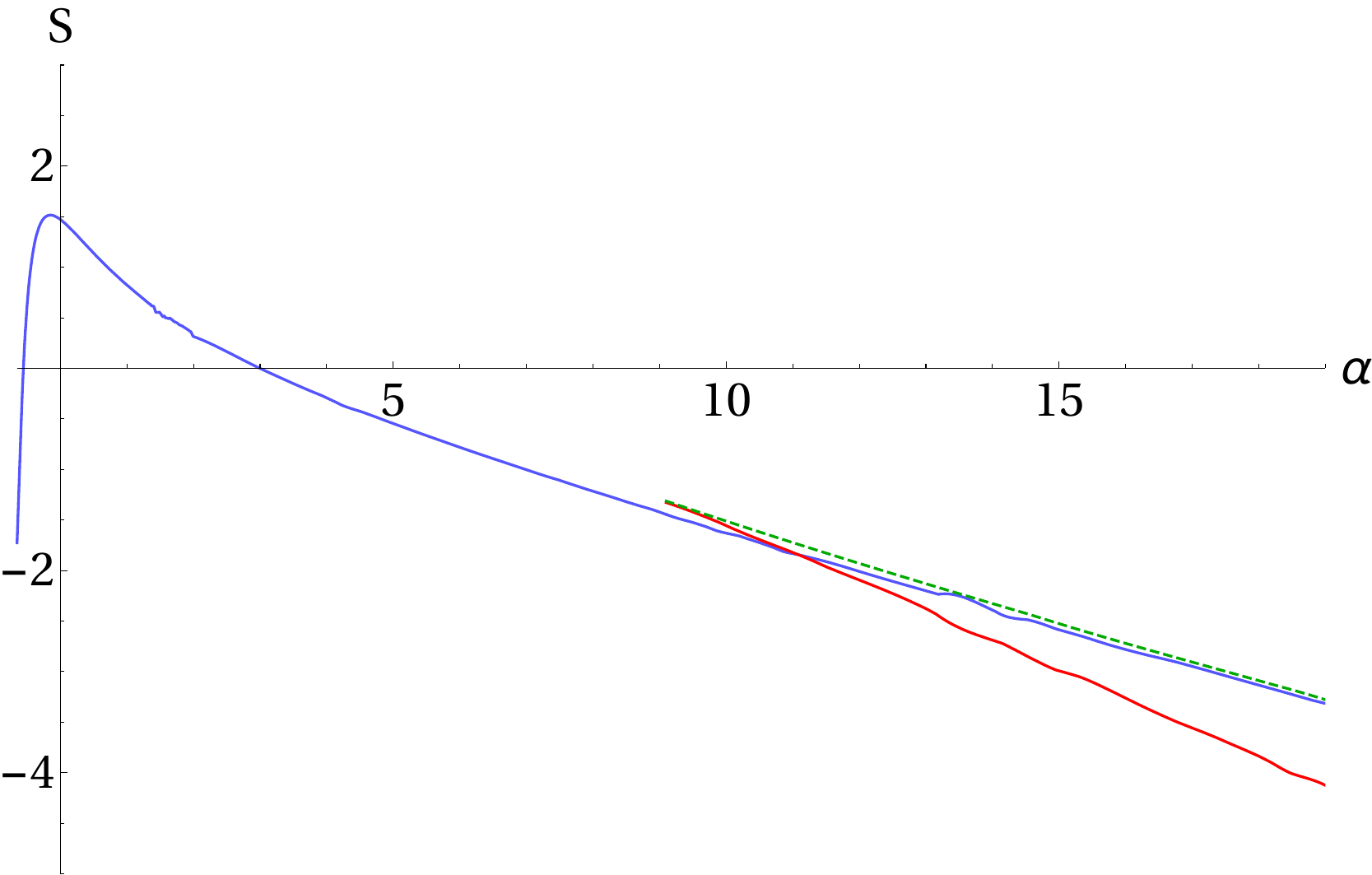}
    \caption{$\beta=-0.2$}
  \end{subfigure}
    \begin{subfigure}[t]{0.32\textwidth}
    \includegraphics[width=\textwidth]{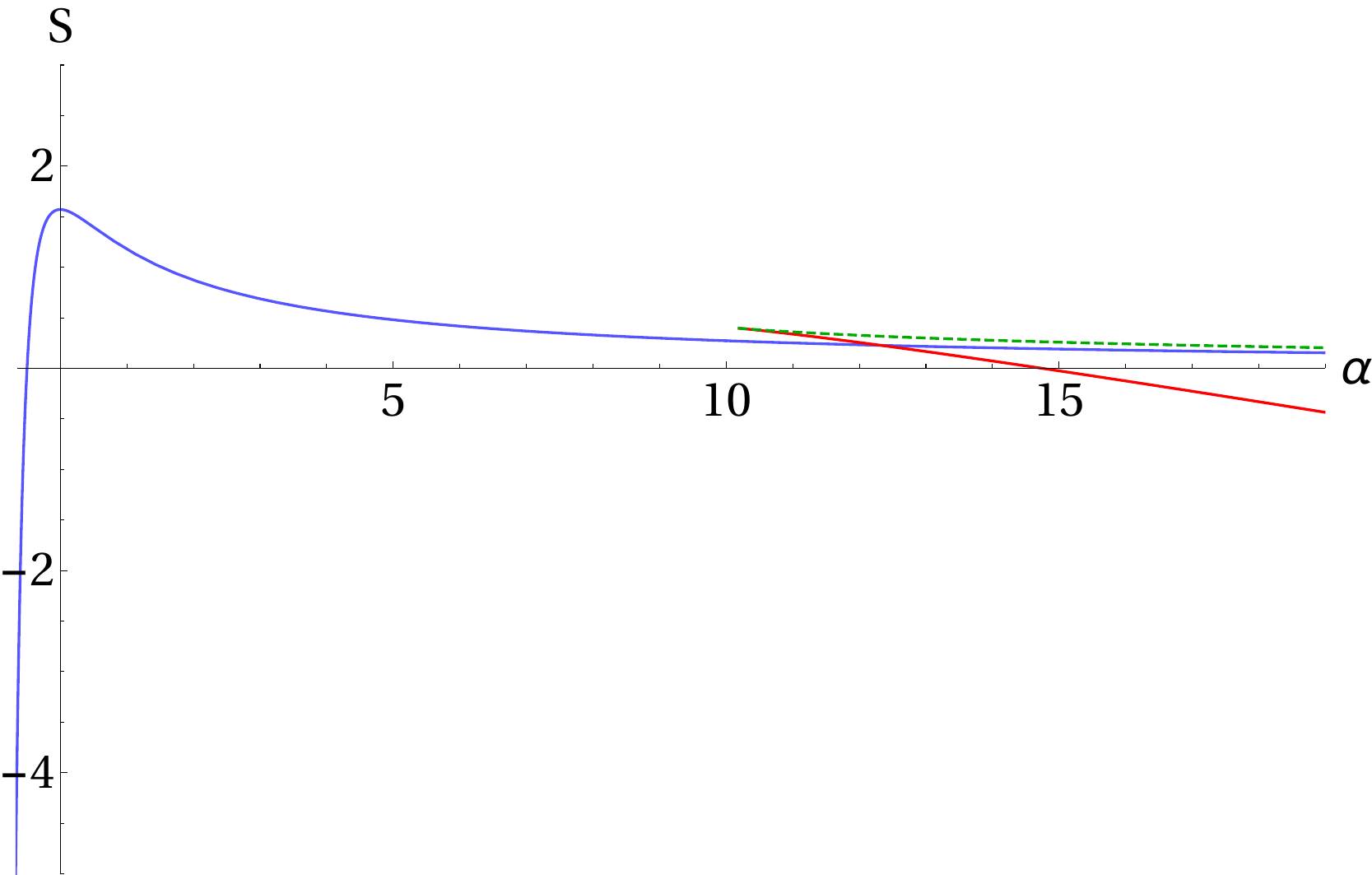}
    \caption{$\beta=0$}
  \end{subfigure}
    \begin{subfigure}[t]{0.32\textwidth}
    \includegraphics[width=\textwidth]{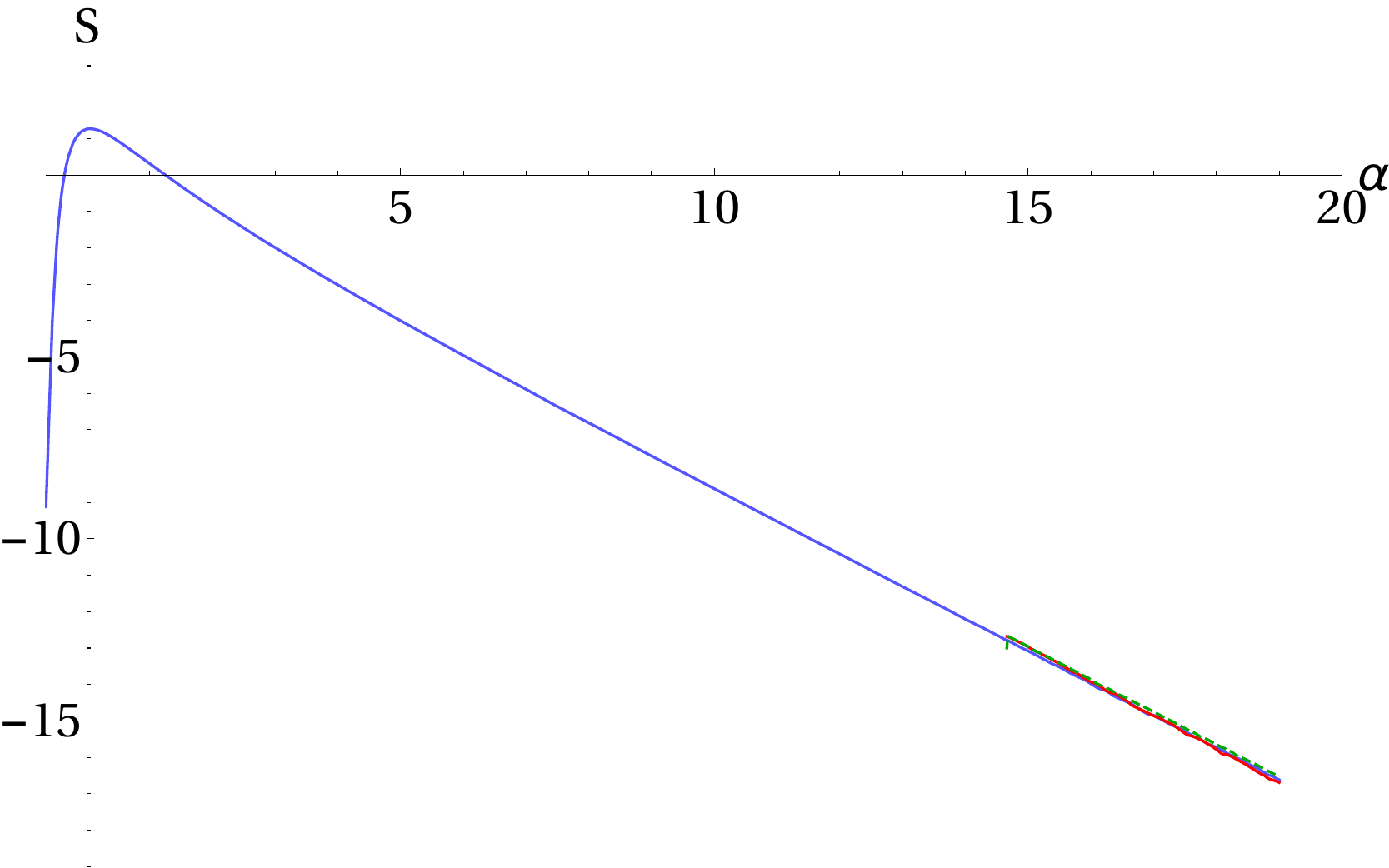}
    \caption{$\beta=0.8$}
  \end{subfigure}
\caption{The regularised on-shell action for the NUT and Bolt solutions as a function of $\alpha$ for slices of constant $\beta$. The NUT action is plotted in blue, the negative branch Bolt action is in dashed green  while the positive branch is in red. }\label{fig:actionsslicesAdS}
\end{figure}

Equipped with the regularised on-shell action we can ask which of the solutions is thermodynamically favoured in various regions of parameter space. This is not immediately clear and to illustrate the result better we have compared the on-shell action for the NUT and Bolt solutions by choosing slices of constant $\beta$ and plotted the action as a function of $\alpha$. Some representative results are presented in Figure \ref{fig:actionsslicesAdS}. What we find is that for any fixed value of $\beta_0$ the Taub-Bolt solutions exist only for values of $\alpha$ larger than a finite critical value $\alpha_{\text{crit}}(\beta_0)$. In addition there is always a Hawking-Page type phase transition at some $\alpha_{\text{HP}}(\beta_0)> \alpha_{\text{crit}}(\beta_0)$. Thus we conclude that the new AdS-Taub-Bolt solutions exhibit a qualitatively similar behaviour to the analytic Bolt solutions with $\beta=0$ across the entire configuration space of boundary geometries parameterised by $(\alpha,\beta)$. It is useful to note that $\alpha_{\text{HP}}(\beta_0)$ increases for increasing $\beta_0$. Finally, notice that for negative $\beta_0$, the maximum value of the action as a function of $\alpha$ moves away from $\alpha=0$. 

\section{The $O(N)$ model on the double squashed sphere} 
\label{sec:CFT}

Studying the holographic dual field theory description of the gravitational solutions described in the previous two sections is a non-trivial problem. One approach could be to embed these solutions as backgrounds in string or M-theory and identify a dual CFT in which to phrase the question. An example of how this could be achieved is to think of the NUT/Bolt solutions as deformations of the $AdS_4\times S^7$ solution of eleven-dimensional supergravity. In this case the dual field theory is the ABJM SCFT and we are faced with the problem of evaluating the partition function of this theory at strong coupling upon a supersymmetry breaking deformation. This is a formidable problem which we will not attempt to solve here. Instead, we will focus on a simplified model of this setup where we consider a free vector-like theory on the double-squashed sphere in \eqref{eqn:metric}. In particular we focus on evaluating the partition functions, or free energy, of the free $O(N)$ model as a function of the two squashing parameters $\alpha$ and $\beta$.\footnote{See also \cite{BBV} for some recent results on the usefulness of the squashed sphere partition function for free theories in odd dimensions.} 

The free $O(N)$ vector model in three dimensions is conjectured to be described holographically by a higher-spin Vassiliev theory in $AdS_4$ with certain specific boundary conditions \cite{Sezgin:2002rt,Klebanov:2002ja,Giombi:2009wh}. This four-dimensional theory is very different from pure Einstein gravity with a negative cosmological constant. Despite this difference we find that there are many qualitative similarities between the behaviour of the free energy of the free $O(N)$ vector model and the action of the gravitational solutions in Section \ref{sec:AdS} as a function of $\alpha$ and $\beta$. Similar results were found in \cite{Hartnoll:2005yc} for the case of $S^3$ boundaries with a single squashing parameter, i.e. $\beta=0$.

\subsection{The method}
\label{subsec:method}

The action for the free three-dimensional $O(N)$ vector model in Euclidean signature is given by
\begin{equation}\label{eqn:ONaction}
S_{O(N)} = \frac{1}{2}\int d^3x \sqrt{g} \left( \partial_{\mu} \phi_a \partial^{\mu}\phi^a +\frac{ R}{8}\phi_a \phi^a+m^2\phi_a \phi^a \right)\ .
\end{equation}
Here we are assuming that the $N$ scalar fields $\phi_a$, $a=1,\ldots, N$, are conformally coupled and have mass $m$. We will assume that the metric is given by the squashed sphere metrc in \eqref{eqn:metric} with Ricci scalar $R$ given by \eqref{Ricci}. The partition function for this model is given by
\begin{align}\label{eqn:Zpart}
Z _{O(N)}= \int \mathcal{D} \phi e^{-S_{O(N)}} \ .
\end{align}
To find the free energy of this theory we have to evaluate the Gaussian integral in \eqref{eqn:Zpart}. This amounts to computing the following determinant
\begin{align}
 F_{O(N)} = - \log Z_{O(N)}=\frac{N}{2}\log \left(\textrm{det}\left[ \frac{-\nabla^2 + m^2+\frac{R}{8}}{r_0^2 \Lambda^2}\right]\right) \ , \label{eqn:LogZGeneral}
\end{align}
where $r_0$ is the radius of the sphere \eqref{eqn:metric} (which we set to 1 from now on) and $\Lambda$ is the cutoff used to regularise the UV divergences in this theory. These divergences arise from infinite covariant counterterms written in terms of the metric and curvature scalar of the squashed sphere. Since we are dealing with a CFT in three dimensions there are no conformal anomalies and thus the divergences for large $\Lambda$ can be schematically written in the form
\begin{align}\label{eqn:UVCFTdiv}
 \textrm{divergences}\approx A\Lambda^3 + B \Lambda^2  + C \Lambda \ .
\end{align}
It is not accidental that we chose the same notation for the coefficients in \eqref{eqn:fitfun} and \eqref{eqn:UVCFTdiv}. The role of $\Lambda$ in the CFT is played by the radial cutoff $e^{r_c}$ used in holographic renormalisation. The cubic and linear terms in \eqref{eqn:UVCFTdiv} arise from the covariant counterterms $\Lambda^3\int d^3x \sqrt{g}$ and $\Lambda\int d^3x \sqrt{g} R$  respectively. There is no covariant counterterm that will lead to a quadratic divergence and thus we should have $B=0$ in \eqref{eqn:UVCFTdiv}.
 
Our goal is to calculate \eqref{eqn:LogZGeneral}. There are two technical obstacles along the way. One has to first find the spectrum of the scalar Laplacian $\nabla^2$ for the metric in \eqref{eqn:metric}. Since this operator is infinite dimensional, to evaluate the determinant in \eqref{eqn:LogZGeneral} one has to perform an infinite sum and regularise the divergences described above.

To address the first problem we can use the fact that the metric in \eqref{eqn:metric} is a homogeneous metric on $S^3$ and thus one can use algebraic techniques to find the spectrum of the Laplacian. When one of the squashing parameters vanishes, say $\beta=0$, the spectrum of the scalar Laplacian can be found in closed analytic form, see for example \cite{PhysRevD.8.1048, Shen1987}. When both squashing parameters are turned on the problem becomes harder and one has to resort to numerical techniques. The procedure to find the eigenvalues of the scalar Laplacian is outlined in Appendix \ref{App:eigen}. The upshot of the analysis is that we are able to determine this spectrum numerically to (in principle) any desired accuracy.

To regularise the infinite sum in \eqref{eqn:LogZGeneral} one may be tempted to use an analytic approach like $\zeta$-function regularisation \cite{Dowker:1998pi,Hartnoll:2005yc, Klebanov:2011gs}. However this method is not well-adapted to situations where the spectrum of the Laplacian is known only numerically. Therefore we will use a heat-kernel type regularisation which can be implemented numerically and was discussed in some detail in \cite{Anninos:2012ft}. Here we briefly summarise this approach.

The regularisation we adopt proceeds by rewriting \eqref{eqn:LogZGeneral} using a heat-kernel\footnote{See \cite{Vassilevich:2003xt} for a review of heat kernel methods.}
\begin{align}
\log Z_{_{O(N)}}= \frac{N}{2} \sum_n \int_{\epsilon}^{\infty}\frac{dt}{t} e^{-t\lambda_n} \ ,
\end{align}
where the sum is over all eigenvalues, $\lambda_n$, of the operator $-\nabla^2 + m^2+\frac{R}{8}$ in \eqref{eqn:LogZGeneral} and the $UV$ cutoff is implemented through the parameter $\epsilon=1/\Lambda^2$. The resulting determinant captures all modes with energies lower than the cutoff $\Lambda$. The contributions of modes with eigenvalues above the cutoff is exponentially small. To see this note that for the low lying modes, i.e. $\lambda_n/ \Lambda^2 \ll 1$, we have
\begin{align}
  \int_{\epsilon}^{\infty}\frac{dt}{t} e^{-t\lambda_n}= \Gamma(0,\lambda_n/\Lambda^2) =-\log(\lambda_n /\Lambda^2)+\mathcal{O}(\lambda_n /\Lambda^2) \ ,\label{eqn:upperincompleteGamma}
\end{align}
while for the higher modes, i.e. $\lambda_n /\Lambda^2 \gg 1$, we find
\begin{align}
 \Gamma(0,\lambda_n /\Lambda^2) =e^{-\lambda_n/\Lambda^2}\left(\frac{1}{\lambda_n /\Lambda^2}+ \mathcal{O}\left(\frac{1}{(\lambda_n /\Lambda^2)^2}\right)\right) \ .
\end{align}
Using this kernel, it can be shown that the divergences are going to appear when $t$ is integrated over small values. To keep track of the divergences we split the integral over $t$ into an $UV$ and an $IR$ part
\begin{align}
 \log Z_{O(N)} = \textrm{det}_{UV} +\textrm{det}_{IR} \ ,
\end{align}
where 
\begin{subequations}
\begin{align}
 \textrm{det}_{UV} &\equiv \frac{N}{2} \int_{\epsilon}^{\delta} \frac{dt}{t} \sum_{n}m_n e^{-t \lambda_{n}} \ , \\
 \textrm{det}_{IR} &\equiv \frac{N}{2} \sum_{n}m_n \int_{\delta}^{\infty} \frac{dt}{t} e^{-t \lambda_{n}}= \frac{N}{2} \sum_{n}m_n \Gamma(0,\lambda_{n}\delta) \ . \label{eqn:detUV}
\end{align} \label{eqn:detUVIR}
\end{subequations}
Here $m_n$ is the degeneracy of eigenvalue $\lambda_n$ and $\delta$ is an arbitrarily chosen small number that we can vary in order to get better convergence of the numerical results. 

While the sum in det$_{IR}$ converges, in general, quite fast, the sum in det$_{UV}$ contains all the divergences and should be treated with care. The approach we adopt, is to numerically evaluate $ \textrm{det}_{UV}$ for many values of $\Lambda$ and fit the diverging results to the function in \eqref{eqn:UVCFTdiv}. The divergencies obtained in this way are removed by hand and the remaining finite result is added to the finite value of det$_{IR}$ to obtain the desired result for the free energy. As a consistency check we find that there is no dependence of the finite result on $\Lambda$ and that the coefficient $B$ in \eqref{eqn:UVCFTdiv} is  vanishing with good numerical accuracy.

To test our numerical method it is instructive to calculate the free energy of the free $O(N)$ model on the round three sphere, i.e. $\alpha=\beta=0$. From now on we also focus on the case of massless scalars so we set $m=0$ in \eqref{eqn:ONaction}. The eigenvalues $\lambda_n$ of the conformal Laplacian in \eqref{eqn:LogZGeneral} and their multiplicities $m_n$ are given by
\begin{align}
	\lambda_n=n^2-\frac14 \ , \qquad m_n=n^2 \ ,
\end{align}
where $n \ge 1$. After setting\footnote{For simplicity all the results we show below are for $N=1$, to obtain the results for higher $N$, one just has to multiply the free energy by $N$.} $N=1$  we can plug this into our numerical machinery and find
\begin{align}\label{eqn:Fscalarnum}
	F = -(\textrm{det}_{UV} +\textrm{det}_{IR}) \approx 0.0638070552 \ .
\end{align}
The free energy of a conformally coupled scalar field on $S^3$ can be computed analytically by using $\zeta$-function regularisation, see for example \cite{Klebanov:2011gs}. The result is
\begin{align}
	F=\frac{1}{16}\left( 2\log 2 - \frac{3\zeta(3)}{\pi^2}\right)\approx 0.0638070548 \ .
\end{align}
Comparing this analytic results with \eqref{eqn:Fscalarnum} we see a very good agreement. This gives us confidence in our numerical methods and in the next section we will apply them for the squashed sphere.

 \subsection{The results}
 \label{subsec:CFTresults}
 
Let us start with the single squashed sphere by setting $\beta=0$. The eigenvalues of the conformal Laplacian for this metric are known analytically, e.g \cite{PhysRevD.8.1048, Shen1987} (see also \cite{BBV} for an extension of this result to squashed spheres in higher dimensions) and are labelled by two integers $n$ and $q$ 
\begin{align}\label{eqn:deg1sq}
	\lambda_{n,q} = \left(n^2+\alpha (n-1-2q)^2 -\frac1{4(1+\alpha)}\right),
\end{align}
where $q=0,1,\ldots , n-1$ and $n \ge 1$ and they have a multiplicity of $m_n=n$. Since these eigenvalues are known analytically one can apply the heat-kernel regularisation procedure described above by using an analytic method to approximate the integral in \eqref{eqn:detUVIR} and subtract the UV divergences \cite{Anninos:2012ft, Anninos:2013rza}. The results of this procedure are captured by the solid red line in Figure \ref{fig:onequashingCFT}.

As described in some detail in Appendix \ref{App:eigen} the eigenvalues of the conformal Laplacian with $\beta\neq 0$ are not known analytically and we have to resort to numerics. To gain even further confidence in our numerical procedure we applied it to the case of the single squashed sphere with $\beta=0$ and compared the results with the semi-analytic approach of \cite{Anninos:2012ft, Anninos:2013rza}. The outcome of this analysis is summarised in Figure \ref{fig:onequashingCFT}. 

\begin{figure}[ht!]
\centering
\includegraphics[width=0.49\textwidth]{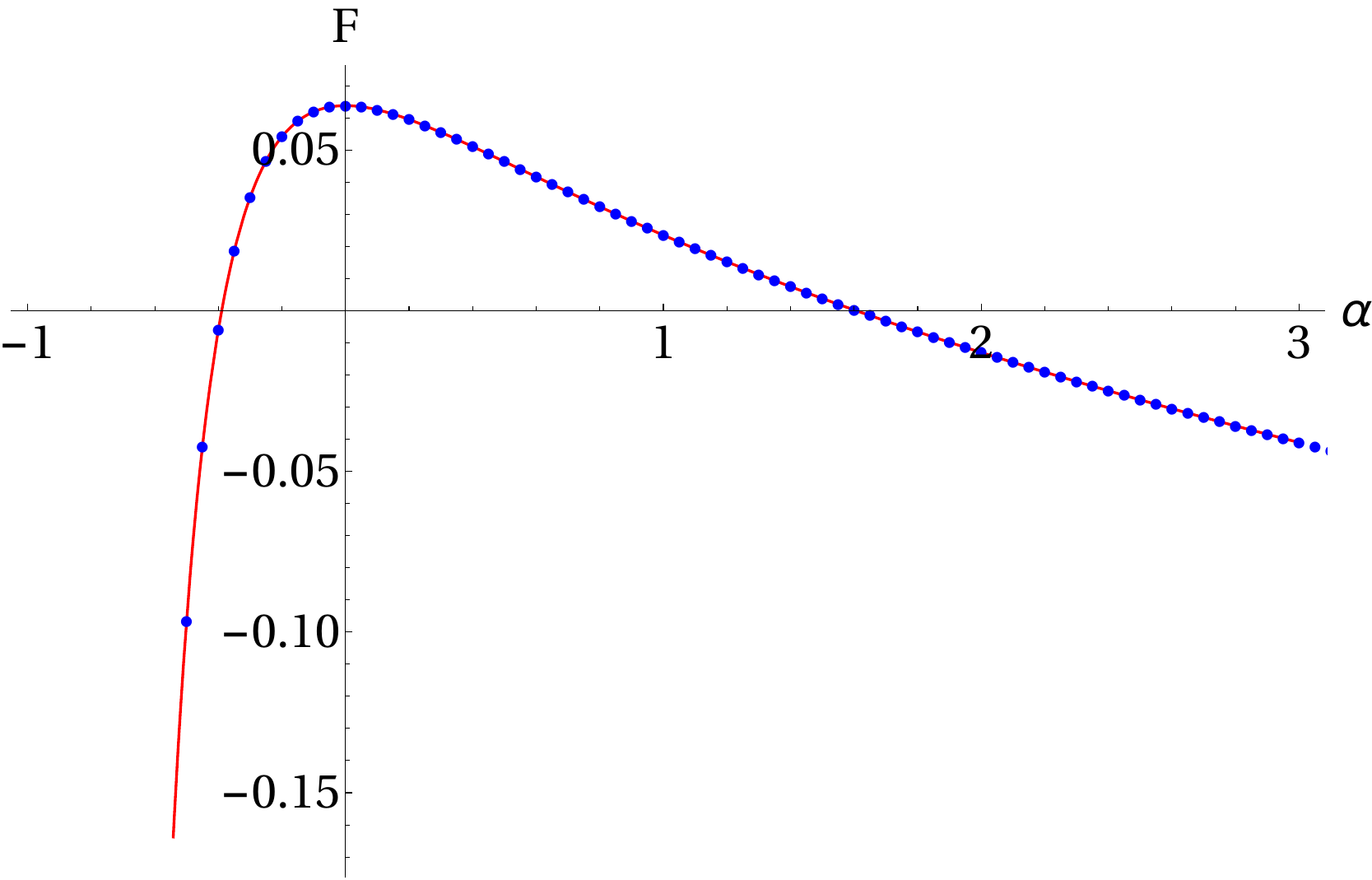}
\caption{The free energy for the squashed sphere with $\beta=0$ using our numerical techniques (in blue dots), compared with the semi-analytic approach of \cite{Anninos:2012ft} (in red). }\label{fig:onequashingCFT}
\end{figure}

\begin{figure}[ht!]
\centering
\includegraphics[width=0.48\textwidth]{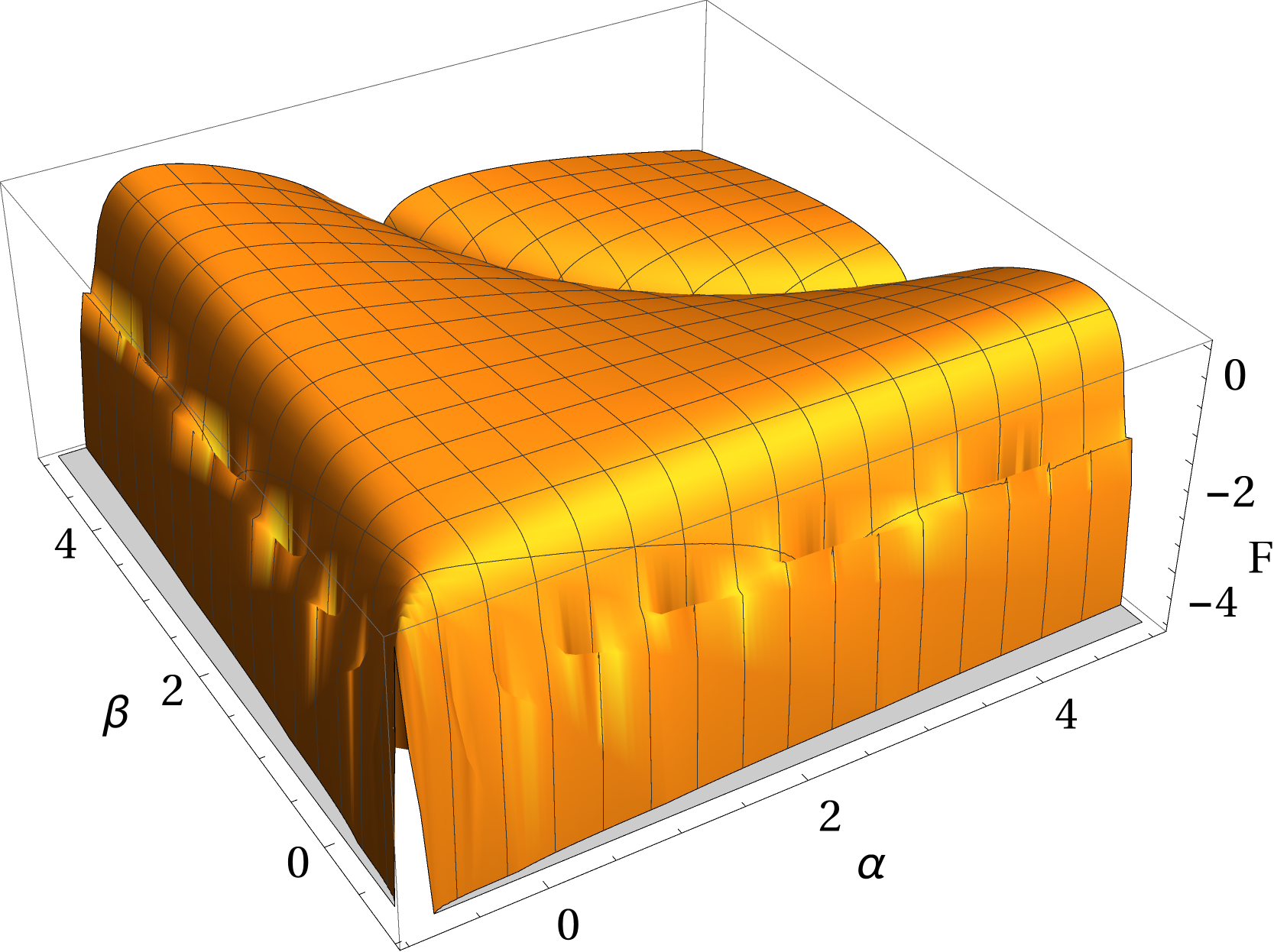}
\caption{The free energy for a free conformally coupled scalar on the double squashed sphere computed using our numerical technique.}\label{fig:minusLogZ}
\end{figure}

The numerical results for the free energy of double squashed three sphere are presented in Figure  \ref{fig:minusLogZ}. One immediately recognizes some similarities with the regularised on-shell action of the new AdS-Taub-NUT solutions constructed in Section \ref{sec:AdS}. For instance, there is a global maximum around $\alpha=\beta=0$, and away from this point the free energy has a local maximum around $\alpha=0$ or $\beta=0$ for positive squashings. For negative squashings the maxima are around $\alpha=\beta$. In the next subsection we analyse the differences and similarities between the gravitational and field theories more thoroughly. 

\begin{figure}[ht!]
\centering
\includegraphics[width=0.45\textwidth]{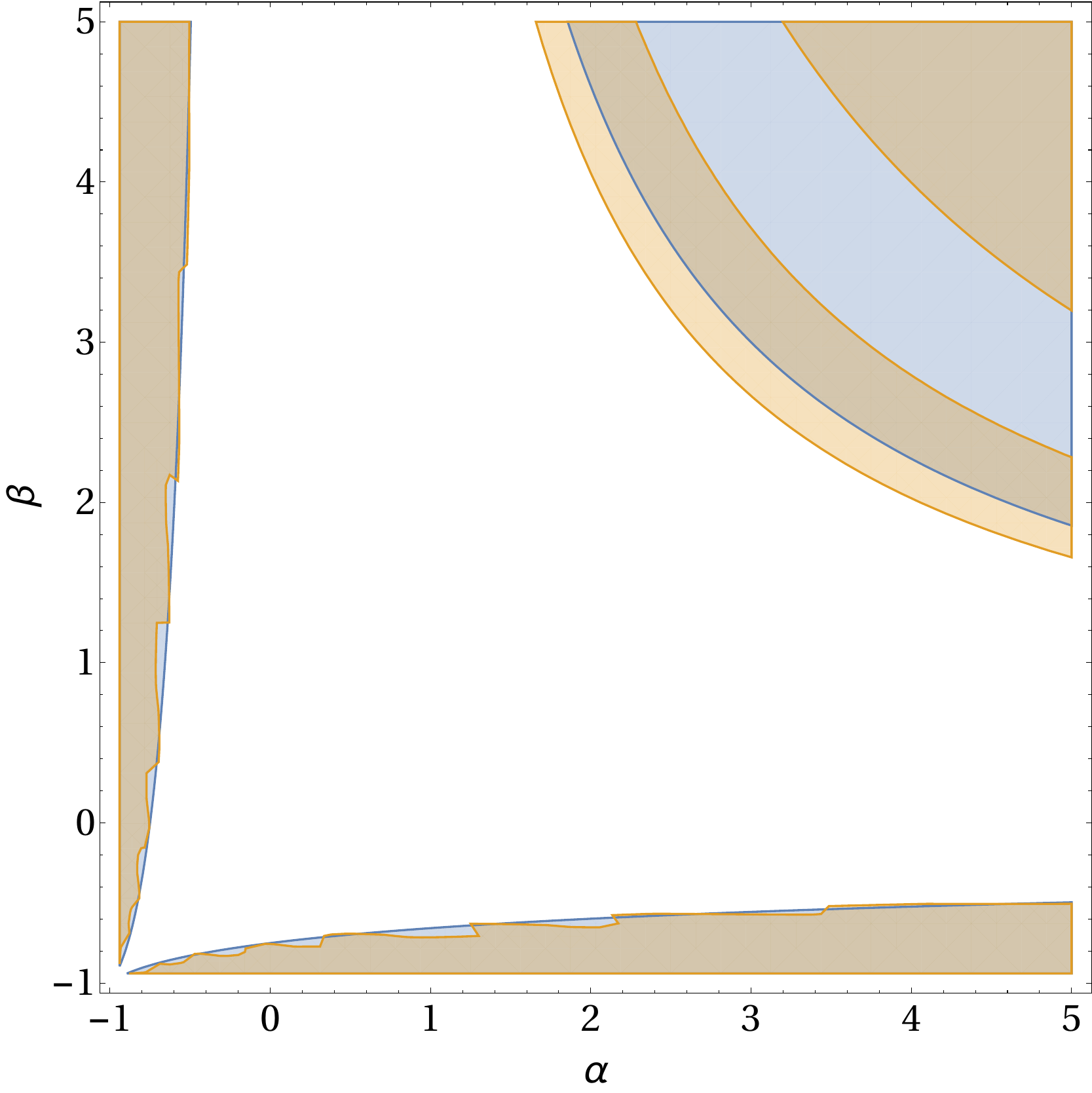}
\caption{Comparison between the region in the $(\alpha,\beta)$ plane where $R$ is less than $0$ (in blue) and the region where the regularised path integral $Z$ vanishes (in orange), which corresponds to a divergent free energy.}
\label{fig:RicciandZSquaredLmax1000}
\end{figure}

There is one more feature in Figure \ref{fig:minusLogZ} we would like to highlight here. The free energy seems to diverge when the Ricci scalar approaches 0. To visualise this better, we plot in Figure \ref{fig:RicciandZSquaredLmax1000} the region where the exponential of the free energy becomes 0 together with the region where $R=0$.  To see what happens we have to inspect the IR behaviour of the free energy \eqref{eqn:detUV}, which is given by a sum over upper incomplete gamma functions  $\Gamma(0,\lambda_{L,k} \delta)$. From the definition of these \eqref{eqn:upperincompleteGamma}, one can see that if $\lambda$ approaches $0$ then $\Gamma(0,\lambda_{L,k} \delta)$ diverges. It is not too difficult to see from the discussion in Appendix \ref{App:eigen} that the lowest eigenvalue of the Laplacian $\nabla^2$ always vanishes. Therefore the first eigenvalue $\lambda_1$  of the conformal Laplacian ($-\nabla^2 +\frac{R}{8}$) becomes zero when the Ricci scalar vanishes. This explains the features in Figure \ref{fig:RicciandZSquaredLmax1000}. We should note that there are also ``higher order" divergences in Figure \ref{fig:minusLogZ} which are less pronounced. These happen when any of the higher eigenvalues of the Laplacian cancels $R\neq0$ to give another zero eigenvalue of the conformal Laplacian. These  divergences always appear in the regions of the $(\alpha,\beta)$ plane where the Ricci scalar $R$ in \eqref{Ricci} is negative. 

\begin{figure}[ht!]
\centering
  \begin{subfigure}[t]{0.32\textwidth}
    \includegraphics[width=\textwidth]{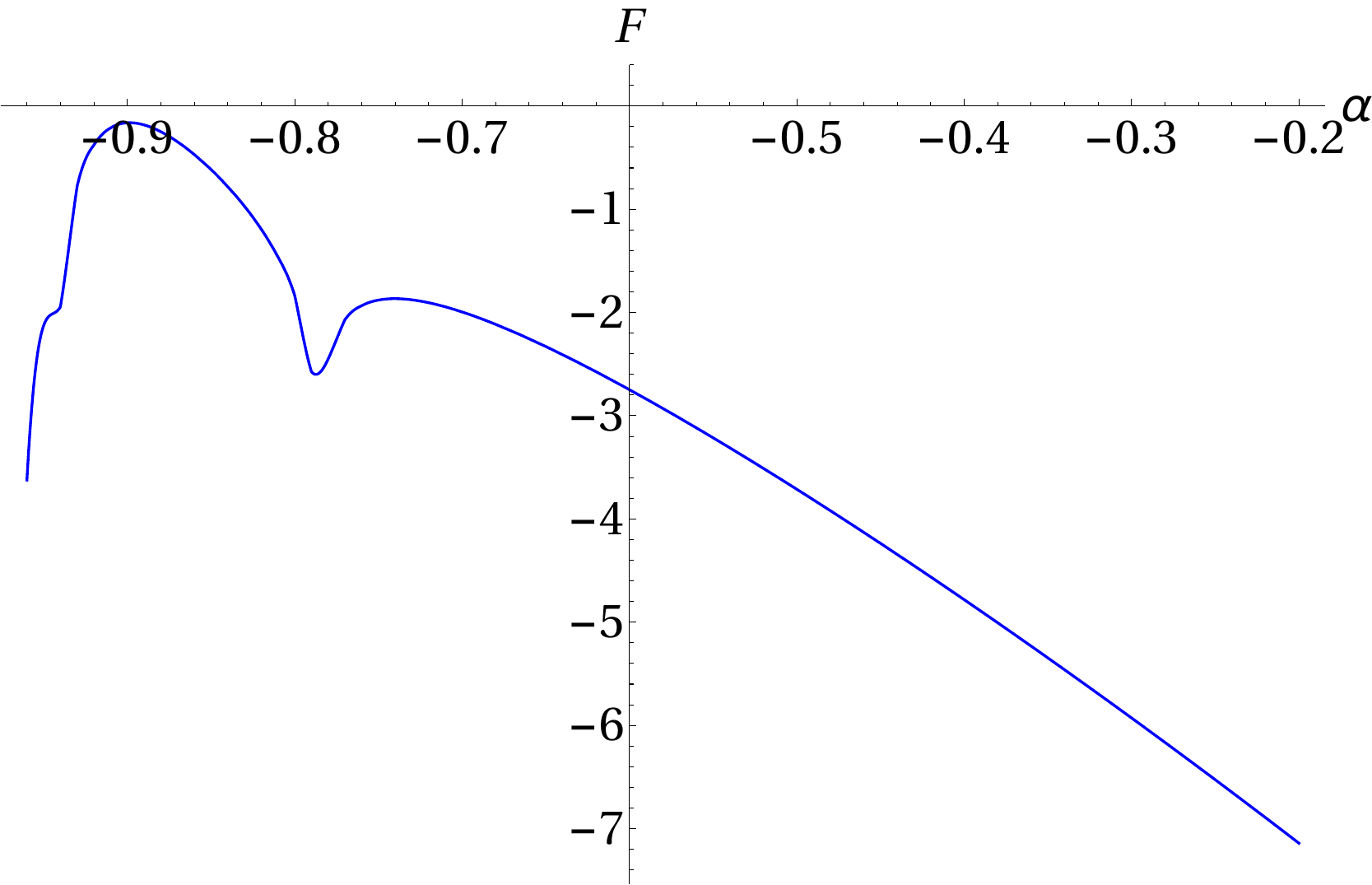}
    \caption{$\beta=-0.9$, small $\alpha$}
    \label{fig:betamin09ActionCFTONZoomed1}
  \end{subfigure}
  \begin{subfigure}[t]{0.32\textwidth}
    \includegraphics[width=\textwidth]{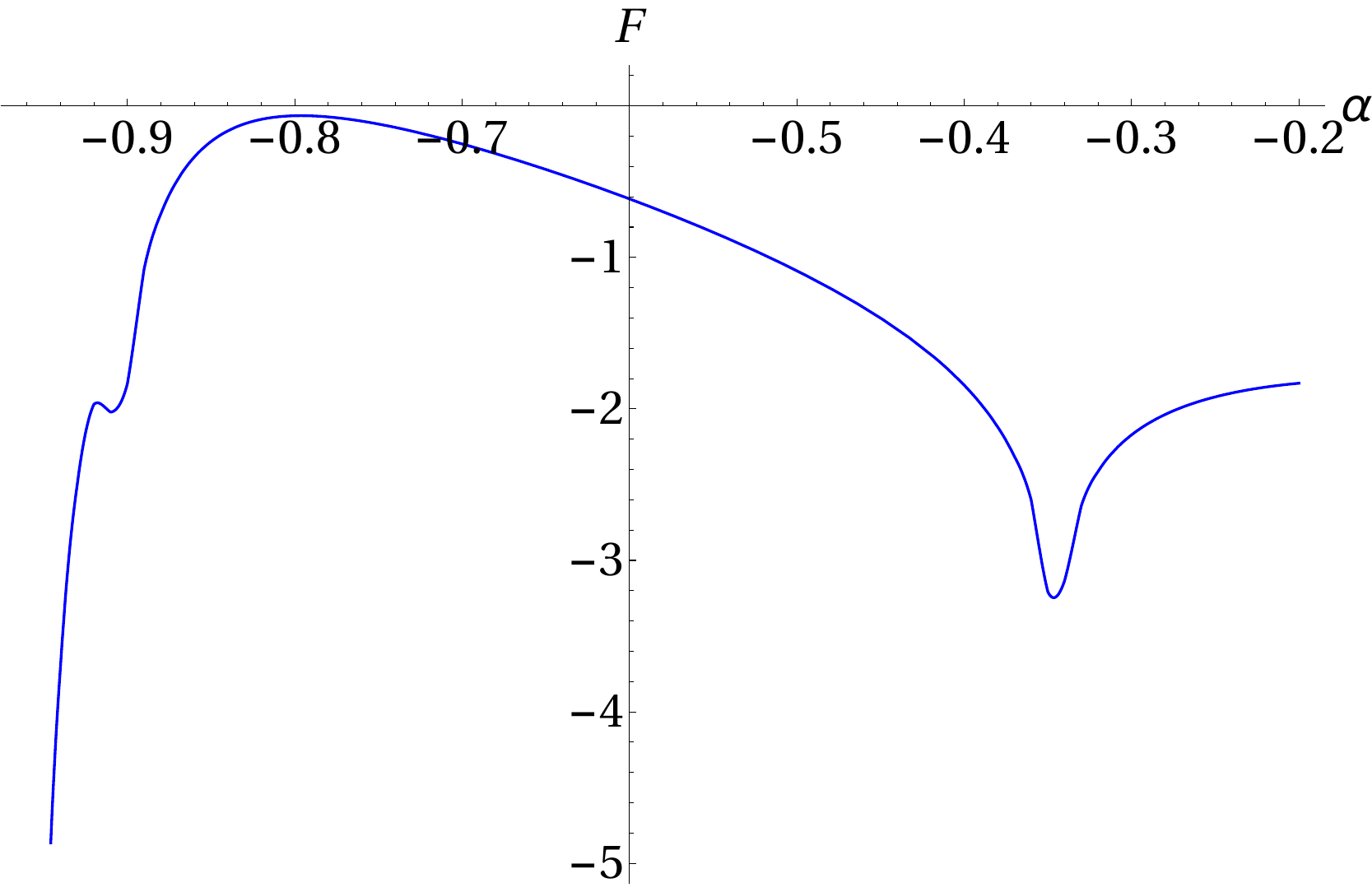}
    \caption{$\beta=-0.8$, small $\alpha$}
    \label{fig:betamin08ActionCFTONZoomed1}
  \end{subfigure}
    \begin{subfigure}[t]{0.32\textwidth}
    \includegraphics[width=\textwidth]{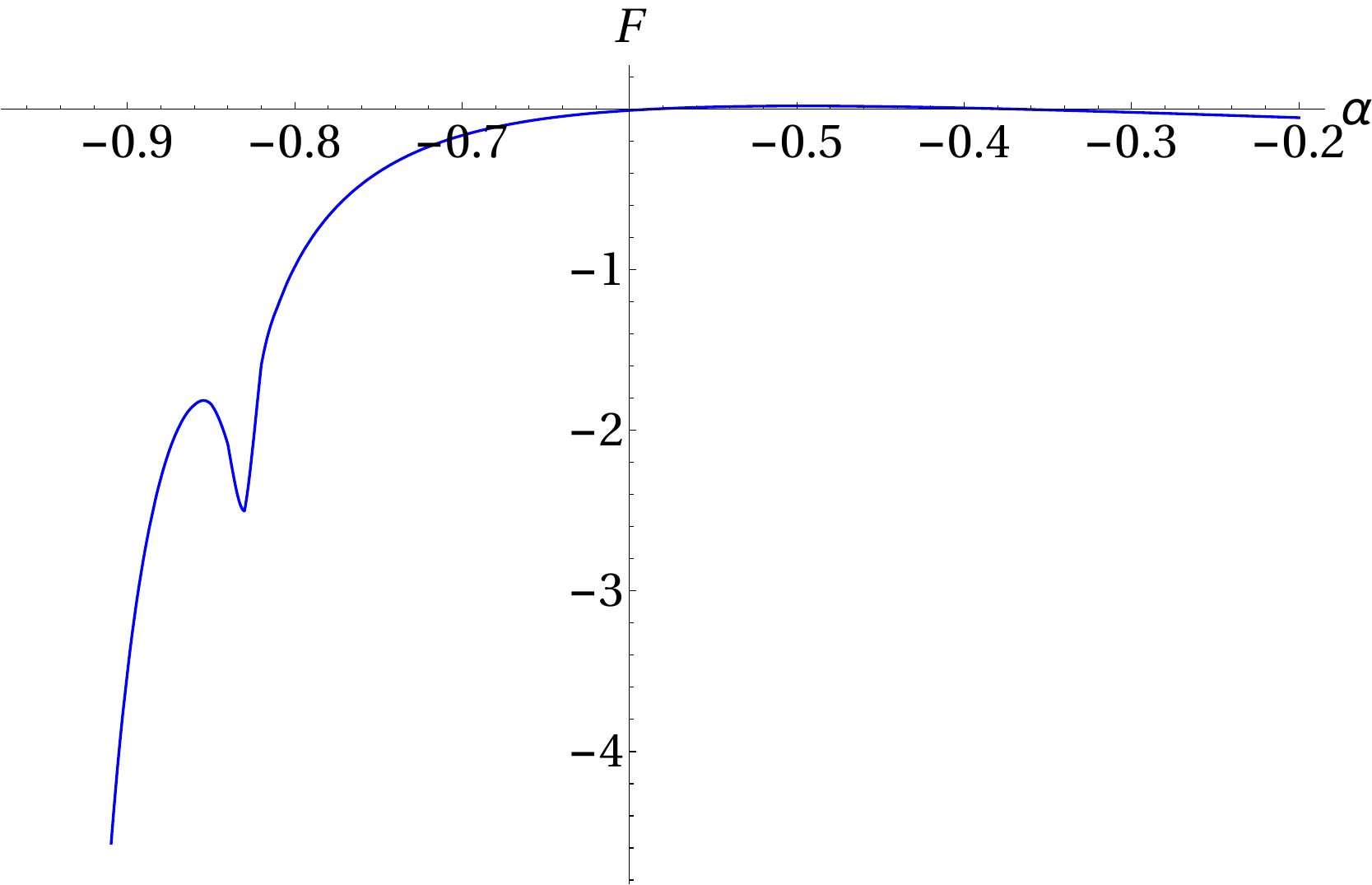}
    \caption{$\beta=-0.52$, small $\alpha$}
    \label{fig:betamin052ActionCFTONZoomed1}
  \end{subfigure} \\
    \begin{subfigure}[t]{0.32\textwidth}
    \includegraphics[width=\textwidth]{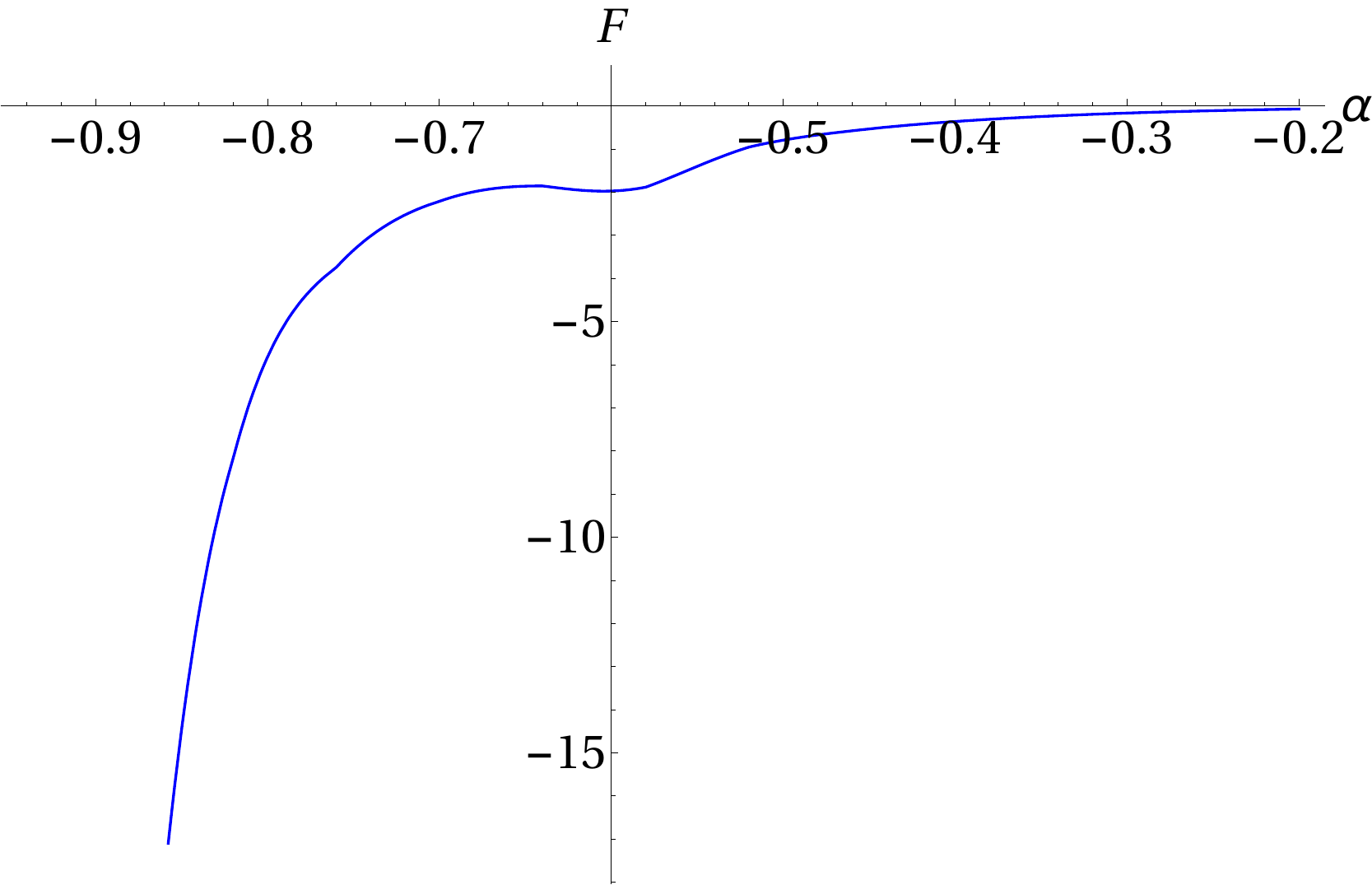}
    \caption{$\beta=2.0$, small $\alpha$}
    \label{fig:betamin20ActionCFTONZoomed1}
  \end{subfigure} \\ 
  \begin{subfigure}[t]{0.32\textwidth}
    \includegraphics[width=\textwidth]{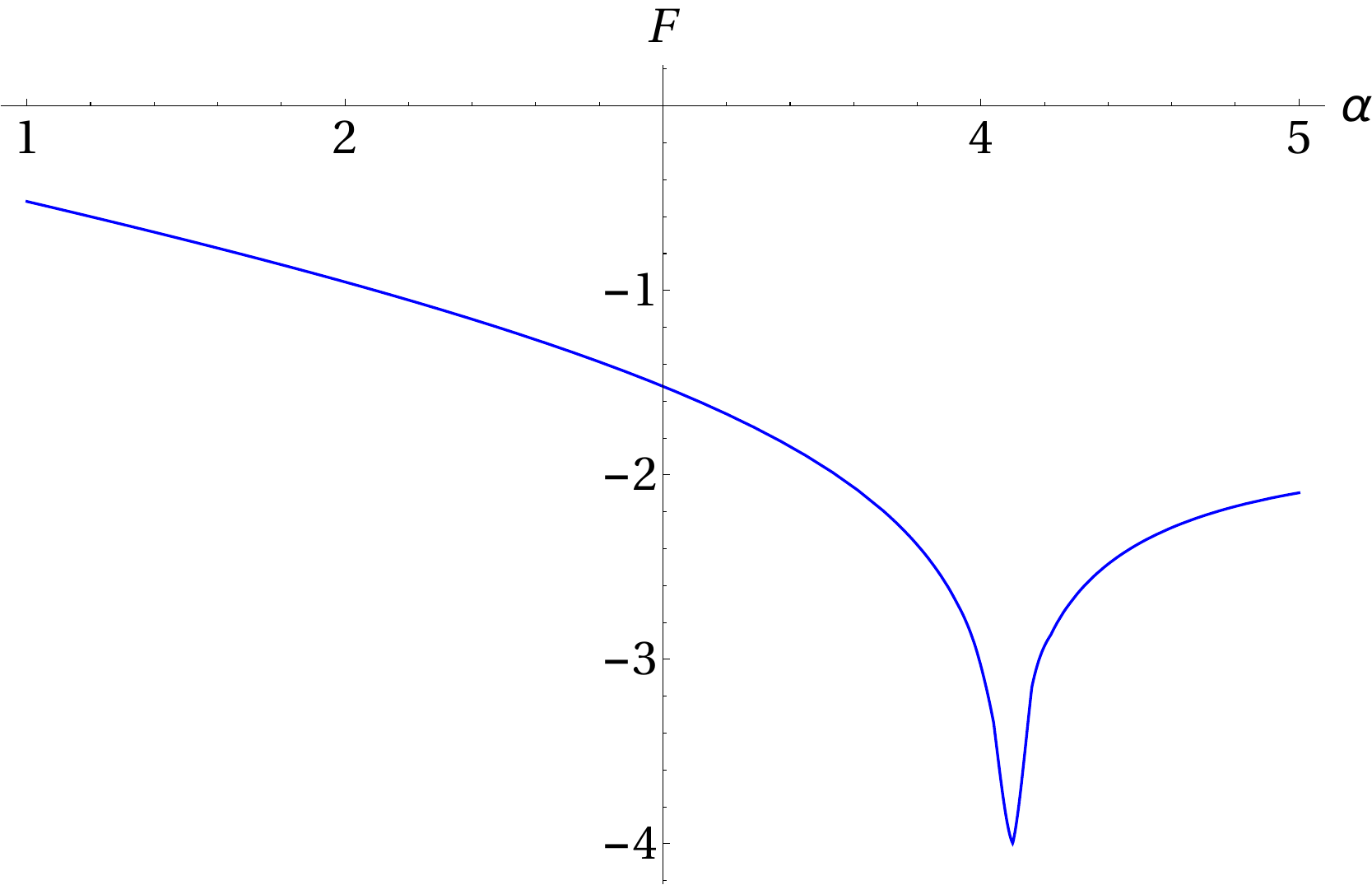}
    \caption{$\beta=-0.52$, large $\alpha$}
    \label{fig:betamin052ActionCFTONZoomed2}
  \end{subfigure}
    \begin{subfigure}[t]{0.32\textwidth}
    \includegraphics[width=\textwidth]{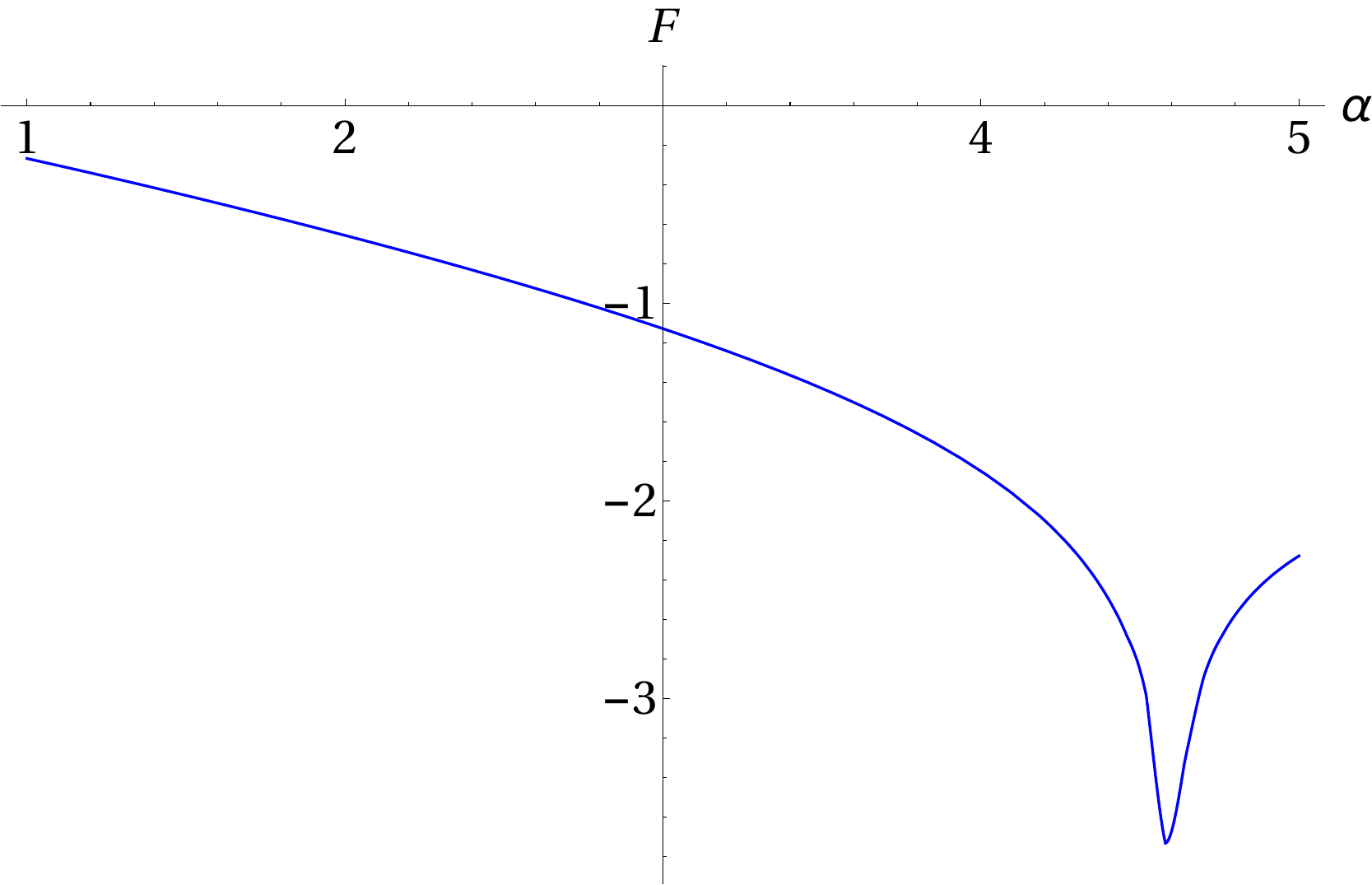}
    \caption{$\beta=2.0$, large $\alpha$}
    \label{fig:betaplus20ActionCFTONZoomed2}
  \end{subfigure} 
  \begin{subfigure}[t]{0.32\textwidth}
    \includegraphics[width=\textwidth]{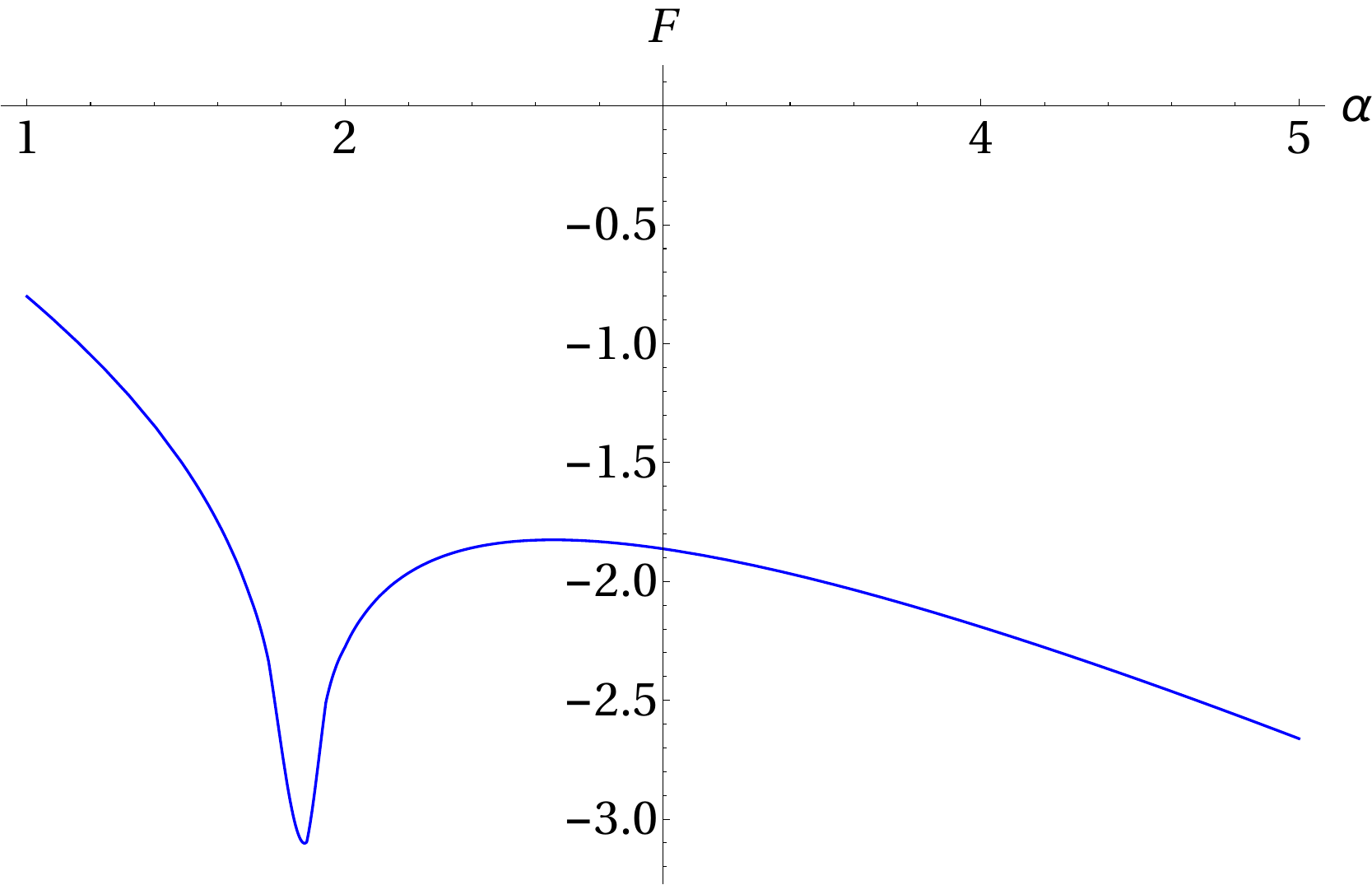}
    \caption{$\beta=5.0$, large $\alpha$}
    \label{fig:betaplus50ActionCFTONZoomed2}
  \end{subfigure}
\caption{ behaviour of the CFT free energy for fixed $\beta$ in the region of small and large $\alpha$. }\label{fig:CFTSlicesBeta}
\end{figure}

The divergences give rise to sharp features in the function $F(\alpha,\beta)$. To illustrate this we show in Figure \ref{fig:CFTSlicesBeta} the free energy as a function of $\alpha$ for a number of different fixed values of $\beta$. We show four plots for a small interval of $\alpha$ around $\alpha=0$. The divergences show up as two sharp negative spikes for small $\beta$ and they both move to the right when $\beta$ is increased. The right spike moves much faster to the right and goes to $\infty$ when $\beta$ becomes $0$ whereas the left spike converges slowly to $\alpha=0$ when $\beta$ diverges. We also show three plots of the behaviour at larger $\alpha$. Here we can only see the right spike, which first moves to $\infty$ when $\beta$ approaches $0$ and then returns for larger $\beta$ eventually going to $\alpha=0$ when $\beta\to\infty$. 

\subsection{Holographic interlude}
\label{sec:AdSCFT}

We can now attempt to compare the regularised on-shell action of our new asymptotically $AdS_4$ solutions of gravity with a double squashed sphere at the boundary with the free energy of the $O(N)$ vector model on the same squashed sphere. We should emphasize from the start that there is no a priori reason to expect that there is any duality between these two theories. The free $O(N)$ model should be dual to a higher-spin Vasiliev theory in $AdS_4$ and this theory is very different from pure Einstein gravity with a negative cosmological constant. Nevertheless the results in \cite{Hartnoll:2005yc} suggest that there are some qualitative similarities between these models which we further explore here.

\begin{figure}[ht!]
\centering
\includegraphics[width=0.49\textwidth]{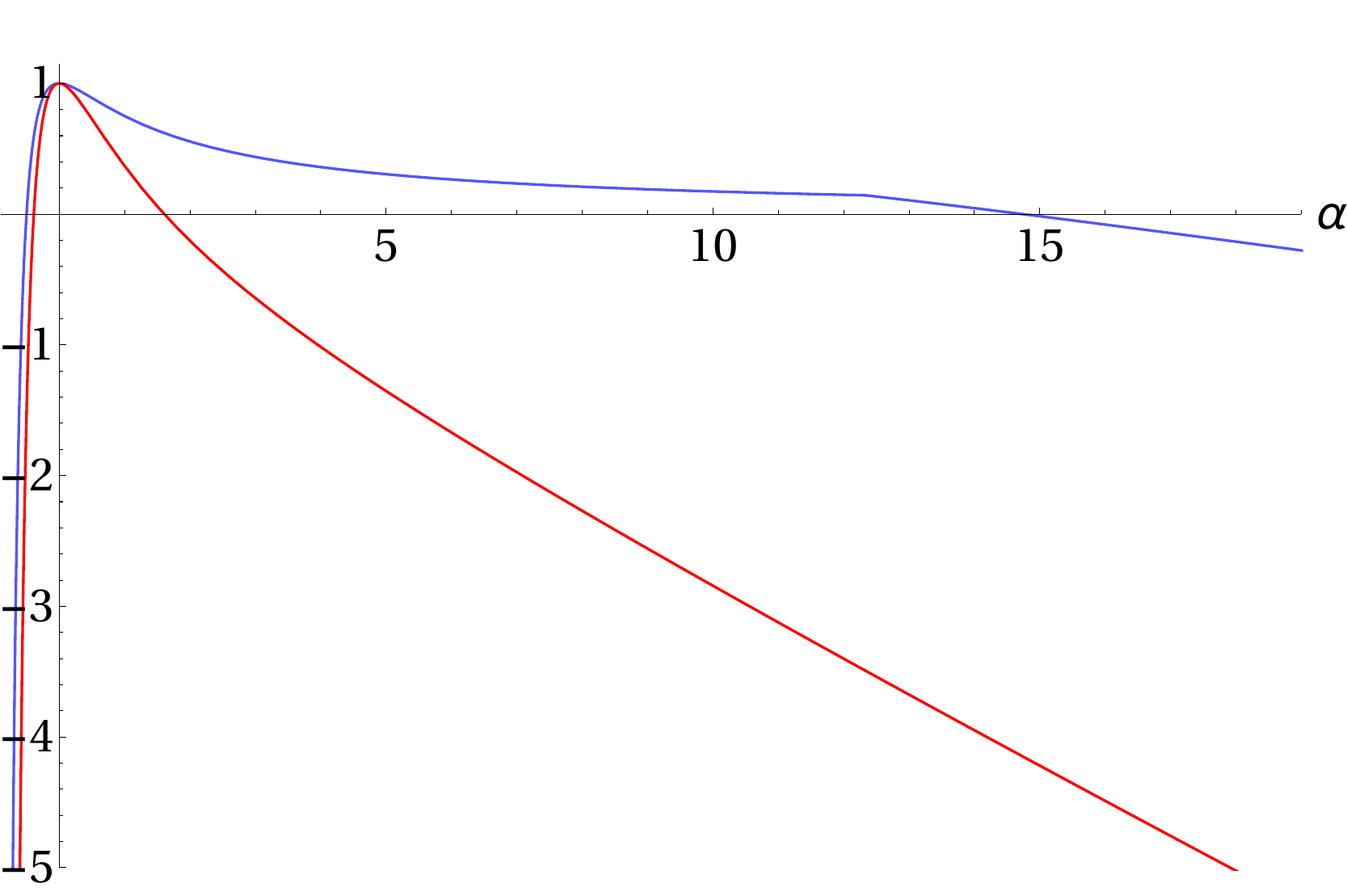}
\caption{Comparison between the free energy of the free $O(N)$ model for $\beta=0$ and as a function of $\alpha$(red) and the on-shell action for the analytic AdS-Taub-Nut/Bolt solutions (blue). For the gravitational results we included the phase transition from NUT to Bolt. All results are normalised to give 1 for $\alpha=\beta=0$.}\label{fig:ancomparisonAdSCFT}
\end{figure}

First let us focus on the case when there is only one nontrivial squashing parameter, i.e. $\beta=0$. In Figure \ref{fig:ancomparisonAdSCFT} we plot the regularised on-shell action of the corresponding gravitational solution and compare it with the free energy of the free $O(N)$ vector model. We chose to normalise both quantities such that for $\alpha=0$ they are equal to 1 and focus on their dependence on the squashing parameter $\alpha$. There are clear similarities between the two functions in the region $\alpha<0$. For $\alpha>0$ the similarity is only qualitative, i.e. both functions decrease as $\alpha$ increases. A notable difference is that the gravitational solutions exhibit a Hawking-Page phase transition for a relatively large positive value of $\alpha$. Such a phase transition is of course absent in a free quantum field theory. Another notable feature is that the free energy of the CFT diverges for $\alpha=-3/4$ due to the simple fact that the lowest eigenvalue of the conformal Laplacian in \eqref{eqn:LogZGeneral} vanishes at this value of the squashing. There is no corresponding divergence in the gravitational on-shell action. We believe that this discrepancy is entirely due to the fact that we are considering a free CFT. Indeed the large $N$ analysis of the free energy of the interacting three-dimensional $O(N)$ vector model in \cite{Hartnoll:2005yc} shows that this divergence in the free energy is removed. For large values of $\alpha$ both functions decrease linearly. From our numerical results we can estimate that the ratio of the slopes of these linear functions is approximately $3.7$.

Let us now compare the gravitational on-shell action and the field theory free energies when both squashing parameters do not vanish. The regularised bulk on-shell action is presented  in Figure \ref{fig:actionsAdStwosquashings} and the field theory free energy is plotted in Figure \ref{fig:minusLogZ}. Some of the similarities between these two figures were already mentioned, the qualitatively similar overall behaviour as well as the global maximum at $\alpha=\beta=0$. However, just like in the case of one squashing parameter, we also have differences between the two quantities: the phase transition in the bulk from NUT to Bolt which doesn't appear in the free CFT, the diverging free energy in the free field theory when $R$ becomes $0$, and the different asymptotic fall-off behaviour for large values of the squashing parameters.

\begin{figure}[ht!]
\centering
  \begin{subfigure}[t]{0.32\textwidth}
    \includegraphics[width=\textwidth]{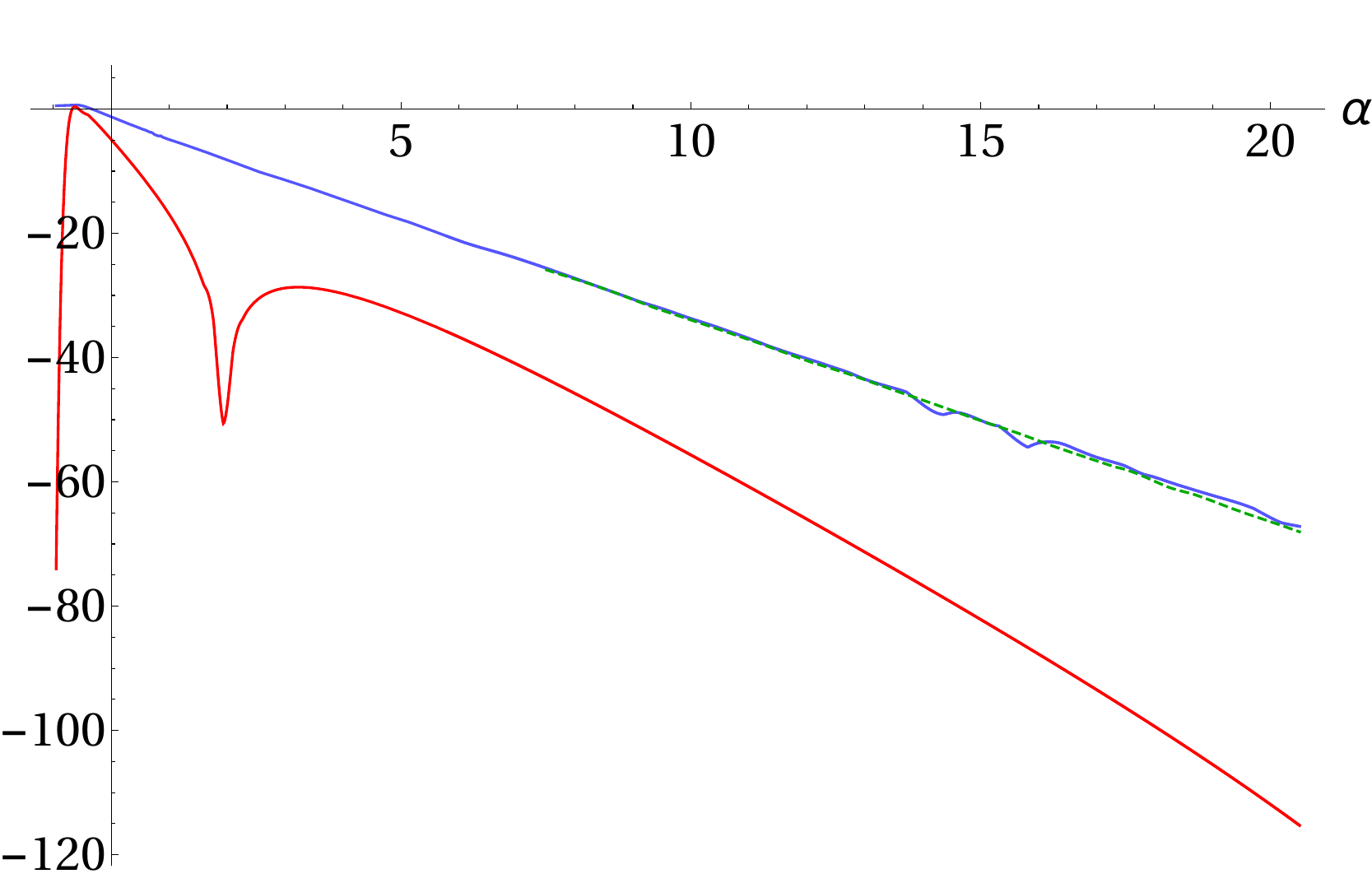}
    \caption{$\beta=-0.6$}
    \label{fig:betamin060ActionComparedCFTBulk}
  \end{subfigure}
  \begin{subfigure}[t]{0.32\textwidth}
    \includegraphics[width=\textwidth]{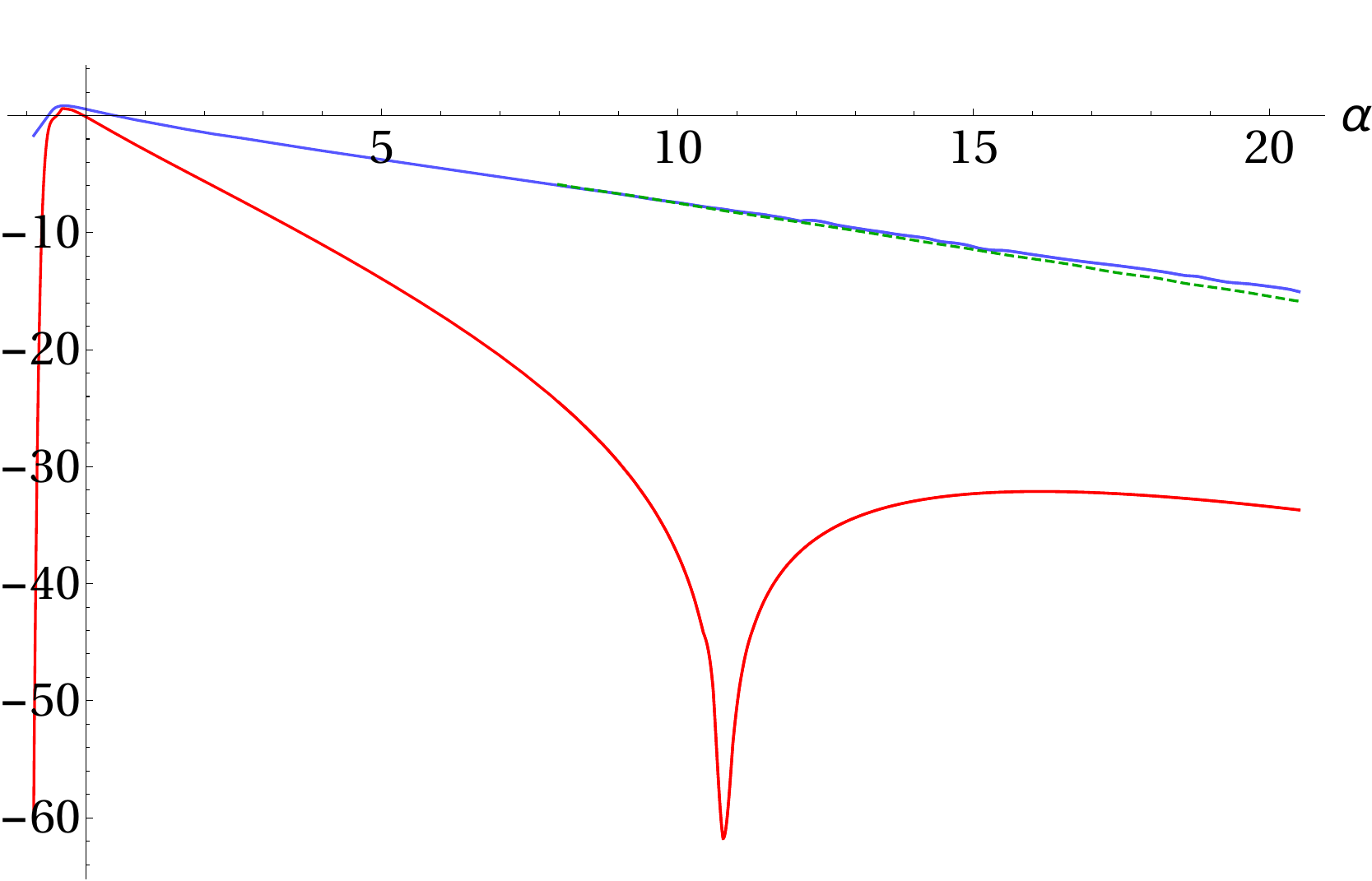}
    \caption{$\beta=-0.4$}
    \label{fig:betamin040ActionComparedCFTBulk}
  \end{subfigure}
    \begin{subfigure}[t]{0.32\textwidth}
    \includegraphics[width=\textwidth]{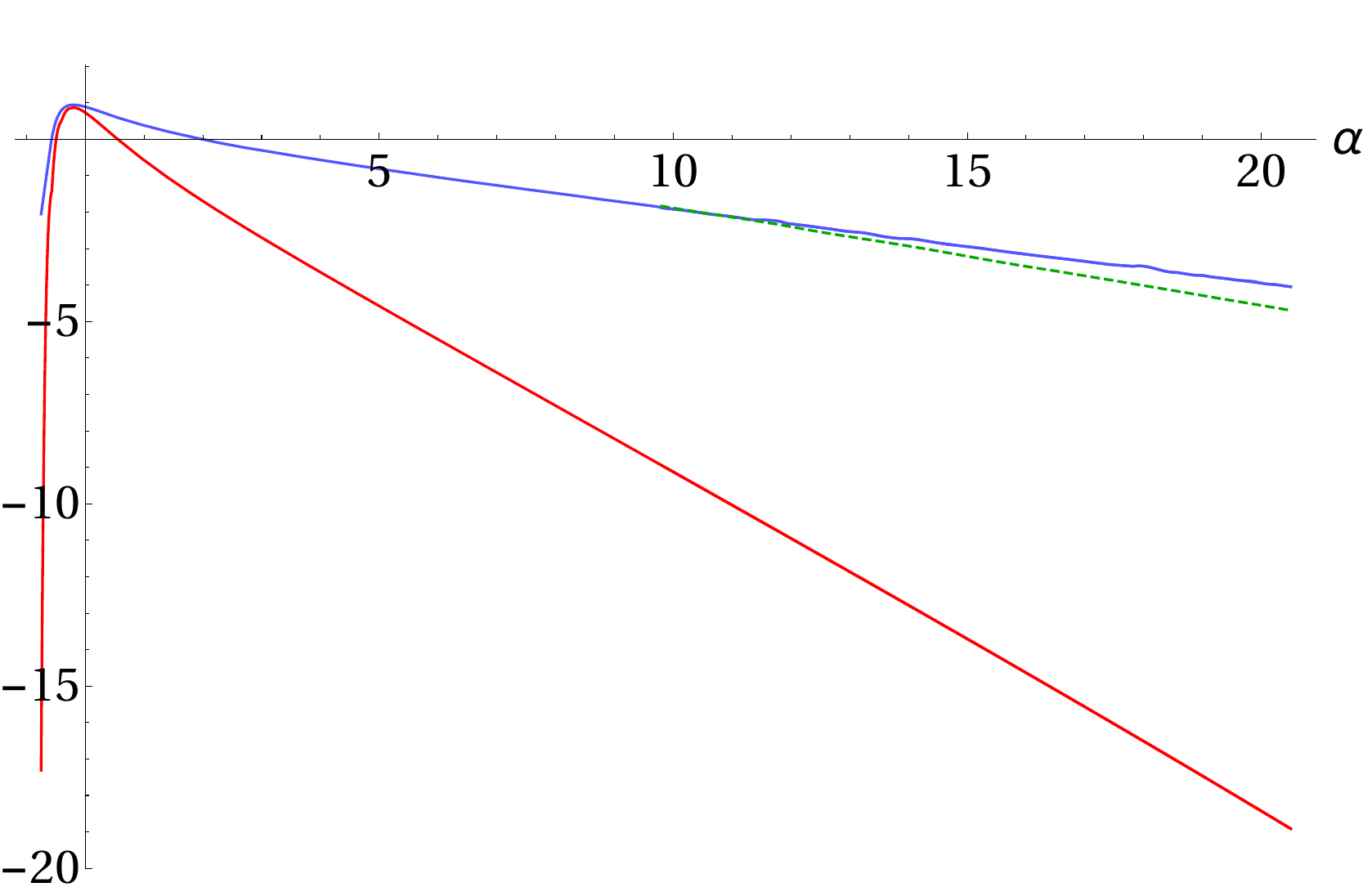}
    \caption{$\beta=-0.25$}
    \label{fig:betamin025ActionComparedCFTBulk}
  \end{subfigure} \\
    \begin{subfigure}[t]{0.32\textwidth}
    \includegraphics[width=\textwidth]{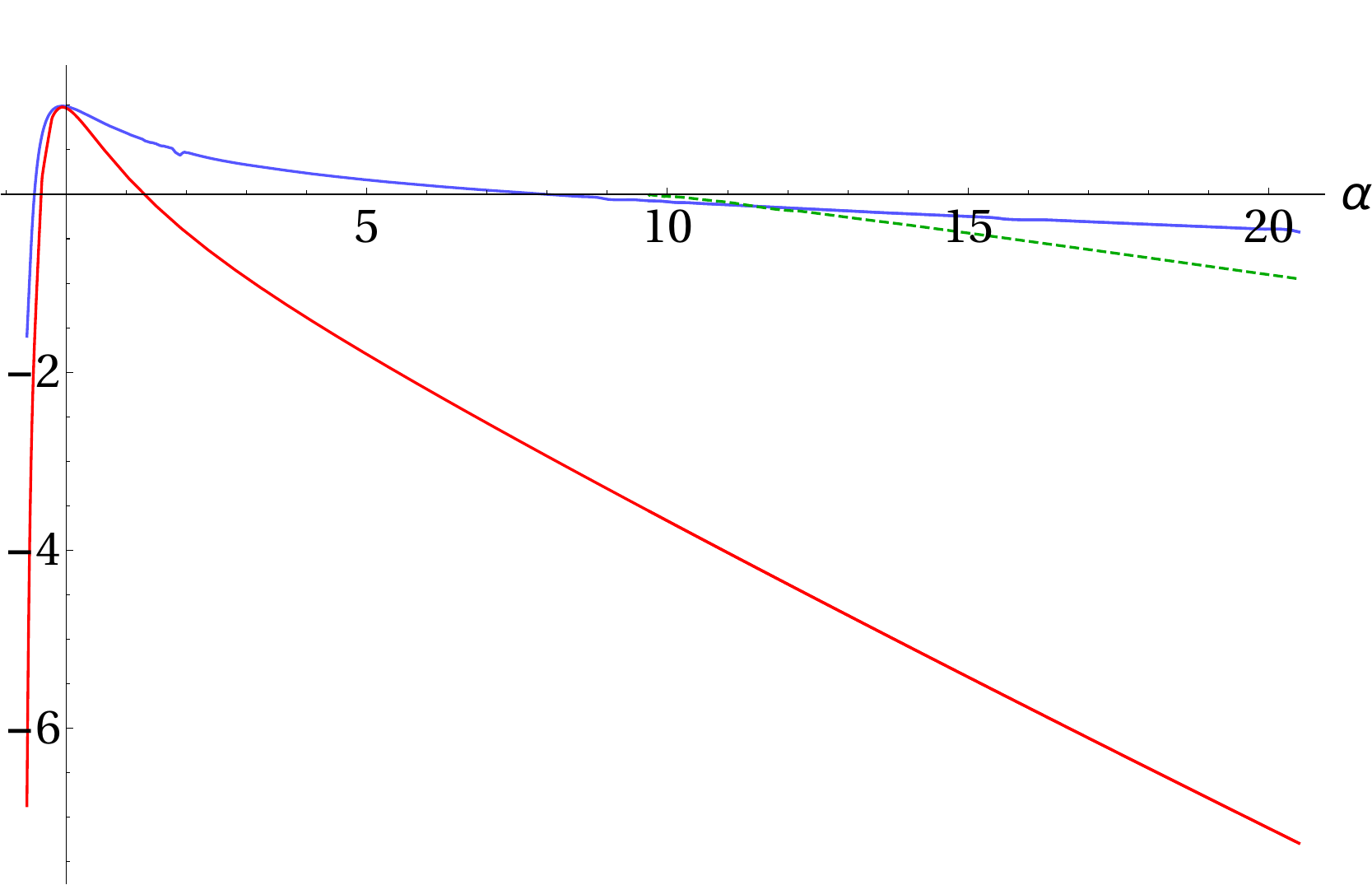}
    \caption{$\beta=-0.1$}
    \label{fig:betamin010ActionComparedCFTBulk}
  \end{subfigure}
  \begin{subfigure}[t]{0.32\textwidth}
    \includegraphics[width=\textwidth]{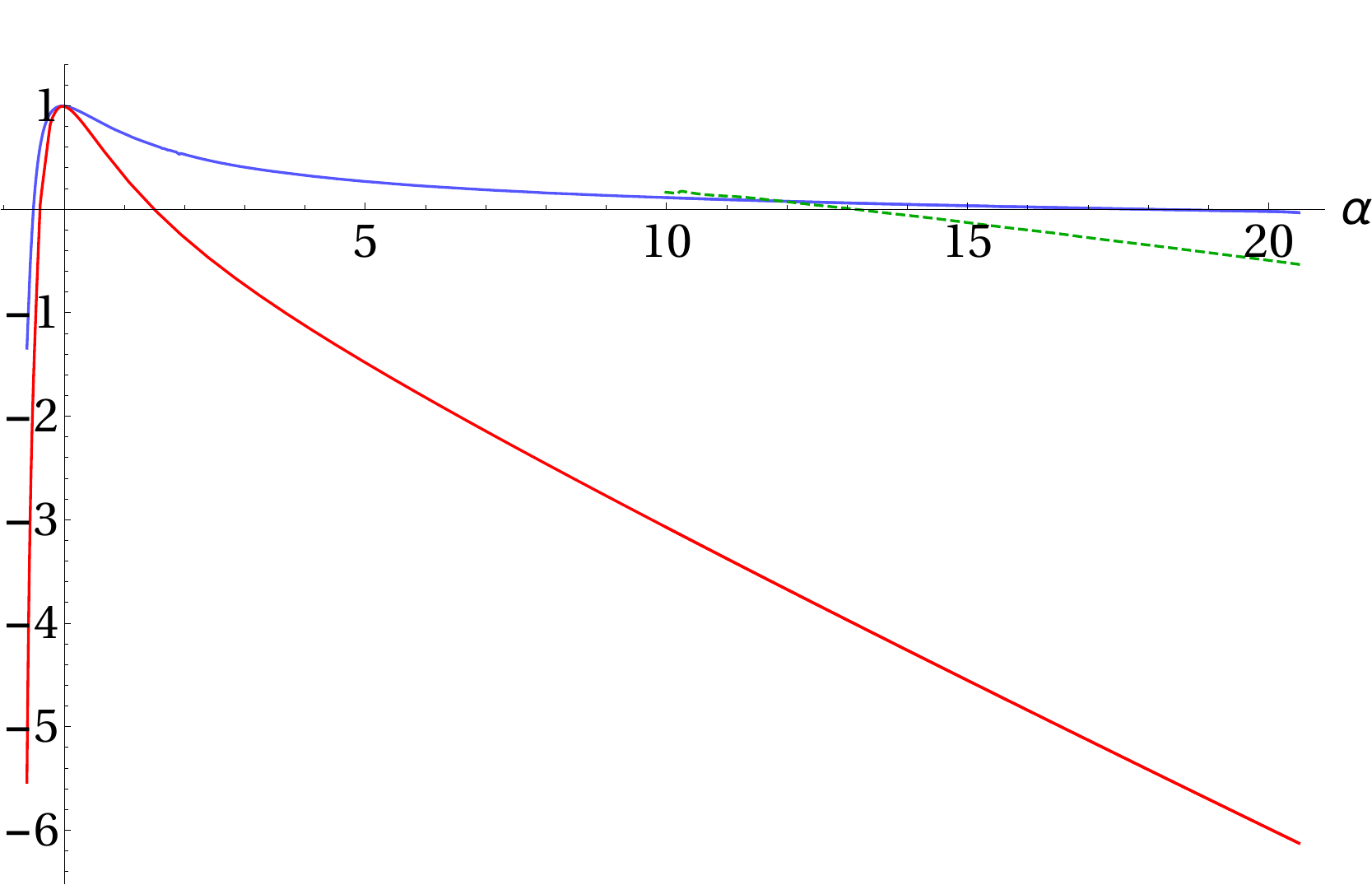}
    \caption{$\beta=-0.05$}
    \label{fig:betamin005ActionComparedCFTBulk}
  \end{subfigure}
    \begin{subfigure}[t]{0.32\textwidth}
    \includegraphics[width=\textwidth]{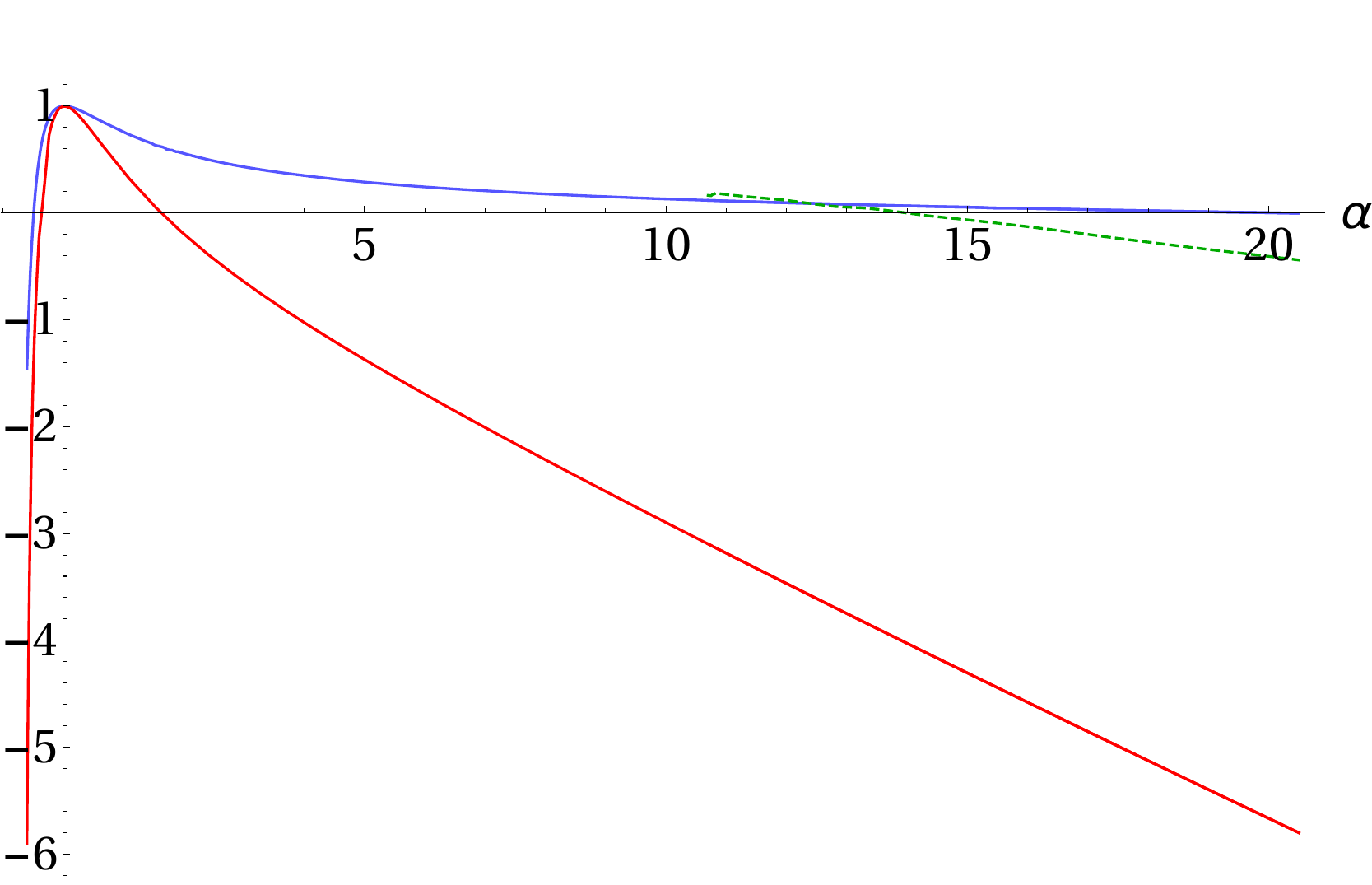}
    \caption{$\beta=0.05$}
    \label{fig:betaplus005ActionComparedCFTBulk}
  \end{subfigure} \\
  \begin{subfigure}[t]{0.32\textwidth}
    \includegraphics[width=\textwidth]{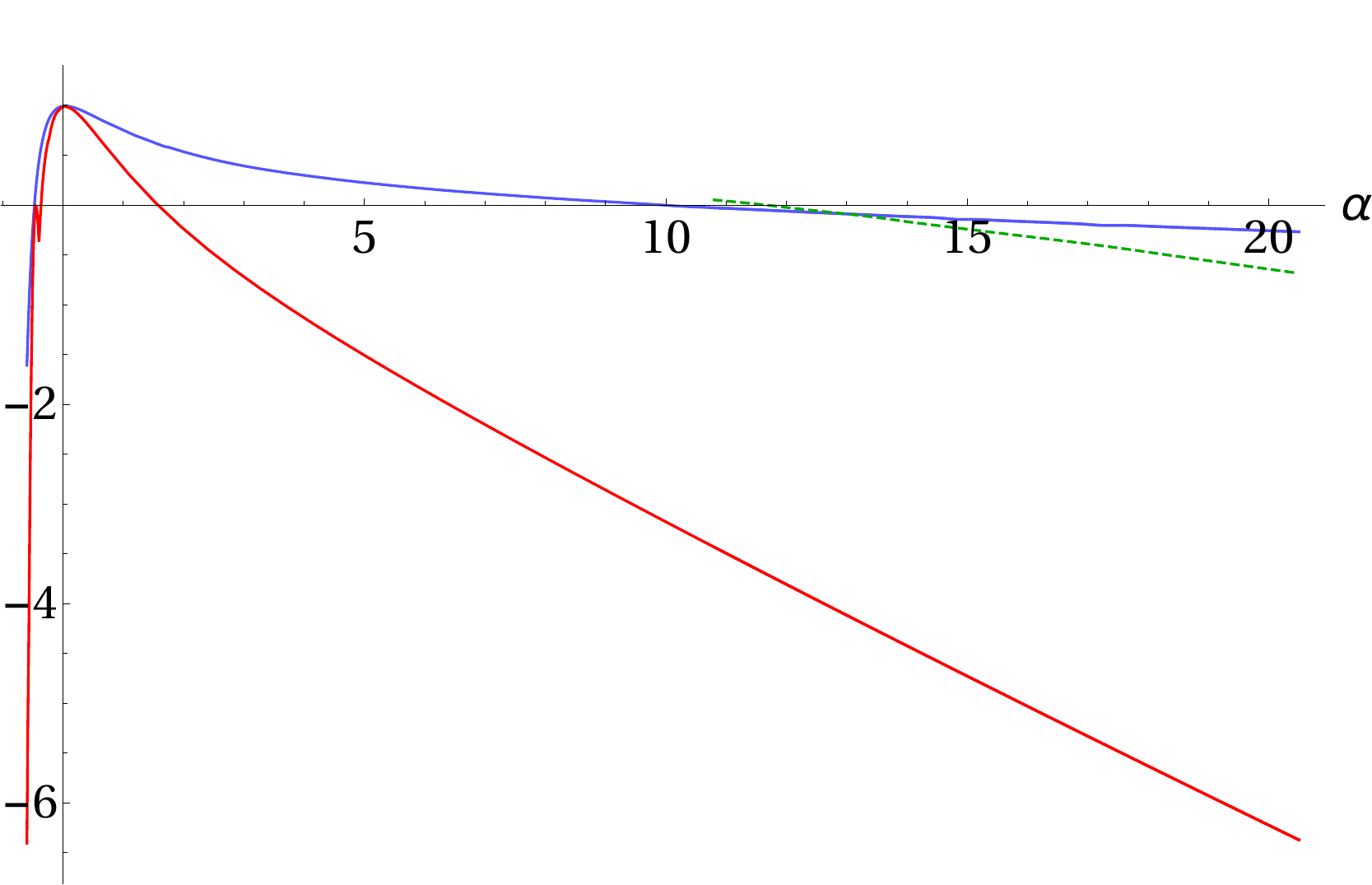}
    \caption{$\beta=0.10$}
    \label{fig:betaplus010ActionComparedCFTBulk}
  \end{subfigure}
  \begin{subfigure}[t]{0.32\textwidth}
    \includegraphics[width=\textwidth]{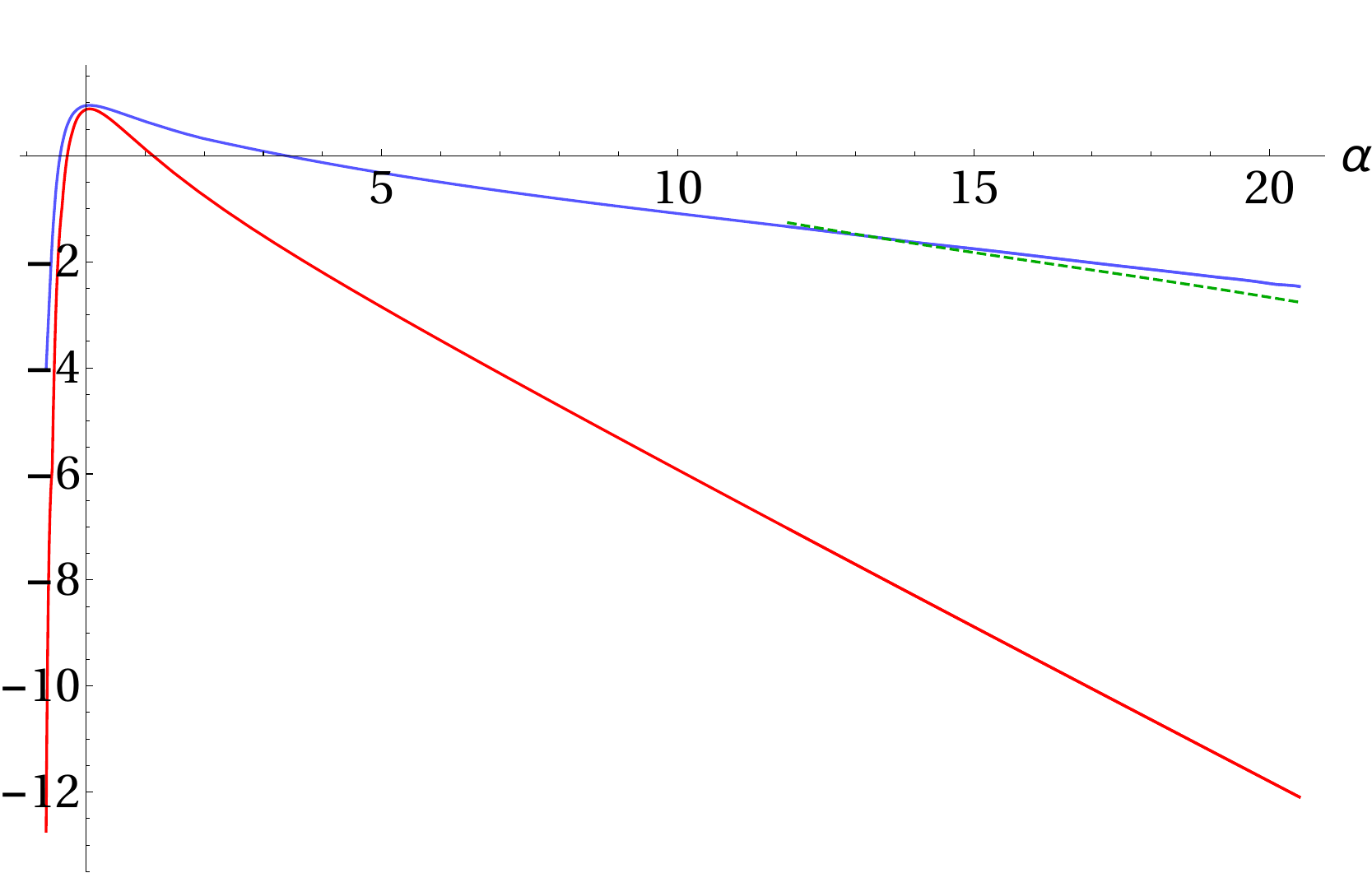}
    \caption{$\beta=0.30$}
    \label{fig:betaplus030ActionComparedCFTBulk}
  \end{subfigure}
  \begin{subfigure}[t]{0.32\textwidth}
    \includegraphics[width=\textwidth]{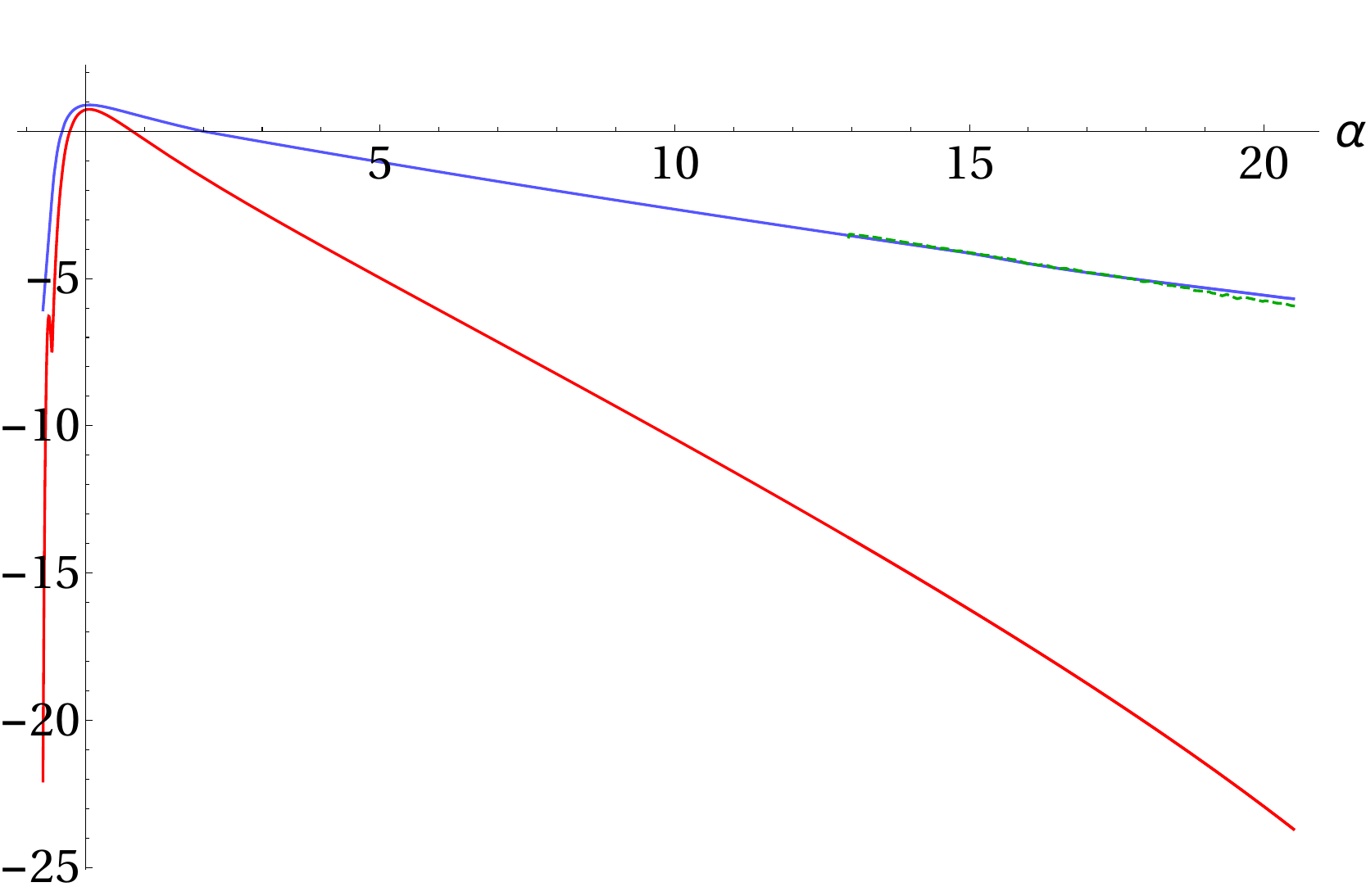}
    \caption{$\beta=0.50$}
    \label{fig:betaplus050ActionComparedCFTBulk}
  \end{subfigure}\\
  \begin{subfigure}[t]{0.32\textwidth}
    \includegraphics[width=\textwidth]{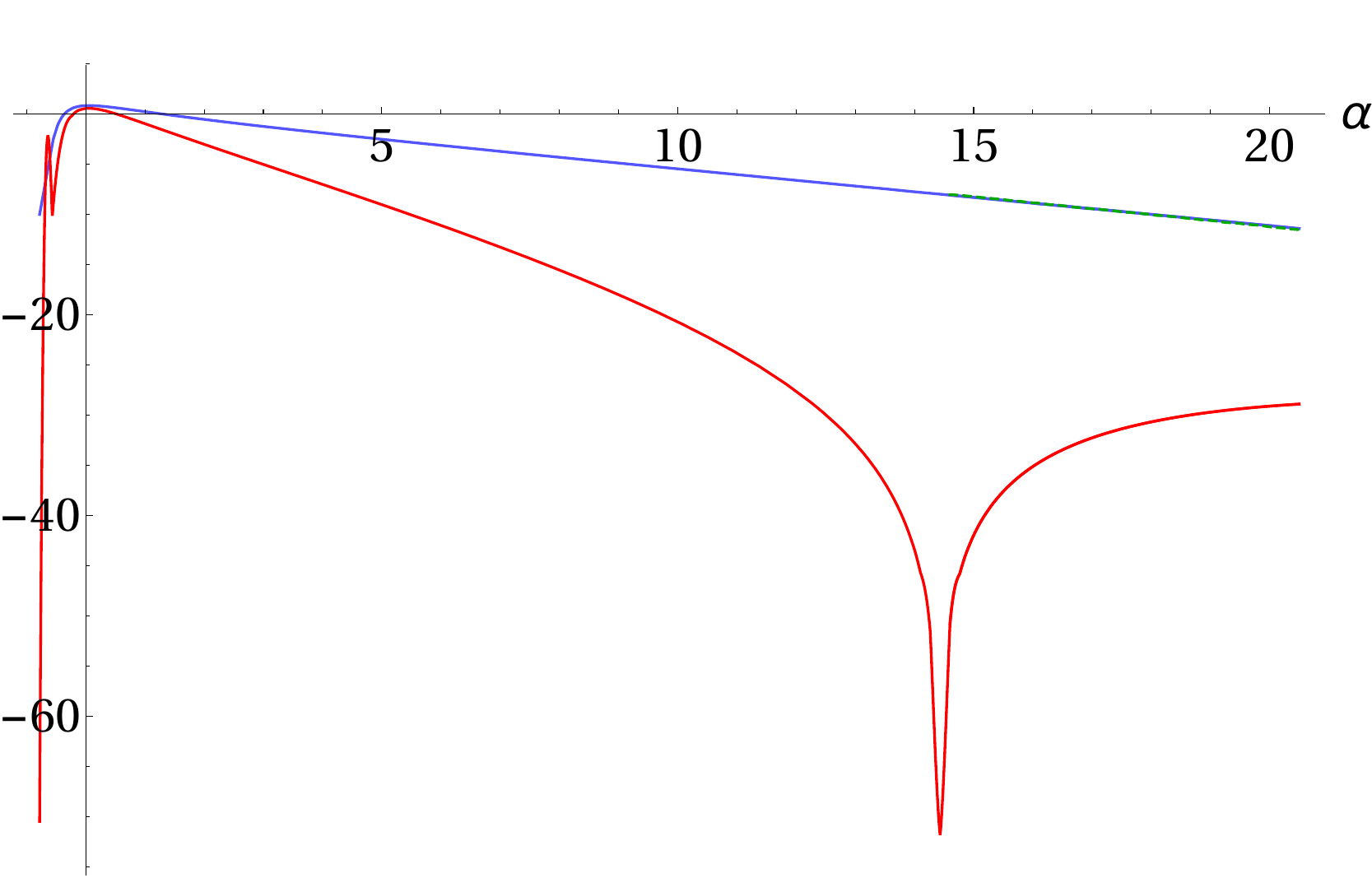}
    \caption{$\beta=0.80$}
    \label{fig:betaplus080ActionComparedCFTBulk}
  \end{subfigure}
  \begin{subfigure}[t]{0.32\textwidth}
    \includegraphics[width=\textwidth]{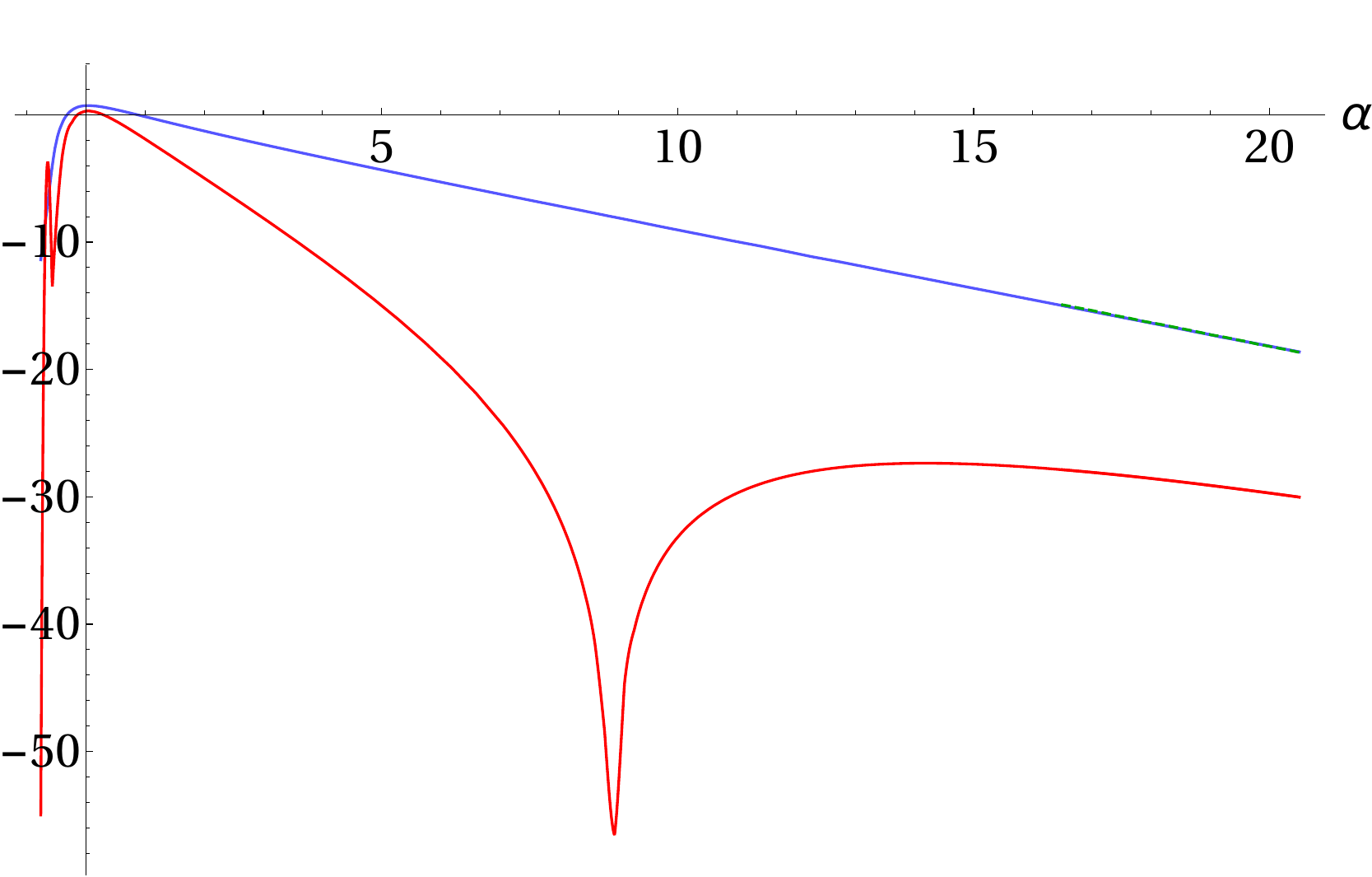}
    \caption{$\beta=1.15$}
    \label{fig:betaplus115ActionComparedCFTBulk}
  \end{subfigure}
  \begin{subfigure}[t]{0.32\textwidth}
    \includegraphics[width=\textwidth]{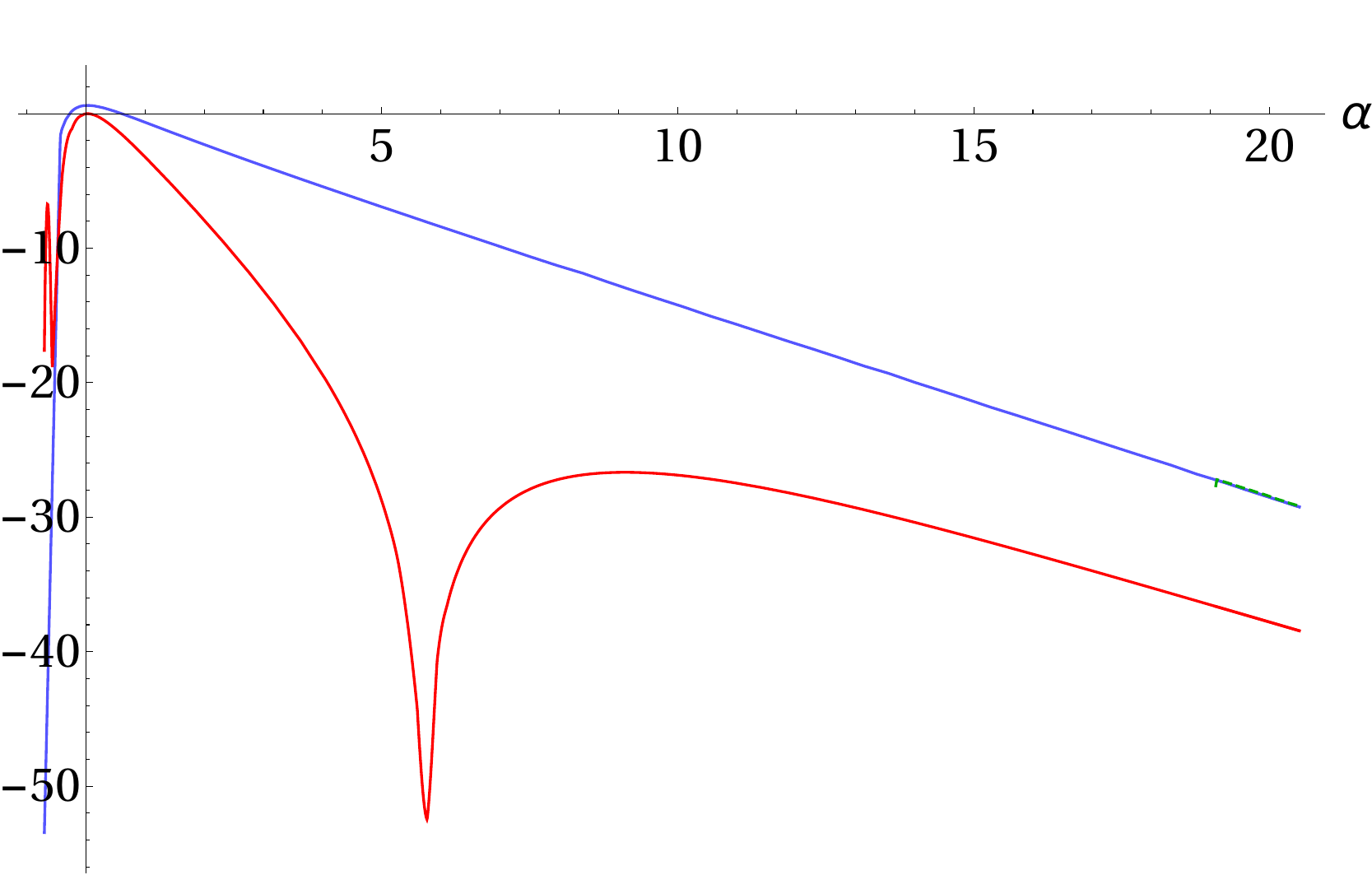}
    \caption{$\beta=1.65$}
    \label{fig:betaplus165ActionComparedCFTBulk}
  \end{subfigure}
\caption{Comparison, for different fixed values of $\beta$, between the free energy of the CFT (in red), the Taub-NUT (in blue) and Taub-Bolt (in dashed green) on-shell actions. All functions are normalised to give 1 when both squashings parameters are zero.}\label{fig:compareSlicesBeta}
\end{figure}

To have a better understanding of the similarities and differences between the gravitational and field theory results we provide plots of the free energy and on-shell action as a function of $\alpha$ for fixed values of $\beta$. This is shown in Figure  \ref{fig:compareSlicesBeta} where the CFT results are plotted in red, the on-shell action of the Taub-NUT solutions is in blue and the one for Taub-Bolt solutions is indicated with a green dashed line. From this figure the conclusions of the previous paragraphs are immediately obvious. We see that for $\beta<0$ the maxima are around $\alpha=\beta$, for positive $\beta$ the maximum is always at $\alpha=\beta=0$. The general behaviour for small and large $\alpha$ is also again comparable, but in general the CFT free energy falls off much faster than the bulk on-shell action. 

Finally we would like to comment on how our results fit into the framework of the F-theorem for three-dimensional CFTs and their holographic duals \cite{Jafferis:2010un,Jafferis:2011zi,Klebanov:2011gs}. The F-theorem states that the free energy of a CFT on $S^3$ decreases along an RG flow triggered by a relevant deformation. We can think of the $O(N)$ model defined on the squashed sphere with the metric in \eqref{eqn:metric} as a CFT perturbed by a relevant deformation. To be more precise the deformation is triggered by the coupling of the energy-momentum tensor of the CFT to the curved background metric. From this perspective the fact that we find that the free energy has a global maximum at $\alpha=\beta=0$ is entirely compatible with the F-theorem, namely the squashing deformations decrease the free energy. The same conclusion can be drawn from the on-shell gravitational action that we computed since we can interpret it holographically as the function that should decrease monotonically under RG flows.

\section{Hartle-Hawking wave function in anisotropic minisuperspace}
\label{sec:Cosmology}

The dS/CFT correspondence \cite{Balasubramanian2001,Strominger2001,Maldacena2002} conjectures that the Hartle-Hawking wave function of the universe $\Psi_{HH}$ with future de Sitter boundary conditions is given at late times in terms of the partition functions of deformations of a Euclidean CFT defined on the future boundary. 

Euclidean AdS/CFT implies a realisation of dS/CFT that is valid in the semiclassical approximation in Einstein gravity and possibly exact in Vasiliev gravity in dS where the duals are Euclidean $Sp(N)$ vector models \cite{Anninos2011}. In this context dS/CFT is often understood as an analytic continuation from Euclidean AdS to Lorentzian dS \cite{Maldacena2002,Harlow2011,McFadden2009,Hartle2012a,Hartle2012b}. However the duality in dS can also be expressed more generally and directly in terms of the AdS/CFT dual partition functions $Z_{QFT}$, as follows
\cite{Hertog2011}, 
\be
\Psi_{HH} [h_{ij}, \phi]= Z^{-1}_{QFT}[\tilde h_{ij},\zeta] \exp(iS_{ct}[h_{ij}, \phi]/\hbar)   \ .
\label{dSCFT}
\ee
where $h_{ij} = a^2 \tilde h_{ij}$ with $\mathrm{Vol}(\tilde h_{ij}) =1$ is the metric on the asymptotic boundary and $a$ is an overall scale factor, and where $\phi$ stands for bulk matter fields, locally related to the sources $\zeta$. We set these to zero in this paper.

In this formulation the connection between AdS/CFT and a cosmological wave function does not involve a continuation. Instead it follows directly from the observation in \cite{Hertog2011} that all complex saddle points of the Hartle-Hawking wave function have a representation in which their interior geometry consists of a Euclidean AdS domain wall that makes a smooth (but complex) transition to a Lorentzian asymptotically dS universe. Moreover the relative probabilities of different asymptotically locally dS boundary configurations are fully specified by the regularised action of the interior AdS regime of the saddle points. The complex transition region merely accounts for the universal phase factor in \eqref{dSCFT}, where $S_{ct}$ are the counterterms in \eqref{eqn:actionct}. This phase factor plays a physical role in dS: it implies the wave function predicts an ensemble of Lorentzian universes that evolve classically at large volume. The reason the inverse of the AdS/CFT dual partition function enters in \eqref{dSCFT} can be traced to the fact that the Hartle-Hawking wave function in cosmology is related to the decaying wave function in AdS whereas Euclidean AdS/CFT is usually concerned with the growing branch of the AdS wave function \cite{Gabriele:2015gca}.

Equation \eqref{dSCFT} shows that dS/CFT relates the argument of the wave function of the universe to external sources in the dual partition functions that turn on deformations of the CFT. The dependence of the partition function on the values of these sources, which include the background geometry, yields a holographic measure on the space of asymptotically locally de Sitter universes. However it has hitherto remained an open question what is the exact configuration space of deformations on which the holographic wave function should be defined. Here we study this question by using \eqref{dSCFT} and our results on the $O(N)$ model above to explore the qualitative behaviour of $\Psi_{HH}$ in new directions in superspace.

\begin{figure}[ht!]
\centering
  \begin{subfigure}[t]{0.32\textwidth}
    \includegraphics[width=\textwidth]{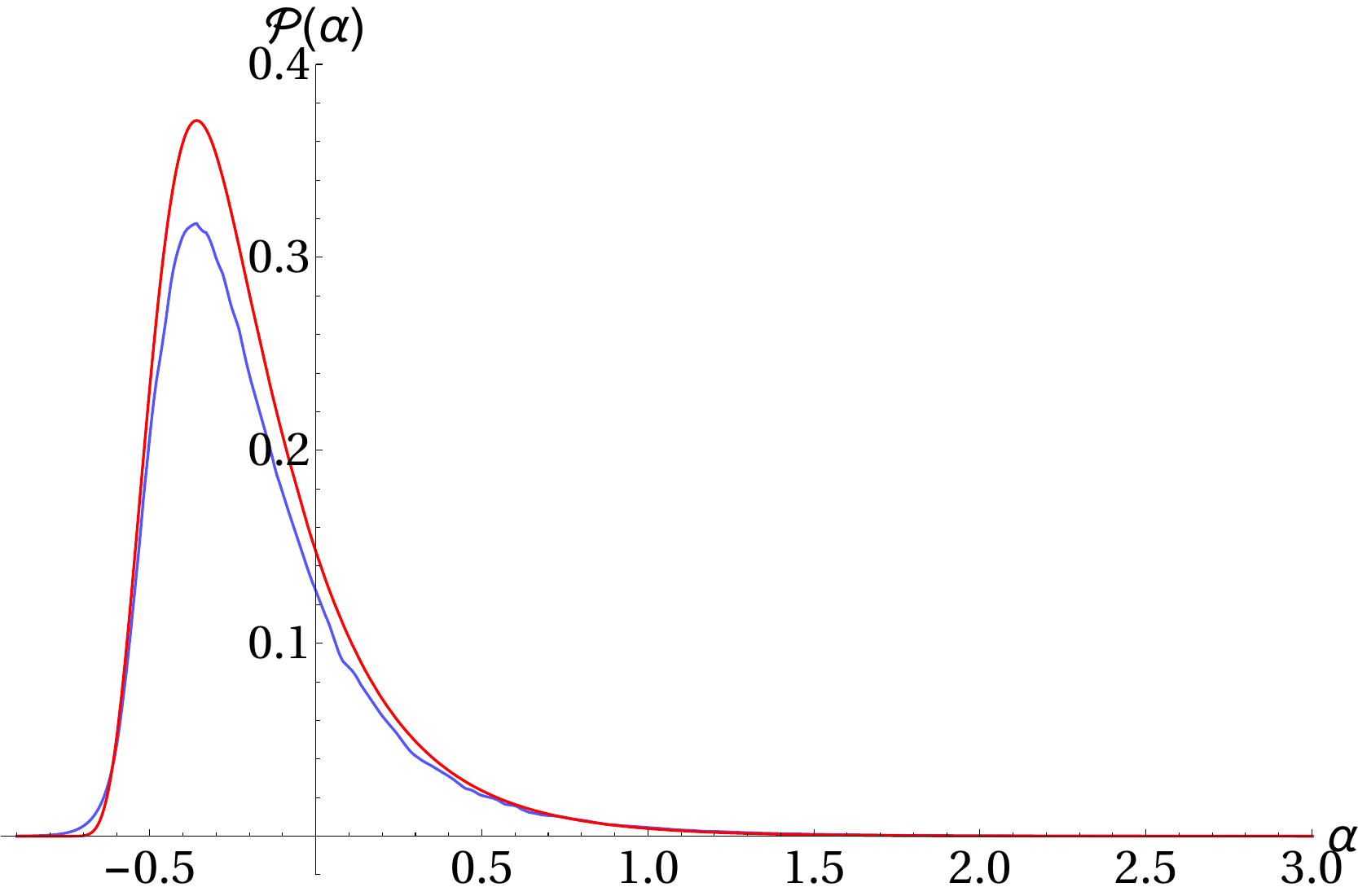}
    \caption{$\beta=-0.4$}
  \end{subfigure}
    \begin{subfigure}[t]{0.32\textwidth}
    \includegraphics[width=\textwidth]{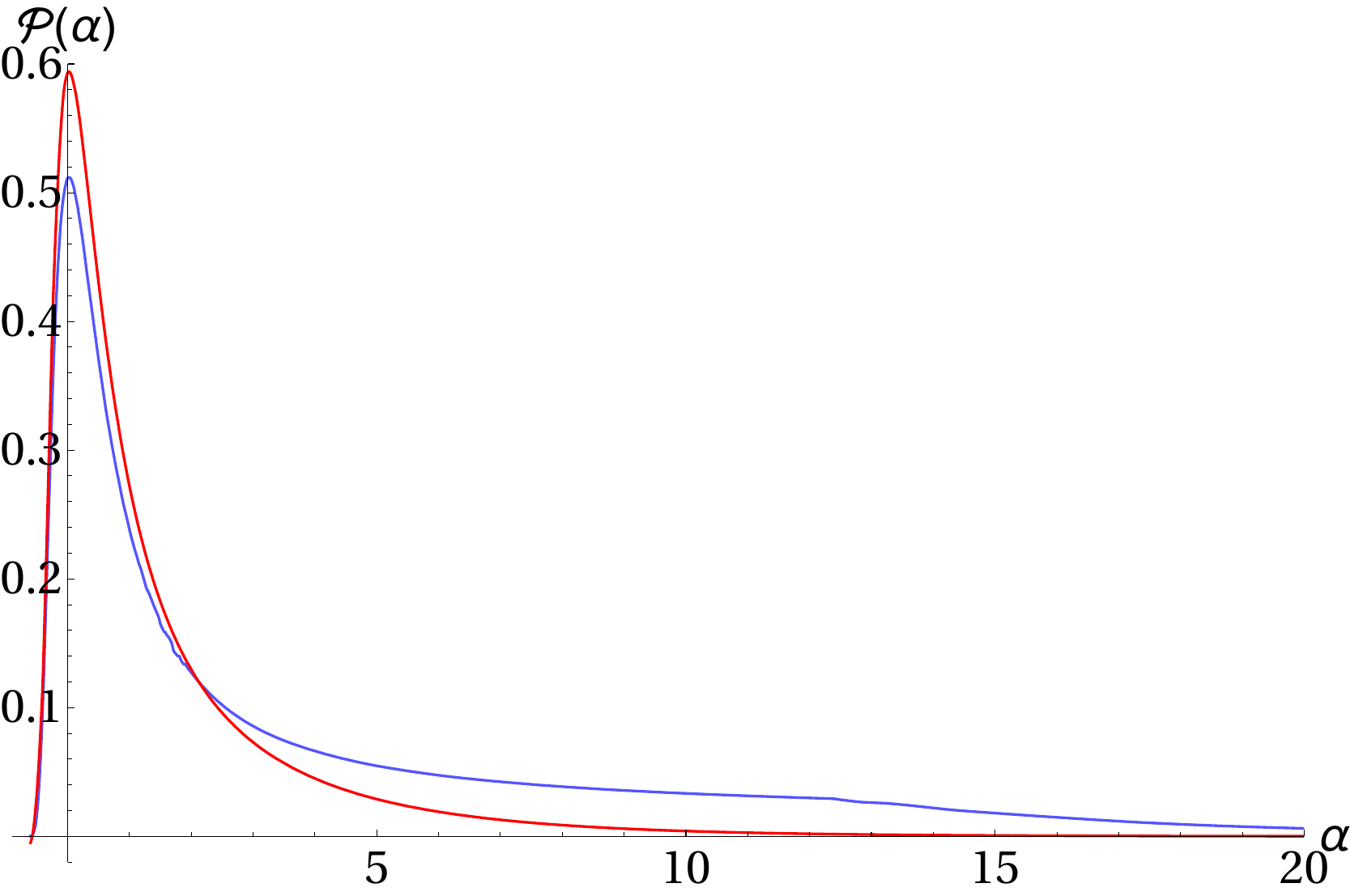}
    \caption{$\beta=0.05$}
  \end{subfigure}
    \begin{subfigure}[t]{0.32\textwidth}
    \includegraphics[width=\textwidth]{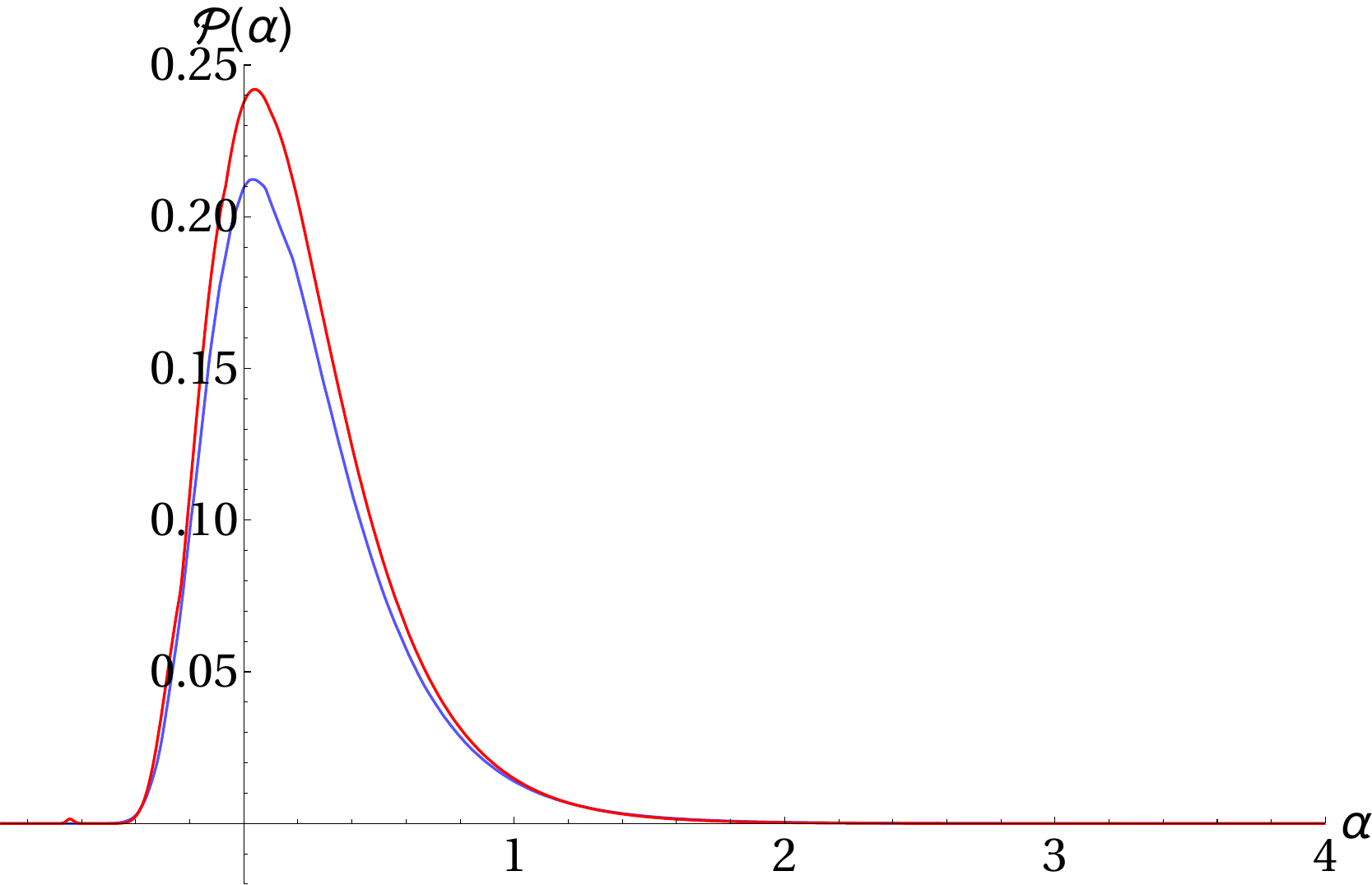}
    \caption{$\beta=1.15$}
  \end{subfigure}
\caption{Three constant $\beta$ slices of the Hartle-Hawking probability distribution ${\cal P}(\alpha,\beta) \equiv \vert \Psi_{HH}(\alpha,\beta) \vert^2$ over a two-parameter family of anisotropic deformations of de Sitter space labeled by the two squashing parameters $\alpha$ and $\beta$ of the future boundary geometry, in Einstein gravity (blue) and in Vasiliev gravity (red). The distribution in Vasiliev gravity is computed invoking the duality with the free $O(N)$ model at large $N$. Both distributions exhibit a qualitatively similar behaviour across the entire minisuperspace of boundary configurations and have a global maximum at $\alpha=\beta=0$ corresponding to dS space. The normalisation of the slices shown here is such that the integral of the probability distribution ${\cal P}(\alpha,\beta)$ over the $(\alpha,\beta)$-plane gives $1$. For the CFT we chose $N=10$.}\label{Probslices}
\end{figure}

The partition functions of $O(N)$ vector models have previously been evaluated in what corresponds in the bulk to a number of minisuperspace models. These include homogeneous isotropic minisuperspace with scalar matter \cite{Anninos:2012ft}, perturbations of this preserving $SO(3)$ invariance \cite{Anninos:2013rza}, and models with a round $S^1 \times S^2$ future boundary \cite{Anninos:2012ft,Conti:2014uda}. Our calculation of the partition function of the free $O(N)$ vector model on boundary geometries that are a two-parameter family of squashed three spheres yields $\Psi_{HH}$ in another kind of minisuperspace model that consists of homogeneous but anisotropic deformations of de Sitter space. 
In this context Figure \ref{fig:minusLogZ} can be interpreted as the logarithm of the probability distribution ${\cal P}(\alpha,\beta)\equiv \vert \Psi_{HH}(\alpha,\beta) \vert^2$ as a function of the two squashing parameters $\alpha$ and $\beta$. The distribution is normalisable and has a global maximum at the round sphere. The corresponding distribution in Einstein gravity computed via bulk methods follows from our results in  Section \ref{subsec:DS} and is shown in Figure \ref{fig:actionsAdStwosquashings} for small values of the squashing parameters and in Figure \ref{fig:actionsAdStwosquashingsB} for the large squashings where the Bolt solutions dominate the probabilities. It is striking that both distributions exhibit a qualitatively similar behaviour across the entire configuration space. This is made more explicit in Figure \ref{Probslices} where we show and compare three slices of these distributions for three different values of $\beta$. This also shows the distributions are significantly broader when $\beta$ (or $\alpha$) is small. On the other hand both distributions differ in specific features such as the NUT to Bolt transition at large positive values of the squashing parameters, which is evidently absent in the dual free theory.  

A particularly interesting region of superspace is the regime of boundary configurations for which the Ricci scalar is negative.
The Ricci scalar of a double squashed three sphere of the form \eqref{eqn:metric} is given by \eqref{Ricci} in terms of $\alpha$ and $\beta$. If one of the squashings is zero then $R<0$ if the remaining squashing parameter is less than $-3/4$. As mentioned earlier, however, adding a second squashing leads to an additional $R<0$ region associated with large positive values of both $\alpha$ and $\beta$. Figure \ref{fig:RicciandZSquaredLmax1000} shows that along all curves in the $(\alpha,\beta)$-plane where $R=0$ the holographic wave function vanishes. This is expected since the Ricci scalar enters as a mass term in the dual theory. When $R \rightarrow 0$ one of the eigenvalues of the scalar Laplacian on the squashed $S^3$ goes to zero, see \eqref{eqn:LogZGeneral}, and since the partition function is proportional to the product of all these eigenvalues it vanishes. This in turn leads to a diverging free energy, and hence the prediction that $\Psi_{HH} \rightarrow 0$ when $R \rightarrow 0$. 

\begin{figure}[ht!]
\centering
\includegraphics[width=0.32\textwidth]{betaplus115ActionComparedCFTSPNBulk.pdf}
\includegraphics[width=0.32\textwidth]{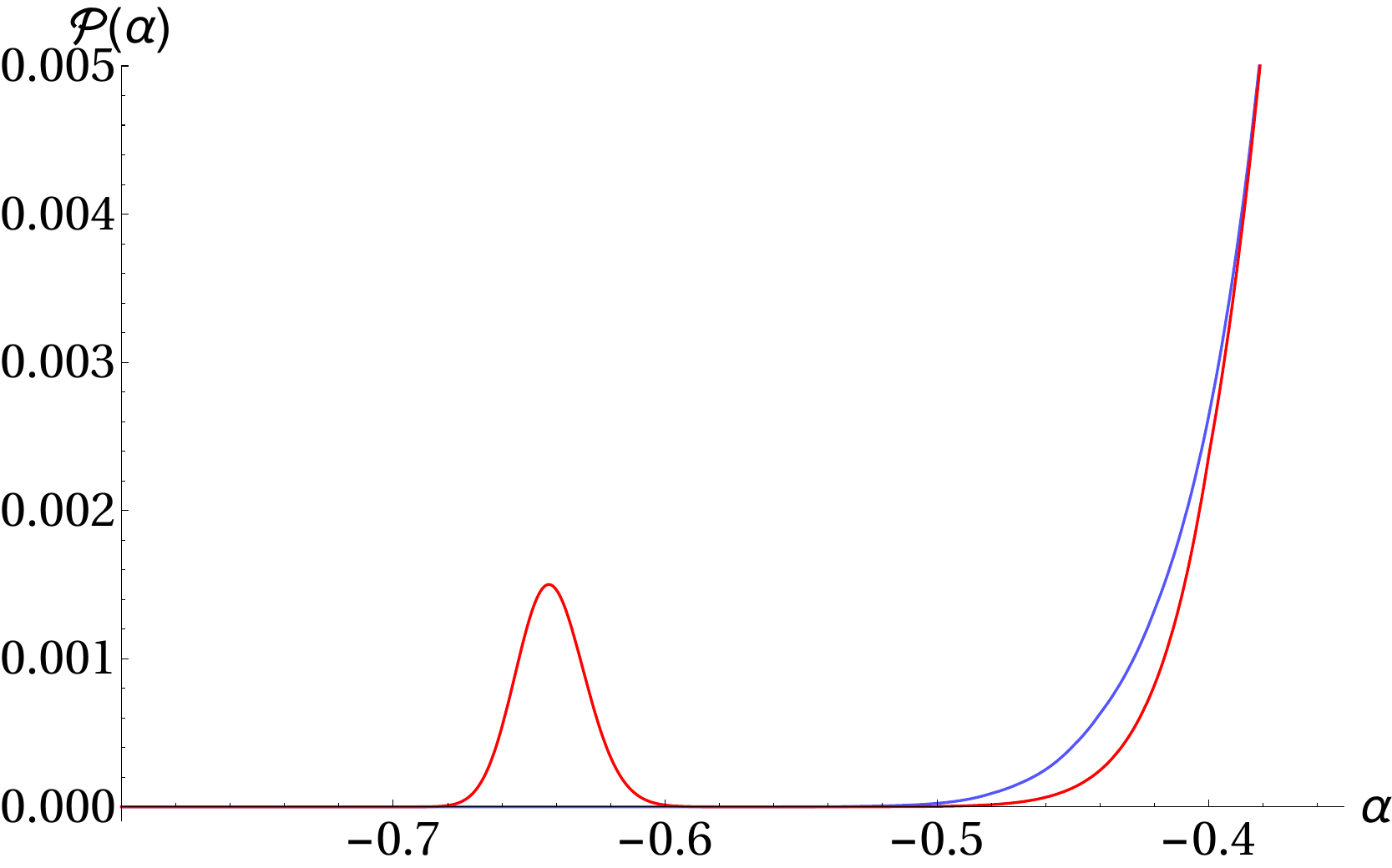}
\includegraphics[width=0.32\textwidth]{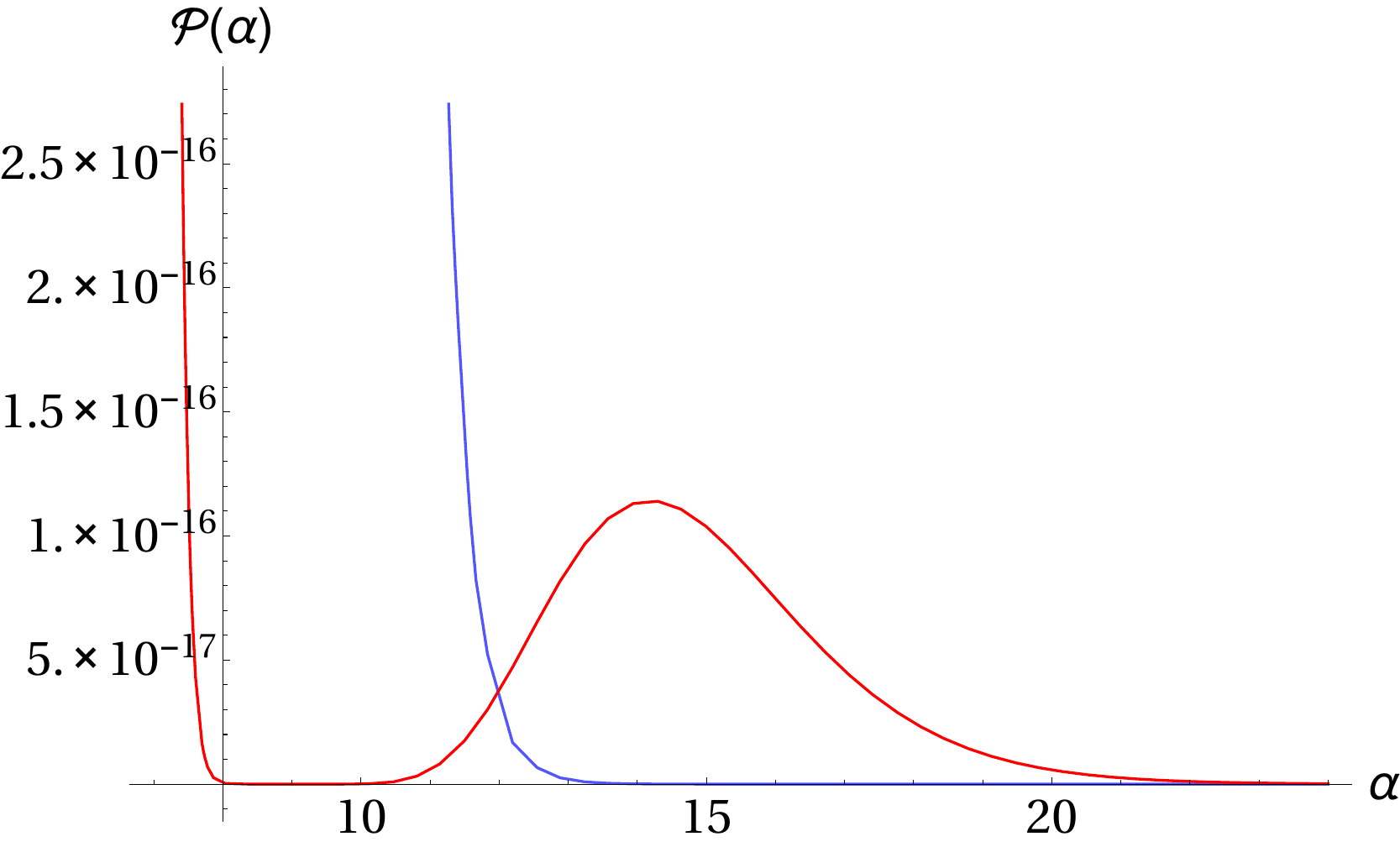}
\caption{A constant $\beta=1.15$ slice of the Hartle-Hawking probability distribution over anisotropic deformations of de Sitter space, in Einstein gravity (blue) and in Vasiliev gravity (red), and details of these distributions in the regime where the curvature of the boundary geometry is negative. The second squashing introduces a second regime where $R(\alpha,\beta) <0$ corresponding to large positive values of both squashing parameters. The overall amplitude of boundary geometries with a negative scalar curvature is exponentially small in this minisuperspace model. The normalisation and $N$ are chosen to be the same as in Figure \ref{Probslices}.}
\label{Probsingle}
\end{figure}

In the region of superspace corresponding to negative curvature boundaries the free boundary theory is unstable. This suggests this region should be excised from the configuration space in order for the wave function to be well-defined and normalisable. In the minisuperspace model we consider here the implications of this instability are nevertheless limited: The overall amplitude of boundary geometries with a negative scalar curvature is exponentially small. This is clear from Figure \ref{Probslices} and Figure \ref{Probsingle} that show representative slices of the probability distribution. However the instability of the boundary theory leads to pathologies when one extends the minisuperspace model to include e.g. a bulk scalar field, both in Einstein gravity and in Vasiliev gravity. In this context it leads to a non-normalisable probability distribution rendering the wave function ill-defined. Evidence for this in Vasiliev gravity was found in \cite{Anninos:2012ft} in homogeneous isotropic minisuperspace with scalar matter. In future work we will show that the instability leads to divergences in other directions in superspace as well, by evaluating the wave function on anisotropic boundary configurations with scalar matter turned on \cite{CHV}. This provides further evidence that the configuration space of boundary geometries on which the wave functional is defined in quantum gravity must be appropriately constrained in order for the resulting probabilities to be well defined and normalisable.

\section{Discussion}\label{sec:discussion}

We have shown that the partition function of the free $O(N)$ model on the double squashed three sphere, as a function of the two squashing parameters, qualitatively reproduces the thermodynamical properties of a new set of Euclidean asymptotically locally $AdS_4$ solutions of Einstein gravity. Using a recent formulation of semiclassical dS/CFT we derived the Hartle-Hawking wave function of the universe in homogeneous but anisotropic minisuperspace from the above partition functions. In this application, the two squashing parameters specify a two parameter set of anisotropic deformations of de Sitter space. We found the resulting probability distribution over cosmological histories is normalisable and globally peaked at isotropic de Sitter space. Strong squashings lead to boundary geometries with negative scalar curvature and the boundary theory becomes manifestly unstable. In this minisuperspace model the overall amplitude of universes with a negative curvature future boundary is nevertheless exponentially small. However, as we discussed in Section \ref{sec:Cosmology}, this does not remain true when other directions in superspace are taken in account, We defer a detailed analysis of this to future work \cite{CHV}.

There are some clear avenues for generalisation of our work. First, it would be interesting to extend our results to higher dimensions. On the gravitational side it should be fairly straightforward to generalise the metric Ansatz in \eqref{eqn:Ansatz2sq} and find new solutions of general relativity in higher dimensions. Based on our experience in four dimensions we expect that these backgrounds can be found only numerically. Extending the results of Section \ref{sec:CFT} to higher dimensions should also be possible. The technical problem here is to find the spectrum of the Laplacian on higher-dimensional squashed spheres. This has been addressed in \cite{BBV} for a generalisation of the metric in \eqref{eqn:metric} with $\beta=0$ to higher odd dimensions. Adding additional squashing parameters to the problem will probably be technically cumbersome. Another extension of our work could be to compute the squashed sphere partition function with two nontrivial squashing parameters for the interacting $O(N)$ model along the lines of the approach in \cite{Hartnoll:2005yc}. This will probably necessitate the use of the large $N$ limit.

We should emphasize again that the free $O(N)$ vector models are not dual to Einstein gravity with a negative cosmological constant. The proper dual theory is the higher-spin Vasiliev theory. In view of the field theory results in Section \ref{sec:CFT} it will therefore be very interesting to construct new solutions of Vasiliev theory which have a squashed $S^3$ metric on the asymptotic boundary. This will provide new opportunities for quantitative checks of the higher-spin/vector model duality. Conversely it will be desirable to construct proper holographic duals to the gravitational solutions we constructed in Section \ref{sec:AdS}. To do this one has to overcome the technical difficulty in dealing with strongly coupled three-dimensional CFTs, like the ABJM theory, on curved manifolds without resorting to the power of supersymmetry.

\bigskip

\noindent \textbf{ Acknowledgements }
\bigskip

\noindent We would like to thank Pablo Bueno, Gabriele Conti, Frederik Denef, Ruben Monten and Balt van Rees for useful discussions. The work of NB is supported in part by the starting grant BOF/STG/14/032 from KU Leuven and by an Odysseus grant G0F9516N from the FWO. The work of TH and YV is supported in part by the FWO grant G.001.12 Odysseus and by the European Research Council grant no. ERC-2013-CoG 616732 HoloQosmos. We are also supported by the KU Lueven C1 grant ZKD1118 C16/16/005, the COST Action MP1210 The String Theory Universe, and by the European Science Foundation Holograv Network.

\begin{appendices}

\appendix

\section{UV and IR expansions}
\label{App:AdSexpansions}

Here we collect some details on the IR and UV asymptotic expansion for the numerical solutions discussed in Section \ref{subsec:AdS2squash} as well as a short discussion on the procedure to construct numerical solutions.

The equations of motion for the Ansatz in \eqref{eqn:Ansatz2sq} arising from the action in \eqref{eqn:GRaction} are given by
\begin{equation} \label{eqn:eom}
\begin{split}
 \frac{ { l_0}   { l_3} }{ { l_1}   { l_2} } & +\frac{ { l_0}   { l_2} }{ { l_1}   { l_3} }+\frac{ { l_0}   { l_1} }{ { l_2}   {l_3} }-\frac{2
    {l_0} }{ {l_1} }-\frac{2  {l_0} }{ {l_2} }-\frac{2  {l_0} }{ {l_3} }+4 \Lambda   {l_0} +\frac{ {l_1}'   {l_2}' }{ {l_1}   {l_2} }+\frac{ {l_1}' 
    {l_3}' }{ {l_1}   {l_3} }+\frac{ {l_2}'   {l_3}' }{ {l_2}   {l_3} }=0\ ,  \\
   -\frac{ {l_0}'   {l_1}' }{ {l_0}   {l_1} }&-\frac{ {l_0}'   {l_3}' }{ {l_0}   {l_3} }-\frac{ {l_0}   {l_3} }{ {l_1}   {l_2} }+\frac{3  {l_0} 
    {l_2} }{ {l_1}   {l_3} }-\frac{ {l_0}   {l_1} }{ {l_2}   {l_3} }-\frac{2  {l_0} }{ {l_1} }+\frac{2  {l_0} }{ {l_2} }-\frac{2  {l_0} }{ {l_3} }+4
   \Lambda   {l_0}
   +\frac{2  {l_1}'' }{ {l_1} }+\frac{ {l_1}'   {l_3}' }{ {l_1}   {l_3} }  \\
   -\frac{ {l_1'}^2}{ {l_1}^2}&+\frac{2
    {l_3}'' }{ {l_3} }-\frac{ {l_3'}^2}{ {l_3}^2}=0\ ,\\
   -\frac{ {l_0}'   {l_2}' }{ {l_0}   {l_2} }&-\frac{ {l_0}'   {l_3}' }{ {l_0}   {l_3} }-\frac{ {l_0}   {l_3} }{ {l_1}   {l_2} }-\frac{ {l_0} 
    {l_2} }{ {l_1}   {l_3} }+\frac{3  {l_0}   {l_1} }{ {l_2}   {l_3} }+\frac{2  {l_0} }{ {l_1} }-\frac{2  {l_0} }{ {l_2} }-\frac{2
    {l_0} }{ {l_3} }    +4 \Lambda   {l_0} +\frac{2  {l_2}'' }{ {l_2} }+\frac{ {l_2}'   {l_3}' }{ {l_2}   {l_3} } \\
    -\frac{ {l_2'}^2}{ {l_2}^2} &   +\frac{2    {l_3}'' }{ {l_3} }-\frac{ {l_3'}^2}{ {l_3}^2}=0\ ,\\
   -\frac{ {l_0}'   {l_1}' }{ {l_0}   {l_1} }&-\frac{ {l_0}'   {l_2}' }{ {l_0}   {l_2} }+\frac{3  {l_0}   {l_3} }{ {l_1}   {l_2} }-\frac{ {l_0} 
    {l_2} }{ {l_1}   {l_3} }-\frac{ {l_0}   {l_1} }{ {l_2}   {l_3} }-\frac{2  {l_0} }{ {l_1} }-\frac{2  {l_0} }{ {l_2} }+\frac{2  {l_0} }{ {l_3} }
    +4   \Lambda   {l_0} +\frac{2  {l_1}'' }{ {l_1} }+\frac{ {l_1}'   {l_2}' }{ {l_1}   {l_2} } \\
    -\frac{ {l_1'}^2}{ {l_1}^2}&+\frac{2 {l_2}'' }{ {l_2} }-\frac{ {l_2'}^2}{ {l_2}^2}=0\  .
\end{split}
\end{equation}

The IR expansion for the NUT solution with two squashings is given in \eqref{eqn:genIRNUT}. Using the equations of motion leads to the following series expansion
\begin{equation}\label{eqn:IRexpansion}
\begin{split}
l_0(r)&=1 \quad ,\\
l_1(r)&=\frac{1}{4} (r-r^*)^2+\beta_4 (r-r^*)^4  \\
   +& \frac{(r-r^*)^6 \left(36\Lambda  (\beta_4-4 \gamma_4)+ \left(-\Lambda ^2\right)+96 \left(-18 \beta_4 \gamma_4+17
   \beta_4^2-18 \gamma_4^2\right)\right)}{480} +\mathcal{O}((r-r^*)^{8}) \ ,\\
   l_2(r)&=\frac{1}{4} (r-r^*)^2+\gamma_4 (r-r^*)^4
  \\
     -&\frac{(r-r^*)^6 \left(36 \Lambda  (4 \beta_4-\gamma_4)+ \Lambda ^2+96 \left(18 \beta_4 \gamma_4+18 \beta_4^2-17
   \gamma_4^2\right)\right)}{480 } +\mathcal{O}((r-r^*)^{8})\ ,\\
   l_3(r)&=(r-r^*)^4 \left(-\frac{1}{12}\Lambda -\beta_4-\gamma_4\right)+\frac{1}{4} (r-r^*)^2 \\
   +& \frac{(r-r^*)^6 \left(354  \Lambda  (\beta_4+\gamma_4)+11 \Lambda ^2+144 \left(52 \beta_4 \gamma_4+17 \beta
   (4)^2+17 \gamma_4^2\right)\right)}{720}+\mathcal{O}((r-r^*)^{8}) \ .
\end{split} 
\end{equation}
As explained in the main text this expansion is controlled by the two real parameters $\beta_4$ and $\gamma_4$ which are ultimately related to the two squashing parameters $\alpha$ and $\beta$, at the asymptotic boundary.

The IR expansion for the Bolt solution with two squashings is given in \eqref{eqn:genIRBolt}. Using the equations of motion in \eqref{eqn:eom} in the gauge $l_0(r)=1$ leads to the following series expansion
\begin{equation}\label{eqn:IRBoltexpansion}
\begin{split}
 l_1(r)&=\gamma _0-\frac{1}{2} \left(\gamma _0 \Lambda
   -1\right) (r-r^*)^2+\frac{\left(12 \gamma _0^2 \Lambda ^2-16 \gamma _0 \left(6 \gamma _4+\Lambda \right)+1\right) (r-r^*)^4}{96 \gamma
   _0} \\
   & +\frac{\left(\gamma _0 \Lambda  \left(4 \gamma _0 \Lambda  \left(14-5 \gamma _0 \Lambda \right)-41\right)+32 \gamma _0
   \gamma _4 \left(9 \gamma _0 \Lambda -16\right)+4\right) (r-r^*)^6}{960 \gamma _0^2} + \mathcal{O}((r-r^*)^8)\ , \\
 l_2(r)&=\gamma _0-\frac{1}{2} \left(\gamma _0 \Lambda
   -1\right) (r-r^*)^2+\gamma _4 (r-r^*)^4 \\
  & +\frac{\left(\gamma _0 \Lambda  \left(12 \gamma _0 \Lambda  \left(2 \gamma _0 \Lambda -7\right)+71\right)+48 \gamma _0
   \gamma _4 \left(16-9 \gamma _0 \Lambda \right)-2\right) (r-r^*)^6}{1440 \gamma _0^2}  + \mathcal{O}((r-r^*)^8)\ , \\
 l_3(r)&=\frac{1}{4} (r-r^*)^2 -\frac{(r-r^*)^4}{12 \gamma _0}+ \frac{\left(36 \gamma _0^2 \Lambda ^2-144 \gamma _0 \Lambda +199\right) (r-r^*)^6}{5760 \gamma
   _0^2} + \mathcal{O}((r-r^*)^8)\ . 
\end{split} 
\end{equation}
We chose to parametrise this expansion by the two independent real parameters $\gamma_0$ and $\gamma_4$ which are again mapped to the squashing parameters $\alpha$ and $\beta$ in the UV.

The general UV expansion is given in \eqref{eqn:genUV}. Plugging this into the equations of motion \eqref{eqn:eom} with $l_0(r)=1$ one finds the following consistent series expansion
\begin{equation}\label{eqn:UVexpansion}
\begin{split}
l_1(r)&=A_0 e^{2 r}+ \frac{3 \left(-2 A_0 \left(B_0+C_0\right)+5 A_0^2-3 \left(B_0-C_0\right)^2\right)}{8 B_0 C_0 \Lambda } +A_3 e^{-r} + \mathcal{O}(e^{-2r} ) \ ,   \\
l_2(r)&= B_0 e^{2 r}  -\frac{3 \left(2 A_0 \left(B_0-3 C_0\right)+3 A_0^2+2 B_0 C_0-5 B_0^2+3 C_0^2\right)}{8 A_0 C_0 \Lambda }+B_3 e^{-r}   +  \mathcal{O}(e^{-3r} ) \;,  \\
l_3(r)&= C_0 e^{2 r}+   \frac{-6 C_0 \left(A_0+B_0\right)-9 \left(A_0-B_0\right)^2+15 C_0^2}{8 A_0 B_0 \Lambda }  + C_0 e^{-r} \left(-\frac{A_3}{A_0}-\frac{B_3}{B_0}\right)   +  \mathcal{O}(e^{-3r} )\ .
\end{split} 
\end{equation}
We have performed this expansion up to eight order and have verified that it is controlled by the five parameters $\{A_0,B_0,C_0,A_3,B_3\}$. Since the equations of motion \eqref{eqn:eom} are invariant under constant shifts of the radial coordinate, one can set $A_0=\frac{1}{4}$ by an appropriate shift of $r$. One can then identify $B_0$ and $C_0$ with the squashing parameters in \eqref{eqn:metric} as follows
\begin{equation}\label{eqn:alphabetaBC}
\alpha = \frac{1}{4C_0} - 1\;, \qquad\qquad \beta = \frac{1}{4B_0} - 1\;.
\end{equation}
The parameters $A_3$ and $B_3$ are independent from the point of view of the UV expansion but are ultimately fixed in terms of $\alpha$ and $\beta$ by the regularity conditions that we imposed for the numerical solutions of the full nonlinear equations of motion.

\begin{figure}[ht!]
\centering
\includegraphics[width=0.45\textwidth]{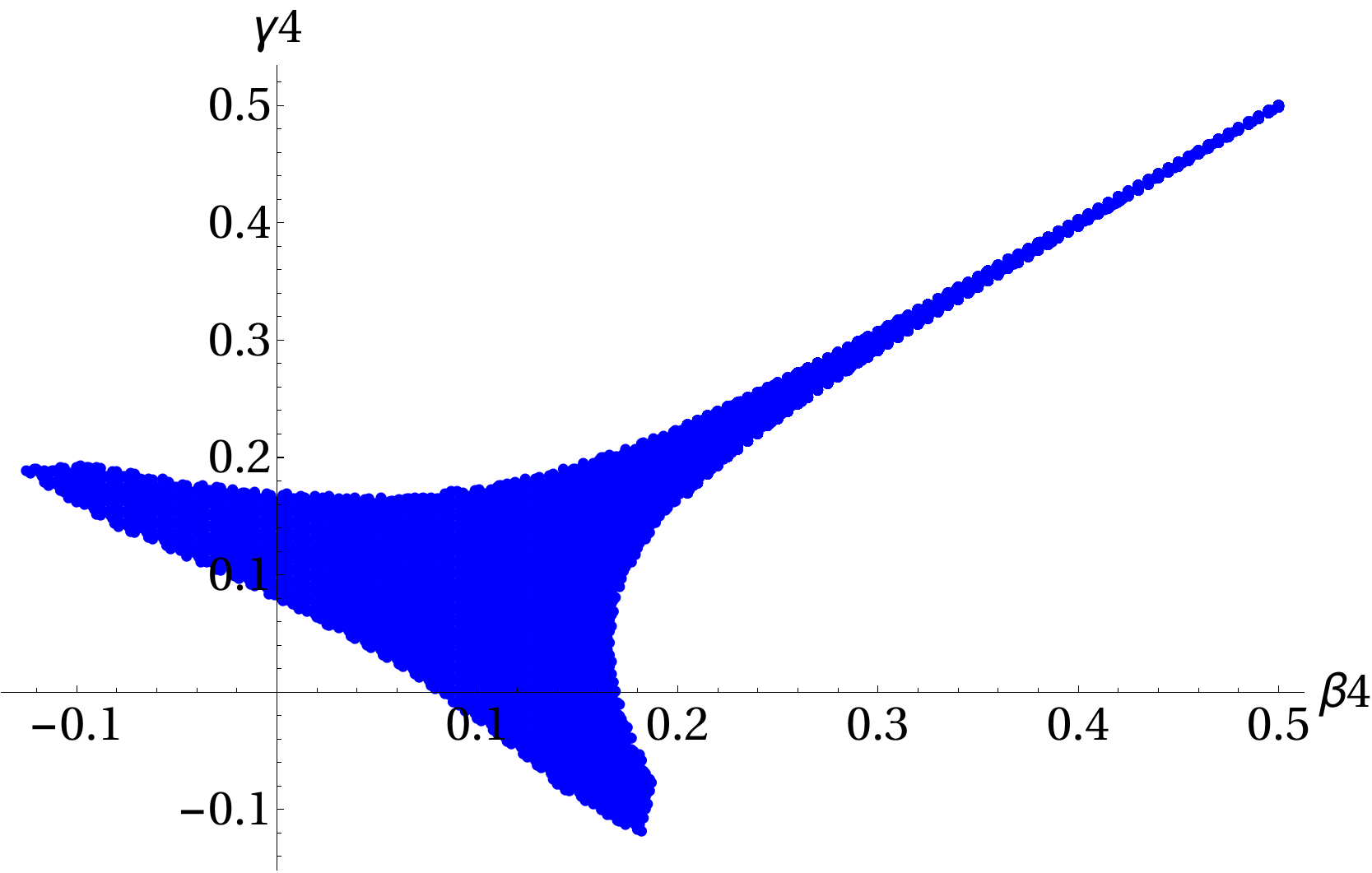}
\includegraphics[width=0.45\textwidth]{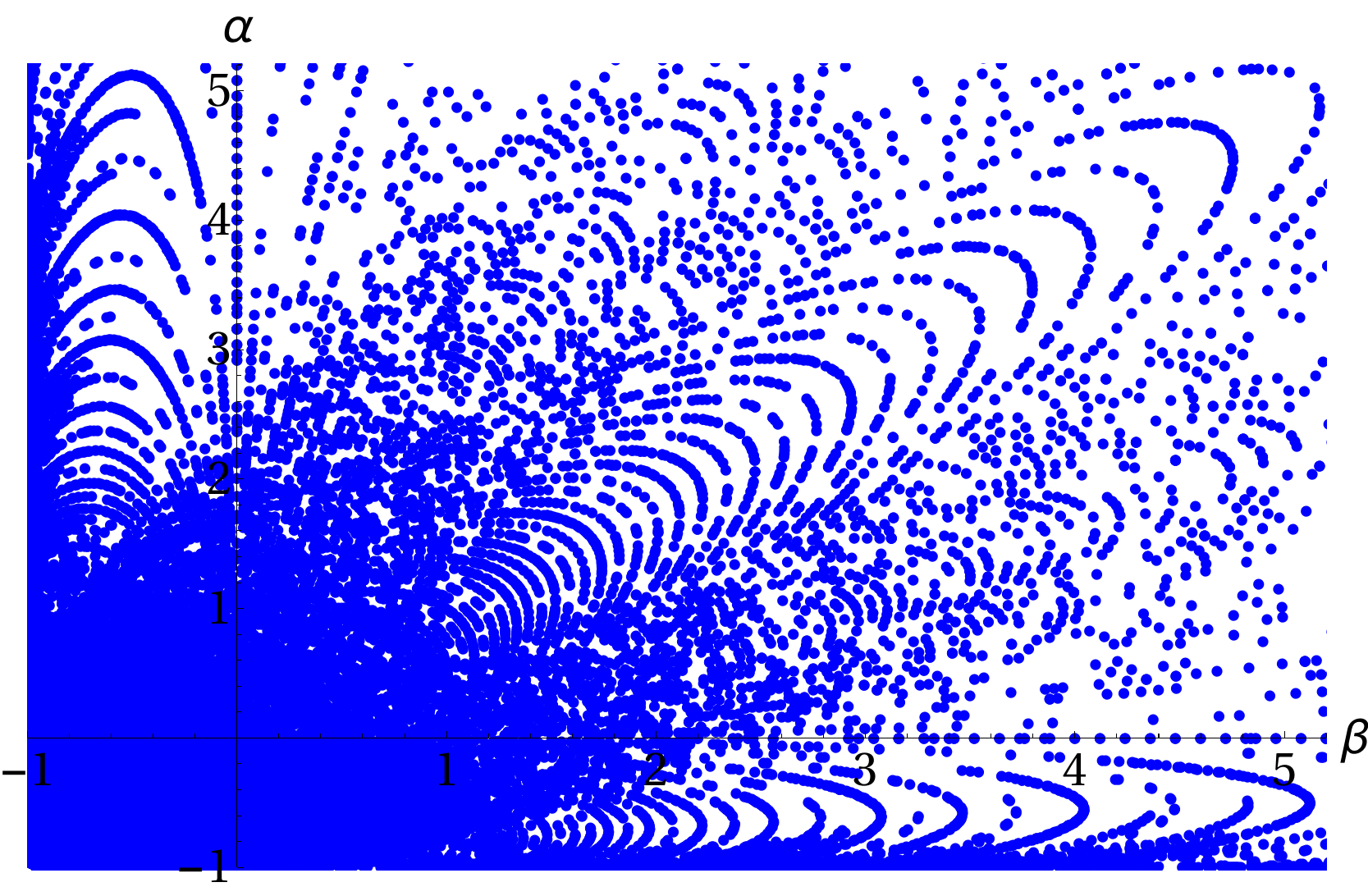}
\caption{The range of parameters for the AdS-Taub-NUT solutions with two squashings. Left: the values of $\gamma_4$ and $\beta_4$ that lead to regular solutions. Right: the resulting values of the squashing parameters $\alpha$ and $\beta$.}\label{fig:rangesolsNUT}
\end{figure}

\begin{figure}[ht!]
\centering
\includegraphics[width=0.45\textwidth]{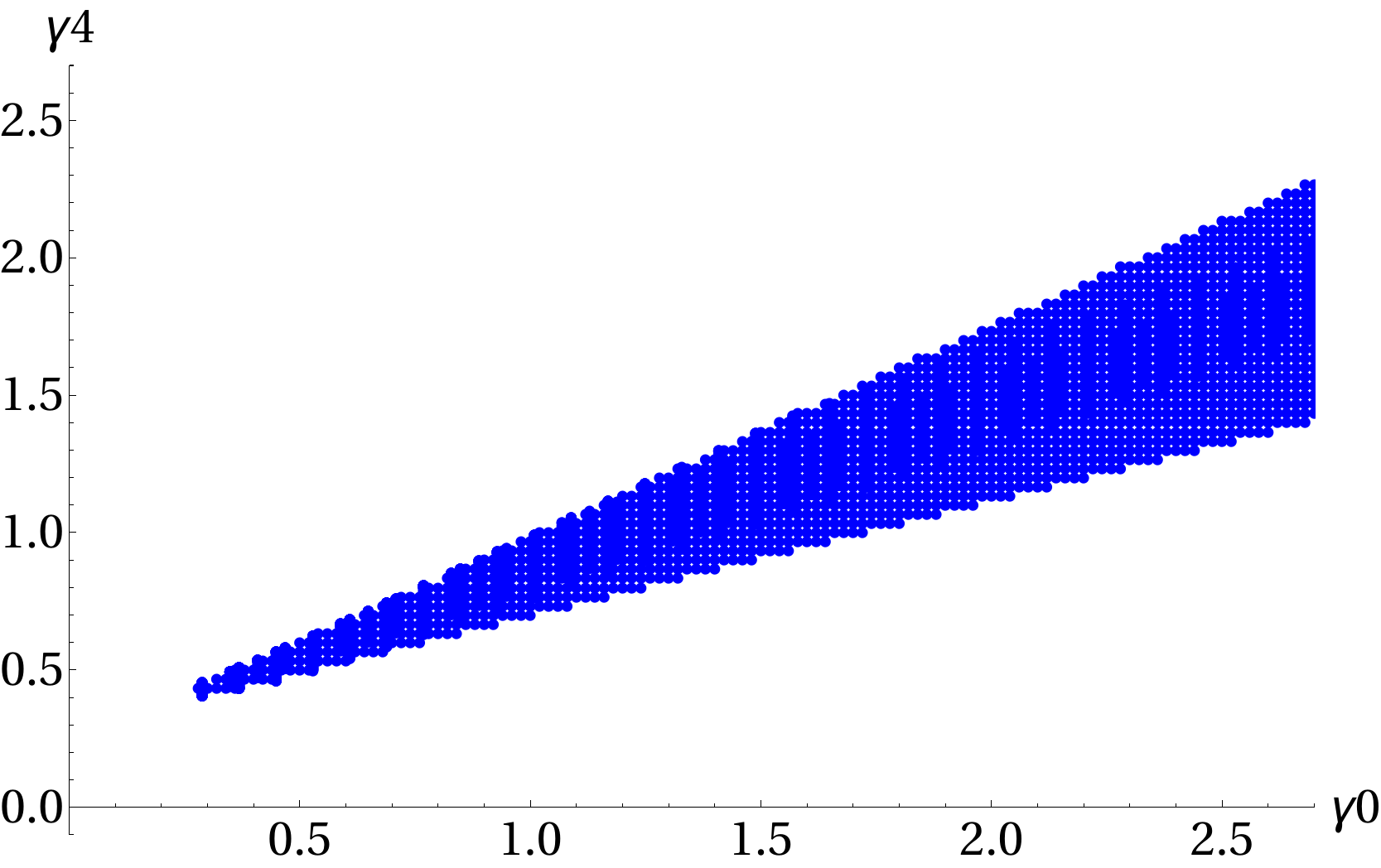}
\includegraphics[width=0.45\textwidth]{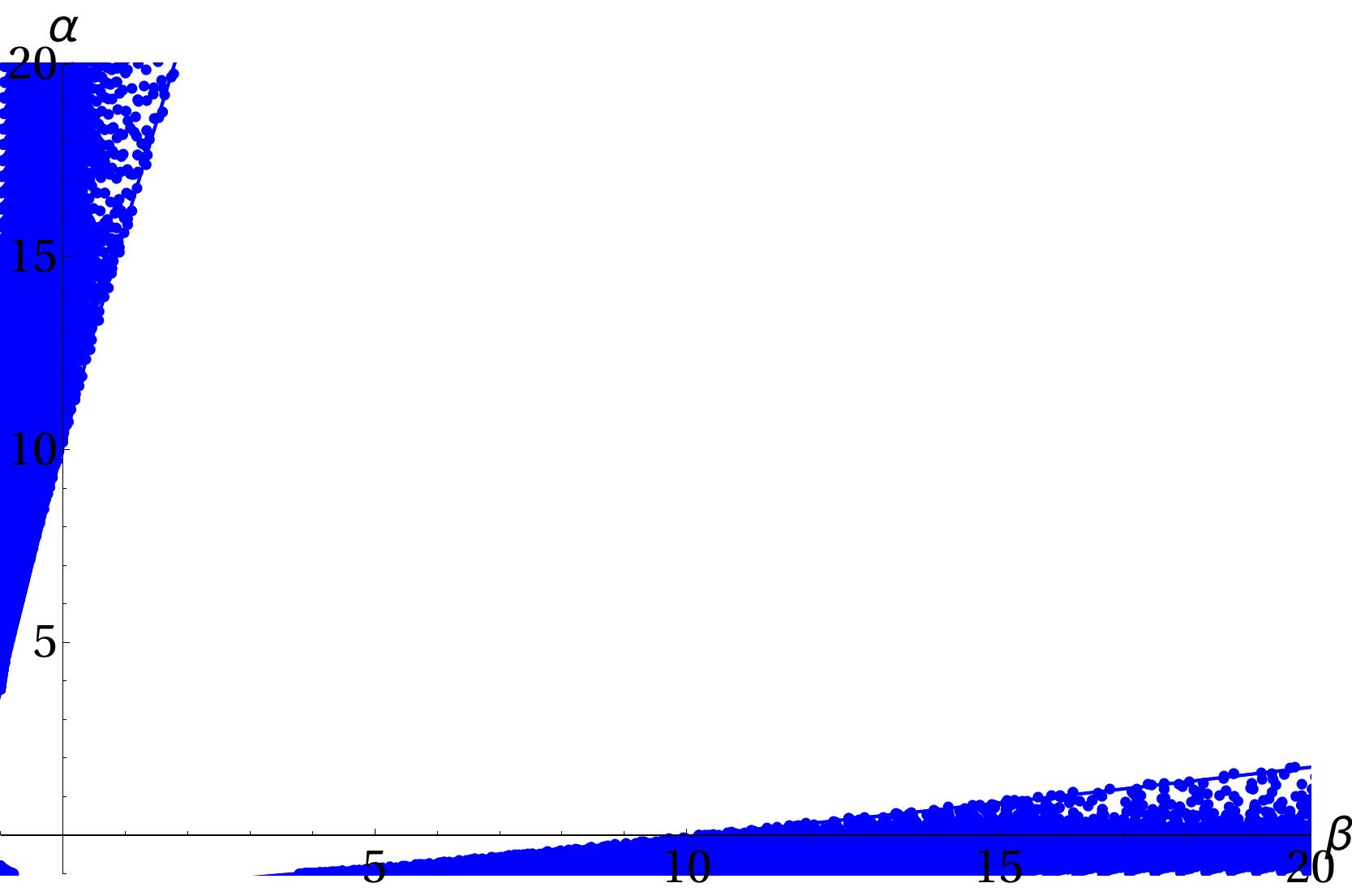}
\caption{The range of parameters for the ``positive branch" AdS-Taub-Bolt solutions with two squashings. Left: the range of parameters $\gamma_0$ and $\gamma_4$ that leads to regular solutions. Right: the resulting asymptotic parameters $\alpha$ and $\beta$.}\label{fig:RangesolsBolt}
\end{figure}

\begin{figure}[ht!]
\centering
\includegraphics[width=0.45\textwidth]{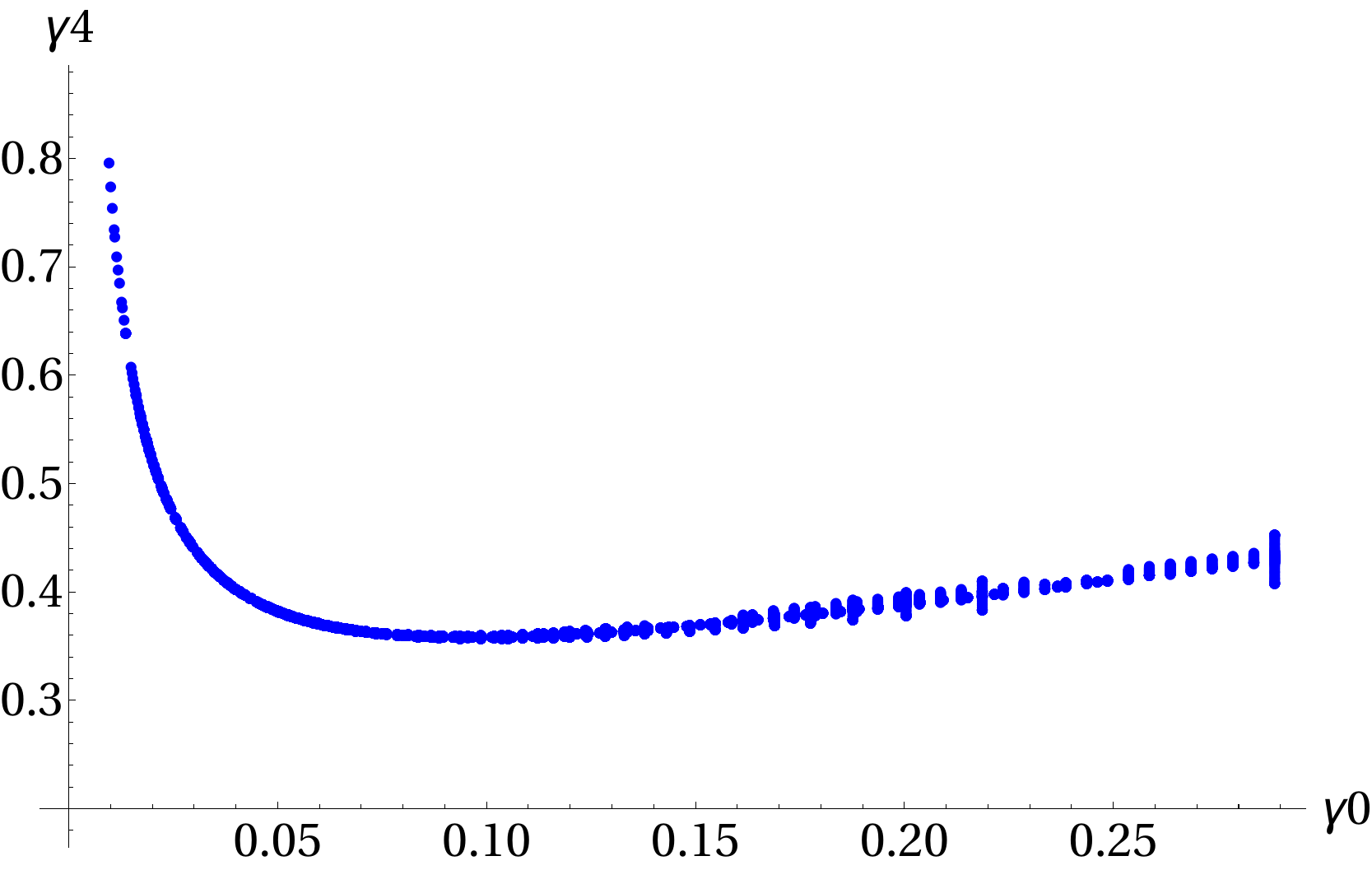}
\includegraphics[width=0.45\textwidth]{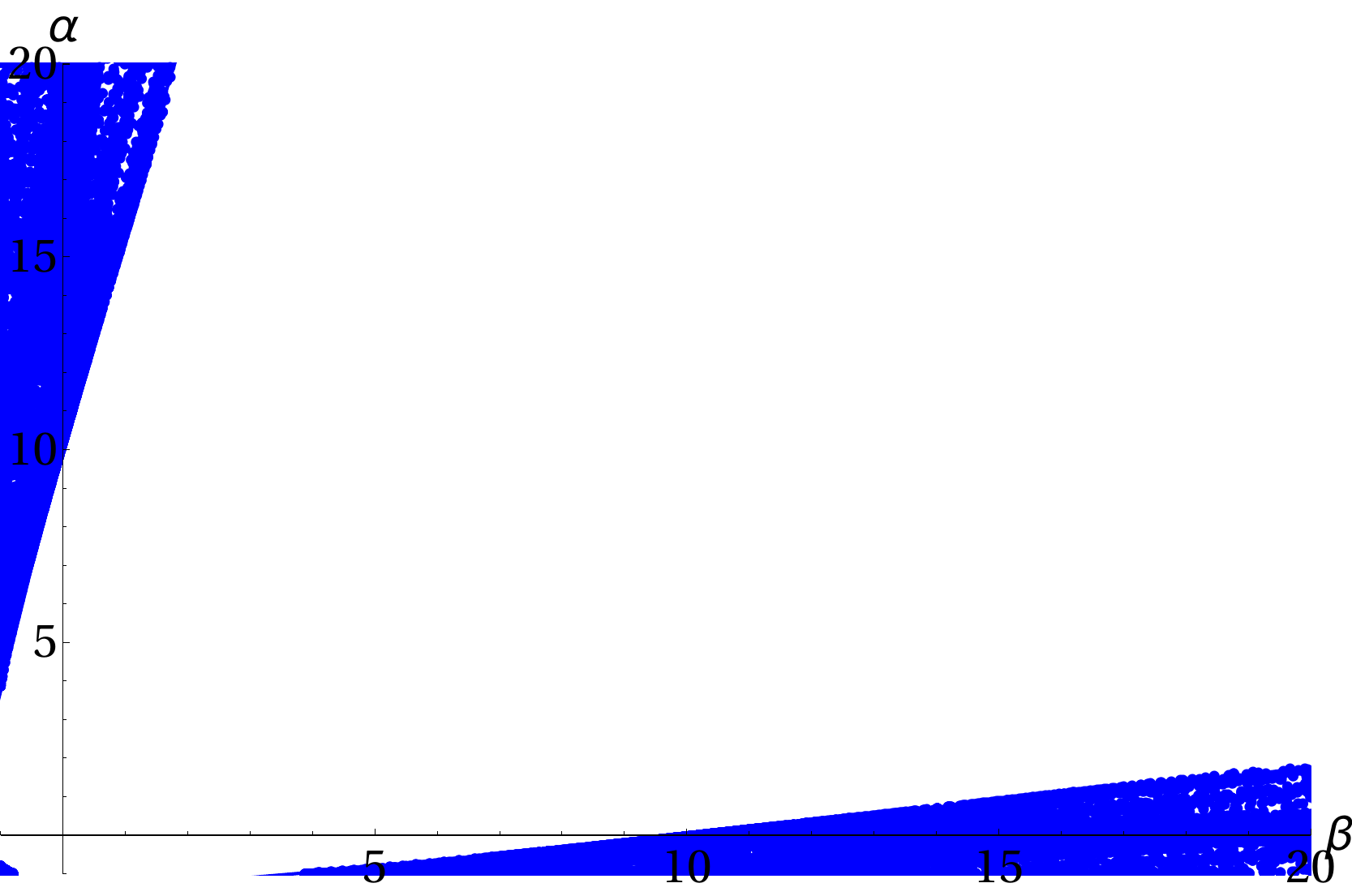}
\caption{The range of parameters for the ``negative branch" AdS-Taub-Bolt solutions with two squashings. Left: the range of parameters $\gamma_0$ and $\gamma_4$ that leads to regular solutions. Right: the resulting asymptotic parameters $\alpha$ and $\beta$.}\label{fig:RangesolsBoltNB}
\end{figure}

It is worth discussing how we constructed the numerical solutions of the full nonlinear equations of motion in \eqref{eqn:eom}. For the AdS-Taub-NUT solutions we picked real values for the parameters $\beta_4$ and $\gamma_4$ in the IR expansion \eqref{eqn:IRexpansion}. For each such value we then numerically integrated the equations of motion from $r=0$ to some large value of $r$. If the resulting numerical solution does not exhibit a singularity at an intermediate value of the radial coordinate $r$ we declared the solution to be regular and read of the asymptotic parameters $B_0$ and $C_0$ in  \eqref{eqn:UVexpansion} which we then related to the squashing parameters $\alpha$ and $\beta$ using \eqref{eqn:alphabetaBC}. As expected we find that there are no restrictions on the parameters $\alpha$ and $\beta$, i.e. as we vary $\beta_4$ and $\gamma_4$ we can explore the whole $(\alpha,\beta)$ plane. This is illustrated in Figure \ref{fig:rangesolsNUT}.
%
%
The procedure we used to construct the AdS-Taub-Bolt solutions is very similar. We start with the IR expansion in \eqref{eqn:IRBoltexpansion}, vary the parameters $\gamma_0$ and $\gamma_4$ and integrate numerically the equations of motion. Finally we read off the asymptotic parameters $B_0$ and $C_0$ from the behaviour of the numerical solutions at large $r$ and deduce the corresponding values of $\alpha$ and $\beta$ using the relation in \eqref{eqn:alphabetaBC}. However, there is an important difference between these solutions and the AdS-Taub-NUT solutions. For a fixed value of $\beta$ there are critical values of $\alpha$ below/above which there are no AdS-Taub-Bolt solutions. This leads to curves in the $(\alpha,\beta)$ plane and the AdS-Taub-Bolt solutions exist only for values of the squashing parameters that are below or above these critical curves.  Furthermore for every value of $(\alpha,\beta)$ for which Bolt solutions exist there are two possible solutions of the equations of motion which we dub ``positive" and ``negative" branch. All of these features are extensions of the familiar behaviour of the analytically known AdS-Taub-Bolt solutions with $\beta=0$ discussed in Section \ref{subsec:AdSTNB}. We illustrate the range of the IR and squashing parameters for the ``positive"  and ``negative" branch AdS-Taub-Bolt solutions in Figure \ref{fig:RangesolsBolt} and Figure \ref{fig:RangesolsBoltNB} respectively.
%
%

\section{Eigenvalues of the Laplace operator}
\label{App:eigen}

To calculate the eigenvalues of the Laplacian on the double squashed three sphere we use some of the results in \cite{PhysRevD.8.1048}. The main observation is that the Laplace operator $-\nabla^2$ on the squashed three sphere corresponds to the Hamiltonian $\hat{H}$ of an asymmetric top
\begin{align}
 -\nabla^2 \to  \hat{H}=\frac{ \hat{L}_1^2}{2I_1} + \frac{ \hat{L}_2^2}{2I_2}+\frac{\hat{L}_3^2}{2I_3} \ , \label{eqn:Hamiltonian}
\end{align}
with $\hat{L}_1$, $\hat{L}_2$ and $\hat{L}_3$ the components of the angular momentum operator $\hat{L}$ along the three principal axes of inertia. $(I_1,I_2,I_3)$ are the principal moments of inertia of the body. These can be mapped to the squashing parameters in \eqref{eqn:metric} in the following way
\begin{align}
I_1=\frac18 \ , \qquad I_2=\frac{1}{8(\beta+1)} \ , \qquad I_3=\frac{1}{8(\alpha+1)} \ .
\end{align}
The round sphere with $\alpha=\beta=0$ is mapped to the spherical top defined by $I_1=I_2=I_3$. The sphere with one squashing parameter, say $\beta=0$ and $\alpha \neq 0$ is the counterpart of the symmetric prolate top with $I_1=I_2\neq I_3$. The problem of finding the eigenvalues of the Laplacian for the sphere with two nontrivial squashing parameters is therefore reduced to finding the eigenvalues of the Hamiltonian of an asymmetric top with three different moments of inertia. This problem is studied to some extent in Section 103 of \cite{landau1977quantum} but since it may not be too familiar we discuss it below in some detail.

Let us start with the simplest case when $I_1=I_2=I_3=I$. The Hamiltonian in this case is reduced to $\hat{H}=\hat{L}^2/(2 I)$ for which the eigenvalues of an eigenvector $\psi$  are 
\begin{align}
 \hat{H}\psi= \frac{L(L+1)}{2I} \psi \ ,
\end{align}
with degeneracy\footnote{This is the degeneracy of the energy levels with respect to the $2L+1$ directions of the angular momentum relative to the body itself. There is another $(2L+1)$-fold degeneracy with respect to a fixed coordinate system. These are not really physical, but need to be taken into account when we calculate the partition function.} $2L+1$. Here $L$ is the rotational quantum number which is related to the quantum number $n$ in \eqref{eqn:deg1sq} by $2L+1=n$. 

It is also possible to find analytic expressions for the eigenvalues of the symmetric prolate top with $I_1=I_2$. In this case the Hamiltonian \eqref{eqn:Hamiltonian} can be rewritten as
\begin{align}
 \hat{H}=\frac{\hat{L}^2}{2I_1} + \frac12 \left(\frac{1}{I_3}-\frac{1}{I_1}\right) \hat{L}_3^2 \ . \label{eqn:Hamiltonianprolate}
\end{align}
Since $\hat{L}_3$ commutes with $\hat{L}^2$ it  has the same eigenvectors as $\hat{L}^2$, but with eigenvalues
\begin{align}
 \hat{L}_3\psi=k\psi, \qquad k=-L, \ldots , L \ .
\end{align}
The relation between the quantum number $k$ above and the quantum numbers $q$ and $n$ used in \eqref{eqn:deg1sq} is 
\begin{equation}
k = q+ \frac{1-n}{2}\;.
\end{equation}
With this at hand one can show that the eigenvalues of $\hat{H}$ for the symmetric prolate top are
\begin{align}
  \hat{H}\psi= \left[ \frac{L(L+1)}{2I_1} + \frac12 \left(\frac{1}{I_3}-\frac{1}{I_1}\right) k^2\right]\psi \ ,
\end{align}
Every such eigenvalue is doubly degenerate.

For the completely asymmetric top it is impossible to solve the eigenvalue problem analytically. This is because $\hat{L}_1$, $\hat{L}_2$ and $\hat{L}_3$ do not mutually commute. The degeneracy that was present in the previous examples, is now completely lifted. A possible resolution is to solve the eigenvalue equation in matrix form. This means that we have to find solutions of a secular equation of degree $2 L+1$. For general values of $L$ we therefore have to resort to numerical methods to find the eigenvalues. Luckily there are some symmetries that reduce the degree of the secular equation making it more tractable to solve numerically.

The matrix elements of the angular momentum operator can be found in many textbooks on quantum mechanics, see for example \cite{landau1977quantum}. The only non-zero matrix elements are
\begin{align}
 (\hat{L}_1)_{k, k-1}= (\hat{L}_1)_{k-1, k}&=\frac12 \sqrt{(L+k)(L-k-1)} \;, \\
  (\hat{L}_2)_{k, k-1}=- (\hat{L}_2)_{k-1, k}&=-i\frac12 \sqrt{(L+k)(L-k+1)} \;, \\
   (\hat{L}_3)_{k, k}&=k.
\end{align}
From these expressions it is not too difficult to see that the only non-zero elements of $\hat{L}_1^2$, $\hat{L}_2^2$ and $\hat{L}_3^2$ are those for which $k\rightarrow k$ or $k\rightarrow k\pm 2$. For a given fixed value of $L$ we have
\begin{align}
 (\hat{H})_{k,k}&= \frac12\left(  \frac{(\hat{L}_1^2)_{k,k}}{I_1}+\frac{(\hat{L}_2^2)_{k,k}}{I_2}+\frac{(\hat{L}_3^2)_{k,k}}{I_3}\right)=\frac14\left(\frac{1}{I_1}+\frac{1}{I_2}\right)\left(L(L+1)-k^2\right)+\frac{k^2}{2 I_3} \;, \\
  (\hat{H})_{k,k+2}&= (\hat{H})_{k+2,k}=\frac18 \left(\frac{1}{I_1}-\frac{1}{I_2}\right)\sqrt{(L-k)(L-k-1)(L+k+1)(L+k+2)} \;.
\end{align}
These matrices are the essential building blocks in our numerical analysis. To continue our simplification of the secular equations, we have to treat the case of integer and half-integer values of $L$ separately.

First, let us consider the case when $L$ can only take integer values. In this case the even and odd values of $k$ will never be mixed. This means that the secular equation for a given $L$ splits into a secular equation of degree $L$ and one of degree $L+1$
\begin{align}
 \textrm{det}(H_{k,k'}-E \delta_{k,k'})= \left. \textrm{det}(H_{k,k'}-E \delta_{k,k'})\right|_{k \textrm{ even}} \times \left. \textrm{det}(H_{k,k'}-E \delta_{k,k'})\right|_{k \textrm{ odd}} \ .
\end{align}
It is possible to lower the degrees of the secular equations even further. To this end we have to consider the matrix elements with respect to a new basis
\begin{align}
 \psi_{L_k}^+ &=\frac{\psi_{L_k}+\psi_{L_{-k}}}{\sqrt{2}} \;,\\
  \psi_{L_k}^- &=\frac{\psi_{L_k}-\psi_{L_{-k}}}{\sqrt{2}}\;, \qquad k\neq 0 \;.
\end{align}
This splits everything up in functions that are symmetric or anti-symmetric under sign change of $k$, which leads to another split in the two secular equations we have. At the end of the day in the new basis we have four independent matrices for which we have to find the eigenvalues. These are denoted by $O^+$, $O^-$, $E^+$ and $E^-$ where $O$, $E$ stands for odd or even respectively. Furthermore with $k^{\pm}$ we will distinguish between the eigenbasis spanned by $\psi^+$ or $\psi^-$.

The matrix elements in the new basis are then given by
\begin{align}
  (\hat{H})_{k^{\pm},k^{\pm}}&=\langle \psi_{L_k}^{\pm}| H|\psi_{L_k}^{\pm}\rangle\\
  &=\frac12\left(\langle \psi_{L_k}| H|\psi_{L_k}\rangle \pm \langle \psi_{L_k}| H|\psi_{L_{-k}}\rangle \pm \langle \psi_{L_{-k}}| H|\psi_{L_{k}}\rangle + \langle \psi_{L_{-k}}| H|\psi_{L_{-k}}\rangle\right)\\
  &=\begin{cases} 
      (\hat{H})_{k,k} & k\neq 1\\
      (\hat{H})_{1,1} \pm (\hat{H})_{1,-1} & k=1
    \end{cases} , 
\end{align}
and 
\begin{align}
 (\hat{H})_{k^{\pm},k+2^{\pm}}=\begin{cases}
                                (\hat{H})_{k,k+2} & k\neq 0,-2 \\
                               \sqrt{2} (\hat{H})_{0,2}
                               \end{cases} .
\end{align}
This means that, for a given $L$, the matrices we have to find the eigenvalues of, are 
\begin{align*}
 O^{\pm}=\begin{pmatrix}
          (\hat{H})_{1,1} \pm (\hat{H})_{1,-1} &  (\hat{H})_{1,3}& 0 & \ldots \\
           (\hat{H})_{1,3} &  (\hat{H})_{3,3} &  (\hat{H})_{3,5} &\ldots \\
           0&   (\hat{H})_{3,5} &  (\hat{H})_{5,5} & \ldots \\
           \ldots & \ldots &\ldots & \ldots  
         \end{pmatrix} ,
\end{align*}
\begin{align}
 E^{+}=\begin{pmatrix}
          (\hat{H})_{0,0} &  \sqrt{2} (\hat{H})_{0,2}& 0 & \ldots \\
           \sqrt{2} (\hat{H})_{0,2} &  (\hat{H})_{2,2} &  (\hat{H})_{2,4} &\ldots \\
           0&   (\hat{H})_{2,4} &  (\hat{H})_{4,4} & \ldots \\
           \ldots & \ldots &\ldots & \ldots 
         \end{pmatrix} \textrm{ and }   
         E^{-}=\begin{pmatrix}
          (\hat{H})_{2,2} &   (\hat{H})_{2,4}& 0 & \ldots \\
          (\hat{H})_{2,4} &  (\hat{H})_{4,4} &  (\hat{H})_{4,6} &\ldots \\
           0&   (\hat{H})_{4,6} &  (\hat{H})_{6,6} & \ldots \\
           \ldots & \ldots &\ldots & \ldots 
         \end{pmatrix}.
\end{align}

We still have to consider the half-integer values of $L$. In this case there is no transition between elements for which $k+1/2$ is even and for which this is odd, thus
\begin{align}
 \textrm{det}(H_{k,k'}-E \delta_{k,k'})= \left. \textrm{det}(H_{k,k'}-E \delta_{k,k'})\right|_{k+1/2 \textrm{ even}} \times \left. \textrm{det}(H_{k,k'}-E \delta_{k,k'})\right|_{k+1/2 \textrm{ odd}} \ .
\end{align}
The secular equation is now split into two secular equations of degree $L+1/2$. It is not difficult to see that the set of $k$'s spanned by one of the secular equations is just minus the one of the other secular equation. On the other hand one can also show that $(\hat{H})_{k,k}= (\hat{H})_{-k,-k}$ and $(\hat{H})_{k+2,k}= (\hat{H})_{-k-2,-k}$. Therefore the two secular equations give the same result, leading to doubly degenerate eigenvalues for the case of half-integer $L$. The matrix for which we have to find the eigenvalues of is then
\begin{align}
 HI=\begin{pmatrix}
      (\hat{H})_{-L,-L} &   (\hat{H})_{-L,-L+2} & 0 &\ldots\\
        (\hat{H})_{-L,-L+2} & (\hat{H})_{-L+2,-L+2} & \hat{H})_{-L+2,-L+4} & \ldots \\
        0&  (\hat{H})_{-L+2,-L+4} & (\hat{H})_{-L+4,-L+4} & \ldots\\
        \ldots & \ldots &\ldots & \ldots 
    \end{pmatrix} .
\end{align}
For the purposes of Section \ref{sec:CFT} we have implemented these eigenvalue problems into a numerical routine which produces all eigenvalues up to a certain quantum number $n=2L+1$. To ensure good convergence properties we had to choose values of $n$ that are of the order of 2000.

\end{appendices}

\bigskip
\bigskip

\bibliography{LiteratureSquashedS3}
\bibliographystyle{JHEP}

\end{document}